\documentclass[twocolumn]{aastex63}
\usepackage{url,amsfonts,epsfig}
\usepackage{graphicx}
\usepackage{amsmath}
\usepackage{amssymb}
\usepackage[section]{placeins}
\usepackage{chngpage}
\usepackage{calc}
\usepackage{subfigure}
\usepackage{mathtools}
\usepackage{comment}
\usepackage{appendix}
\usepackage{natbib}
\usepackage{graphicx}
\usepackage{color}
\usepackage{booktabs}

\makeatletter 
\renewcommand\@biblabel[1]{}

\begin{document}
\def\gap{\;\rlap{\lower 2.5pt
\hbox{$\sim$}}\raise 1.5pt\hbox{$>$}\;}
\def\lap{\;\rlap{\lower 2.5pt
 \hbox{$\sim$}}\raise 1.5pt\hbox{$<$}\;}

\title{Galaxy Core Formation by Supermassive Black Hole Binaries: the Importance of Realistic Initial Conditions and Galaxy Morphology
}

\author{Fani~Dosopoulou}\thanks{PCTS and Lyman Spitzer, Jr. fellow 
}\email{fanid@pricneton.edu}
\affil{Princeton Center for Theoretical Science, Princeton University, Princeton, NJ 08544, USA}
\affil{Department of Astrophysical Sciences, Princeton University, Princeton, NJ 08544, USA }
\author{Jenny~E. Greene}
\affil{Department of Astrophysical Sciences, Princeton University, Princeton, NJ 08544, USA }
\author{Chung-Pei~Ma}
\affil{Department of Astronomy and Department of Physics, University of California, Berkeley, CA 94720, USA}

\begin{abstract}
The binding energy liberated by the coalescence of supermassive black hole (SMBH) binaries during galaxy mergers is thought to be responsible for the low density cores often found in bright elliptical galaxies. We use high-resolution $N$-body and Monte Carlo techniques to perform single and multi-stage galaxy merger simulations and systematically study the dependence of the central galaxy properties on the binary mass ratio, the slope of the initial density cusps, and the number of mergers experienced. We study both the amount of depleted stellar mass (or ``mass deficit'), $M_{\rm def}$, and the radial extent of the depleted region, $r_{\rm b}$. We find that $r_{\rm b}\simeq r_{\rm SOI}$ and that $M_{\rm def}$ varies in the range $0.5$ to $4M_{\bullet}$, with $r_{\rm SOI}$  the influence radius of the remnant SMBH and  $M_{\bullet}$ its mass. The coefficients in these relations depend weakly on the binary mass ratio and remain remarkably constant through subsequent mergers. 
 We conclude that  the core size and mass deficit do not scale linearly with the number of mergers, making it hard to infer merger histories from observations.
On the other hand, we show that both $M_{\rm def}$ and $r_{\rm b}$ are sensitive to the morphology of the galaxy merger remnant, and that adopting spherical initial conditions, as done in early work, leads to misleading results.
Our models reproduce the range of values for $M_{\rm def}$ found in most observational work, but span nearly an order of magnitude range around the true ejected stellar mass. 
 \end{abstract}

\section{Introduction}
Observations reveal the existence of tight correlations between the SMBH mass and the bulge luminosity \citep{1993nag..conf..197K,1995ARA&A..33..581K,1998AJ....115.2285M} and   velocity dispersion  \citep{2000ApJ...539L...9F,2000ApJ...539L..13G}. This leads to the widespread belief that SMBHs and bulges coevolve by regulating each other’s
growth \citep{2013ARA&A..51..511K}, while the above and other correlations  support the growing evidence that SMBH growth and galaxy
evolution are intertwined.

Modelling the dynamical evolution of SMBH binaries inside galactic nuclei is central to the
astrophysical interpretation of galaxy properties and the observed correlations between core and global
properties of galaxies and for the interpretation of  gravitational waves from inspiraling SMBHs \citep{2013MNRAS.429.3155A}. Preceding the final coalescence, the interaction of the SMBH binary with the surrounding stellar environment has various
observable consequences, possibly explaining properties
of both quiescent and active galaxies \citep{2003AIPC..686..161K}, including
AGN variability \citep{2000ApJ...531..744V}, the bending
and precession of radio jets \citep{1993ApJ...409..130R,2000A&A...360...57R} and
X- and
Z-shaped radio lobes \citep{2002Sci...297.1310M,2003ApJ...594L.103G}. The evolution
driven by SMBH binaries could also provide a plausible explanation – albeit one that is difficult to prove – of an
almost ubiquitous feature observed in the brightest elliptical galaxies and bulges, namely a slowly rising or flat density profile within their central regions.
 The suggested explanation is that the core of these galaxies  may be scoured by the SMBH binaries that
form during mergers \citep{1980Natur.287..307B,1991Natur.354..212E,1996ApJ...465..527M,1997NewA....2..533Q,2001ApJ...563...34M,2002MNRAS.331L..51M,2006ApJ...648..976M,2007MNRAS.374.1227B,2018ApJ...864..113R,2019ApJ...872L..17R}.

\textcolor{black}{
One finding of most observational studies is the
existence of scaling relations between nuclear properties of galaxies and SMBHs.
Most studies find that the missing mass from the galaxy core, or ``mass deficit'', $M_{\rm def}$,  correlates well with the mass of the central SMBH, $M_{\bullet}$, with
$M_{\rm def}\sim M_{\bullet}$ \citep[e.g.,][]{2003AJ....125.2951G,2006ApJ...644L..21F,2007ApJ...662..808L,2014MNRAS.444.2700D}.
Furthermore, \citet{2016Natur.532..340T} found 
 a tight correlation between the radius, $r_{\rm b}$, of the galaxy core in the observed light profile and the radius of the SMBH sphere of influence, $r_{\rm SOI}$. The best fit relation is consistent with $r_{\rm b}= r_{\rm SOI}$, which \citet{2016Natur.532..340T}  interpret as the dynamical imprint of the SMBHs. However, it is hard to rigorously test whether these correlations are consistent with a SMBH scouring hypothesis.
}

{One complication is that we do not know the original light profile shape, so we do not know whether the power-law slope was depleted into a core by black hole scouring, or alternatively whether  cores were present in all galaxies initially, and that power-law cusps were generated by the growth of central SMBHs \citep{1999AJ....117..744V}.  Furthermore, it is hard to get stable measurements of $M_{\rm def}$, and different observational studies often find different values, likely largely because they make different assumptions about the original profile \citep{2003AIPC..686..161K,2013ApJ...768...36D,2014MNRAS.444.2700D}. }

On the theoretical side, the formation of cores during galaxy mergers has been studied by many authors \citep[e.g.,][]{2001ApJ...563...34M,2002MNRAS.331L..51M,2006ApJ...648..976M,2007MNRAS.374.1227B,2018ApJ...864..113R,2019ApJ...872L..17R,2021MNRAS.502.4794N}. 
\citet{2006ApJ...648..976M} used $N$-body simulations to follow the evolution of a spherical galaxy model containing a central SMBH binary. \citet{2006ApJ...648..976M} found that the mass deficit at the end of the SMBH evolutionary phase is proportional to
the total mass of the binary, $M_{\rm def} \approx 0.5M_{\bullet}$, with only a weak
dependence on the merger mass ratio or the initial density profile. Running additional simulations of multistage galaxy mergers, \citet{2006ApJ...648..976M}  finds that the effect of a SMBH binary  on a spherical nucleus is cumulative, scaling roughly in proportion both to the number of mergers, $N$, and to the final SMBH mass, $M_{\rm def} \approx 0.5 N M_{\bullet}$.

Several alternative models for the formation of depleted cores are recently gaining attention due to their ability to produce
even larger cores, which could explain the extremely large cores and hence, depleted masses,  found in a handful of galaxies  \citep[i.e., core radius $>1 \rm Kpc$; e.g.,][]{2003AJ....125..478L,2007ApJ...662..808L,2008MNRAS.391.1559H,2016ApJ...829...81B,2019ApJ...886...80D}.
These models include  the  “recoiled
SMBH” scenario  \citep[e,g,][]{1989ComAp..14..165R,2004ApJ...613L..37B,2004ApJ...607L...9M,2008ApJ...678..780G,2021MNRAS.502.4794N}; 
  the “stalled perturber” scenario \citep{2006MNRAS.373.1451R,2010ApJ...725.1707G};
the “multiple-SMBH scouring” scenario  \citep{2012MNRAS.422.1306K}; and the combined “sinking SMBH—AGN feedback” scenario \citep{2012MNRAS.420.2859M}.

Because most of the literature relies on simplified 
model assumptions (e.g., spherical and isotropic initial conditions, isolated galaxy models), 
the effect of  mergers on the central regions of galaxies has not been fully understood. The main reason
for this is that the numerical techniques employed in most work are based either on  direct $N$-body simulations \citep[e.g.,][]{1991Natur.354..212E,2001ApJ...563...34M,2006ApJ...648..976M,2012ApJ...744...74G}, which are still too slow to conduct a full parameter space exploration, or on Fokker-Planck and other statistical methods, which have limited applicability due to their basic assumptions \citep{2007ApJ...671...53M}.
Here we build on former studies 
 and use a combination of $N$-body and Monte Carlo simulations
 to rapidly and accurately explore the destruction of density cusps around SMBHs in realistic mergers of elliptical galaxies. We aim to systematically determine how the remnant core properties depend on the density profile of the galaxy progenitors, the initial galaxy orbits, the mass of their central SMBHs, the number of mergers, and the remnant galaxy morphology. We  use these models as a test to the scouring hypothesis by comparing our results to the observed $M_{\rm def}-M_{\rm \bullet}$ and 
  $r_{\rm b}-r_{\rm SOI}$ correlations.

This paper is structured as follows. In Section 2 we describe the initial conditions of our $N$-body models and the different algorithms followed to study each evolutionary stage of the merger process. In
Section 3 we  describe  the  different methods we use to calculate the core radius and mass deficit in our models, and compare to methods used in previous work.
In Section 4 and 5 we discuss our results and compare them to previous literature and to observations.  In Section 6 we present our conclusions and a short summary of our results.

\section{Models and methods}
\label{sec:models}

\begin{figure*}
 \includegraphics[angle=0,width=\textwidth]{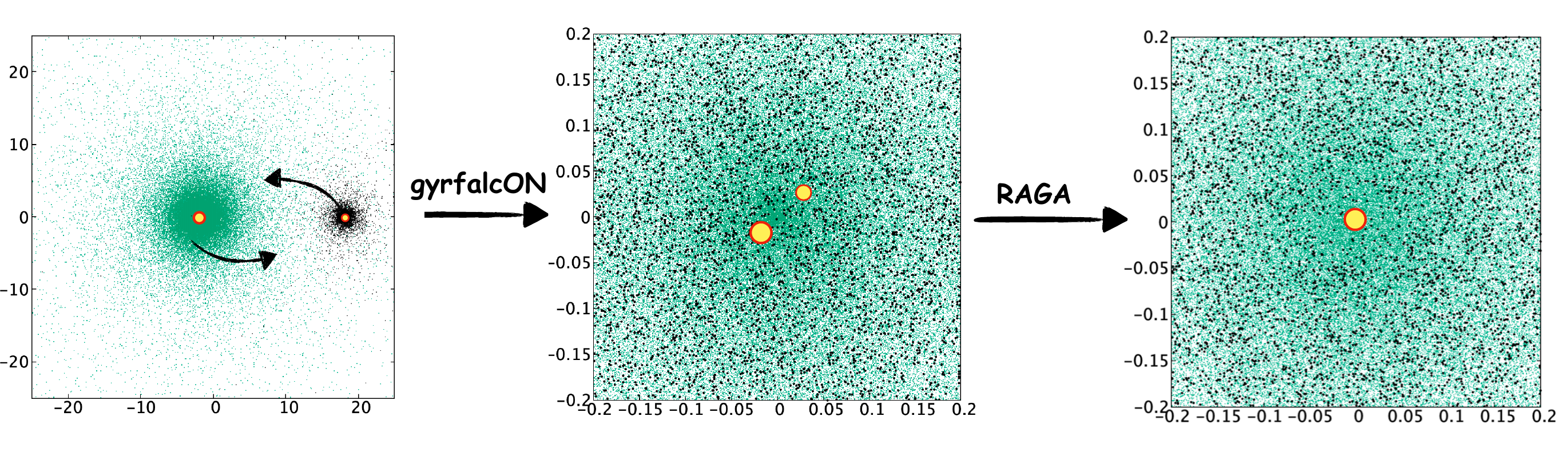}
\caption{Schematic representation of the two-step numerical technique adopted in this study. We first use the tree code {\tt gyrfalcON} to simulate the galaxy merger from its earliest stages (left panel) to the formation of a close SMBH binary (middle panel). We then continue the evolution of the massive binary using the Monte Carlo code {\tt RAGA} until the binary reaches approximately $a \approx a_{\rm h}/10$,  with $a_{\rm h}$ defined in equation (\ref{hard}), at which point the integration is terminated and the central properties of the galaxy are examined (right panel). Green (black) dots are particles belonging to the primary (secondary) galaxy. Yellow filled red circles represent the SMBHs. In this example $\gamma=1.5,e_{0}=0.6,q=0.1$. }
\label{snapshot}
\end{figure*}
In this section we describe the initial conditions of our $N$-body models and the numerical methods we adopt. 

We simulate the merger of two galaxies, each of which contains a central point mass representing a SMBH. We follow the details of the merger from its earliest stages, when the two galaxies are distinct, to its late stages, when the SMBHs have formed a hard binary and the binary has decayed to one tenth of the binary hardening radius via energy exchange with surrounding stars.

For reasons that we will describe below we break the merger evolution in two parts, before and after the formation of the SMBH binary, and use different algorithms for each evolutionary stage. The initial models are self-consistent realization of galaxies with central density cusps of the form $\rho \propto r^{-\gamma}$. Two important free parameters in this problem are the mass-ratio $q=M_{2}/M_{1}$ of the binary black hole and the slope $\gamma$ of the initial density cusp surrounding the black holes. Here $M_1$ and $M_2$ refer to the mass of the primary and secondary SMBH particles, respectively. Results are presented for several combinations of $q$ and $\gamma$ (Table \ref{tab:table1}).

\subsection{Galaxy models}
\label{sec:initial}
Initially,  steady state galaxy models are constructed with Monte Carlo realizations using Dehnen's (1993) density law with an additional central point mass representing a SMBH. Stellar positions and velocities are generated from the unique isotropic phase-space distribution function that reproduces Dehnen's $\rho(r)$ in the combined gravitational potential of the stars and the central point mass. 
This phase space stellar distribution ensures that the initial conditions are in equilibrium and do not evolve significantly in time.

The Dehnen model density $\rho(r)$ is given by
\begin{equation}
\rho(r)=\frac{(3-\gamma)M_{\rm tot}}{4\pi} \frac{r_{0}}{r^{\gamma} (r+r_{0})^{4-\gamma} },
\end{equation}
where $M_{\rm tot}$ is the total galaxy mass, $\gamma$ the central density slope and $r_{0}$ the  scale length.

We set the mass of the central SMBH  to 0.01 of the total galaxy mass. There are three main reasons why we picked this value:
(i) it is similar to what found for massive ellipticals  \citep[e.g., bottom panel of Figure 18 in ][]{2013ARA&A..51..511K,2016AAS...22711901R}, although it is larger than the {\it mean} observational value of $0.2-0.3\%$ 
 \citep[e.g.,][]{2013ApJ...764..184M,2016AAS...22711901R}; (ii)
it allows us to have enough particles to resolve well the sphere of influence of the SMBHs in our models. For $N=1.0 \times 10^{6}$ particles, if we had set $M_{1}/M_{\rm tot}=0.001$, we would only have 1000 (100) particles within the sphere of influence of the central SMBH for the q=1 (0.1) model; and (iii) this is the $M_{\rm 1}/M_{\rm tot}$
value adopted in many other theoretical papers \citep{2006ApJ...648..976M,2008ApJ...678..780G,2018ApJ...864..113R}. Thus, this choice allows us to make a one-to-one comparison to previous literature. In the future, we plan to study whether and how our results would be affected by choosing a different value for $M_{1}/M_{\rm tot}$.

\begin{table*}
  \begin{center}
    \caption{INITIAL PARAMETERS OF GYRFALCON MODELS}
    \label{tab:table1}
    \begin{tabular*}{\textwidth}{@{\extracolsep{\fill}}|llll|llll|llll|}
      \hline 
      \textbf{Run} & \textbf{$\gamma$} & \textbf{$q$} & \textbf{$r_{\rm h}$} & \textbf{$v_{y}$} & \textbf{$r'_{\rm h}$} & \textbf{$a_{\rm h}$}  & \textbf{$t_{\rm b}$} & \textbf{$v_{y}$} & \textbf{$r'_{\rm h}$} & \textbf{$a_{\rm h}$}  & \textbf{$t_{\rm b}$}\\ 
      \hline 
      
       
       &    &      &       & & & ${e}_{0}=0.9$ &  &     &  & ${e}_{0}=0.6$  &  \\     
         \hline

      1 & 0.5 & 0.1 & 0.265 & 0.07  & 0.360 & 0.0074  & 260 &  0.15  & 0.360 & 0.0074  & 600\\
      2 & 0.5 & 0.25 & 0.265 & 0.08   & 0.304 & 0.0117    & 170 &  0.16  & 0.306 & 0.0122  & 350\\ 
      3 & 0.5 & 0.5 & 0.265 & 0.09    & 0.372 & 0.0207   & 143 &  0.18  & 0.392 & 0.0218  & 265\\
      4 & 0.5 & 1.0 & 0.265 & 0.1 & 0.413 & 0.0258   & 135 &  0.2  & 0.440 & 0.0275  & 210\\
      5 & 1.0 & 0.1 & 0.164 & 0.07   & 0.239 & 0.0049  & 243 &  0.15  & 0.242 & 0.0050  & 600\\
      6 & 1.0 & 0.25 & 0.164 & 0.08    & 0.241 & 0.0096  & 153 &  0.16  & 0.245 & 0.0074  & 340\\
      7 & 1.0 & 0.5 & 0.164 & 0.09   & 0.264 & 0.0147  & 141 &  0.18  & 0.262 & 0.0145  & 262\\
      8 & 1.0 & 1.0 & 0.164 & 0.1    & 0.279 & 0.0174  & 110 &  0.2  & 0.296 & 0.0185  & 200\\
      9 & 1.5 & 0.1 & 0.0798 & 0.07   & 0.125 & 0.0026  & 210 &  0.15  & 0.128 & 0.0026  & 613\\
      10 & 1.5 & 0.25 & 0.0798 & 0.08   & 0.132 & 0.0053  & 150 &  0.16  & 0.132 & 0.0053  & 365\\
      11 & 1.5 & 0.5 & 0.0798 & 0.09    & 0.140 & 0.0078  & 120 &  0.18  & 0.144 & 0.0080  & 280\\
      12 & 1.5 & 1.0 & 0.0798 & 0.1   & 0.163 & 0.0102  & 103 &  0.2  & 0.165 & 0.0103  & 208\\
      \hline
    \end{tabular*}

\begin{flushleft} 
Initial and final parameters of the {\tt gyrfalcON} runs for $e_{0}=0.9$ and $e_{0}=0.6$. The initial relative velocity of the secondary SMBH is given by $v_{y}$. The {\tt gyrfalcON} simulations are stopped at $t=t_{\rm b}$ when a close massive binary forms and the galaxies are well merged. The influence radius of the primary galaxy SMBH, $r_{\rm h}$, is given at $t=0$. The influence radius of the massive binary, $r'_{\rm h}$, and its hardening radius, $a_{\rm h}$, are given at $t=t_{\rm b}$.
\end{flushleft} 
  \end{center}
\end{table*}

\begin{table*}
  \begin{center}
    \caption{INITIAL PARAMETERS OF RAGA MODELS}
    \label{tab:tableraga}
    \begin{tabular*}{\textwidth}{@{\extracolsep{\fill}}|lll|lll|lll|}
      \hline
      \textbf{Run} & \textbf{$\gamma$} & \textbf{$q$} & \textbf{$a$} & \textbf{$e$}  & \textbf{$t_{\rm f}$} & \textbf{$a$} & \textbf{$e$}  & \textbf{$t_{\rm f}$}\\ 
      \hline 
       
         &    &      &       & ${e}_{0}=0.9$ &     &  & ${e}_{0}=0.6$  &  \\     
         \hline
      
      1 & 0.5 & 0.1  & 0.04  & 0.90  & 2015 &  0.03  & 0.20  & 3005\\
      2 & 0.5 & 0.25  & 0.01  &  0.89    & 415 & 0.01  & 0.22 & 350\\ 
      3 & 0.5 & 0.5  &  0.03   &  0.90  & 130 & 0.02  & 0.37  & 225\\
      4 & 0.5 & 1.0  & 0.02 & 0.90  & 105 &  0.02 & 0.44 & 220\\
      5 & 1.0 & 0.1  & 0.01   & 0.91  & 1405 & 0.02 & 0.25  & 1150\\
      6 & 1.0 & 0.25  & 0.03  & 0.90  & 160 & 0.02   & 0.25  & 350\\
      7 & 1.0 & 0.5  & 0.01  & 0.90  & 90 & 0.02  & 0.21 & 165\\
      8 & 1.0 & 1.0  & 0.02  & 0.57  & 60 & 0.02 & 0.24  & 160\\
      9 & 1.5 & 0.1  & 0.01  & 0.86  & 545 &  0.01  & 0.24  & 575\\
      10 & 1.5 & 0.25  & 0.01  & 0.89  & 80 & 0.01   & 0.53 & 120\\
      11 & 1.5 & 0.5  & 0.01   & 0.90 & 35 & 0.01  & 0.32 & 55\\
      12 & 1.5 & 1.0  & 0.01  & 0.90 & 30 &  0.01  & 0.39  & 50\\
      \hline
    \end{tabular*}

\begin{flushleft} 
Initial and final parameters of the {\tt RAGA} runs for $e_{0}=0.9$ and $e_{0}=0.6$. The initial binary semi-major axis and eccentricity is given by $a$ and $e$. The {\tt RAGA} simulations are stopped at $t=t_{\rm f}$ when $a$ has reached a separation $\approx a_{\rm h}/10$. 
\end{flushleft} 
  \end{center}
\end{table*}

Four parameters suffice to define the initial models: the SMBH binary mass ratio $q$, the central density slope of the Dehnen models, $\gamma$, the number of particles $N$, and the initial orbital eccentricity, $e$. 
 We choose units in which the total mass in stars of the primary galaxy, its  scale length and the gravitational constant $G$ are all equal to one, i.e., $M_{\rm tot}=r_{0}=G=1.0$.
In these units,  the secondary galaxy mass and scale radius are set to  $M_{\rm tot,2}= q$ and $r_{0,2}= \sqrt{q}$, respectively. 
The primary and secondary galaxies in our merger simulations have the same value of $\gamma$, for which we consider three values: $\gamma=(0.5, 1.0, 1.5)$. For each choice of $\gamma$ four values of $q$ are used: $q=(0.1, 0.25, 0.5, 1.0)$. Finally, each of these 12 models is integrated using a  number of particles $N_{1}= 1.0\times 10^{6}$ and  $N_2= q\times 10^{6}$ in the primary and secondary galaxies, respectively.

Initially, in the merger models the two galaxies are placed at the apocenter of an elliptical orbit with a separation between their centers of $d=20$. The secondary galaxy is given an initial relative velocity $v_{y}$ perpendicular to the line joining the two central black holes and such that  the initial eccentricity of the  orbit is either $e_{0}=0.9$ or $e_{0}=0.6$. Table \ref{tab:table1} gives the initial conditions of our {\tt gyrfalcON} models as well as the influence radius of the  SMBH in the initial model of the primary galaxy, defined as
\begin{equation}
M(r_{\rm h})=2M_1~ .
\end{equation}

We break the problem into two parts, using the tree code {\tt gyrfalcON}  \citep{2000ApJ...536L..39D,2002JCoPh.179...27D} for the early stages of the merger until the formation of the SMBH binary  and the Monte-Carlo code {\tt RAGA} \citep{2015MNRAS.446.3150V} for the later stages. We use {\tt RAGA} in the second stage of evolution for three reasons: (i) it allows us to completely turn off the effect of two-body gravitational encounters between stars (i.e., two-body relaxation), which for the  galaxies considered is negligible; (ii) we can study the binary evolution very efficiently without being limited by numerical resolution or a finite softening length; and (iii) we can easily explore the effect of the galaxy shape on the evolution of the binary and the galaxy core properties.
Note that the combination of all these elements is a unique feature of the present work.

We note finally that our simulations preclude non-dynamical processes such as gas-driven dissipation and do not include a dark-matter component. We believe that both approximations are reasonable. We are looking at the central region of gas-free ellipticals, where the enclosed baryonic  component mass is many times larger than the dark matter mass - but see \cite{2019ApJ...872L..17R,2021MNRAS.502.4794N} for detailed models including a dark matter component.
Note, also, that if the seeds of the current SMBHs were present at large redshifts, the evolutionary history  of a bright galaxy could involve tens of gas-rich mergers. But,
star formation in a dissipational merger can easily regenerate a density cusp after its destruction by a binary if a large amount of gas is present \citep[e.g.,][]{2009ApJS..181..135H}. It is therefore reasonable to think that the  
 more relevant quantity is the number of mergers since the era at which most of the gas was depleted.

A schematic representation of the two-step numerical method adopted  by us is shown in Figure \ref{snapshot}.

\begin{figure*}
\begin{center}
 \includegraphics[angle=0,width=2.24in]{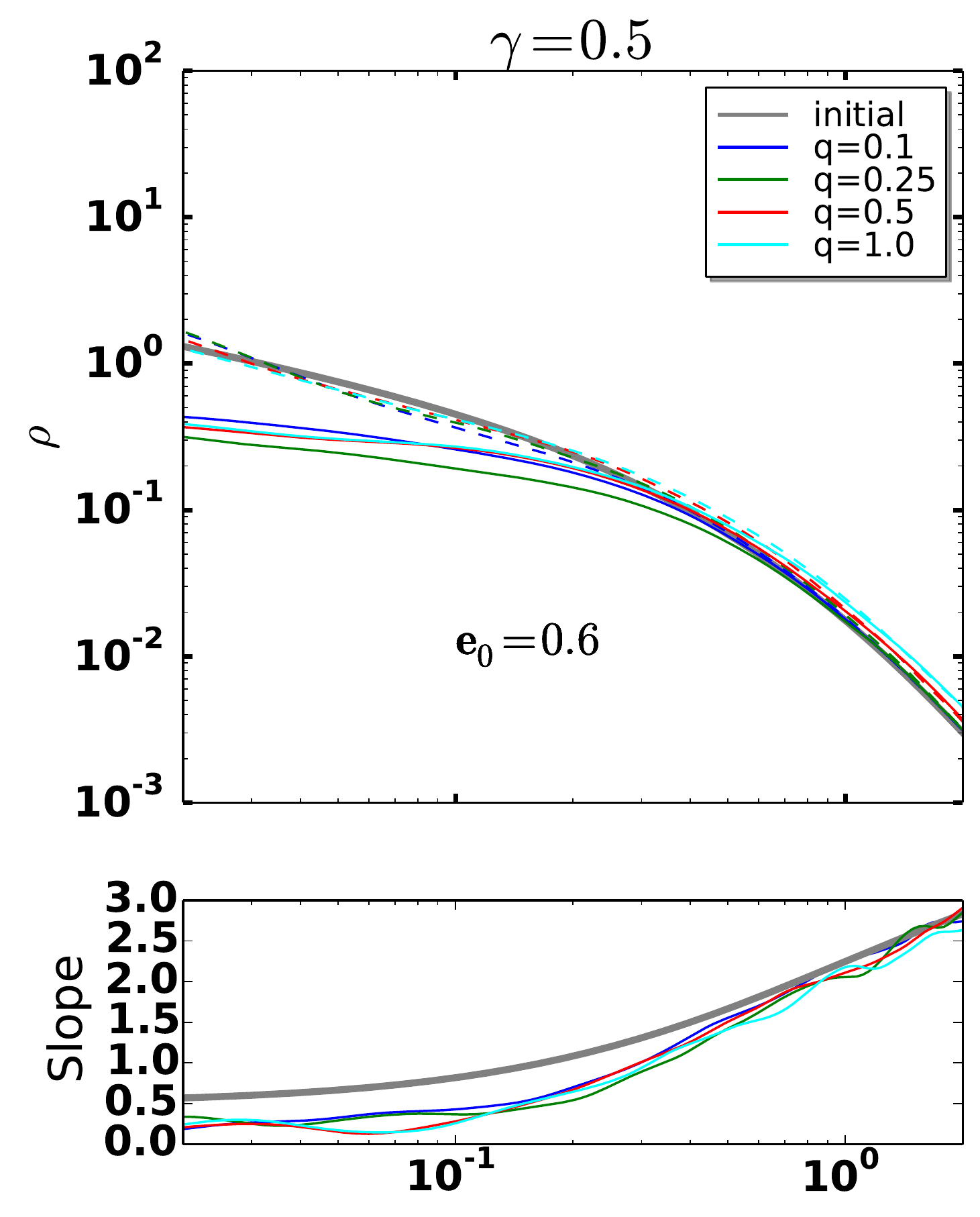}
   \includegraphics[angle=0,width=2.24in]{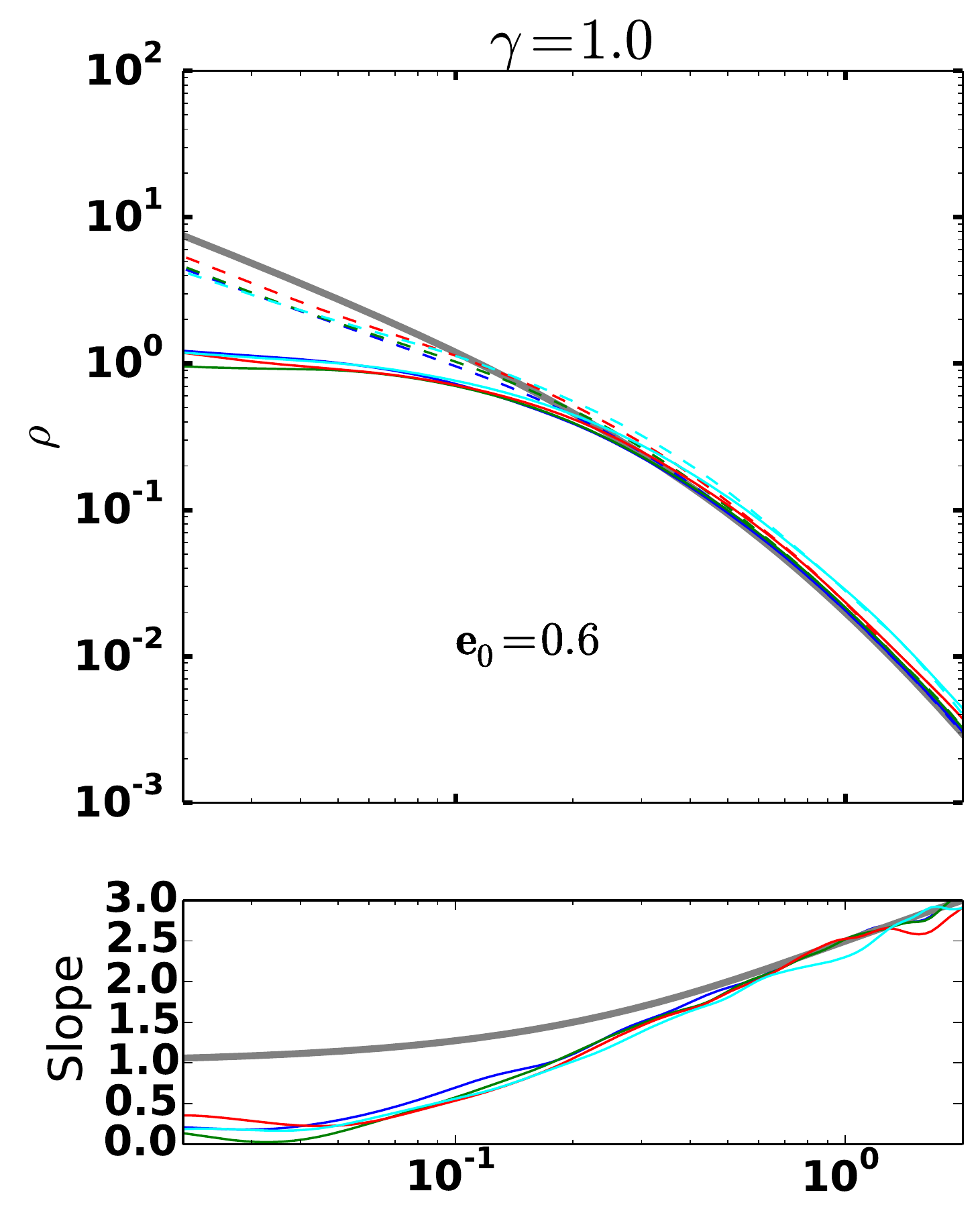}
   \includegraphics[angle=0,width=2.24in]{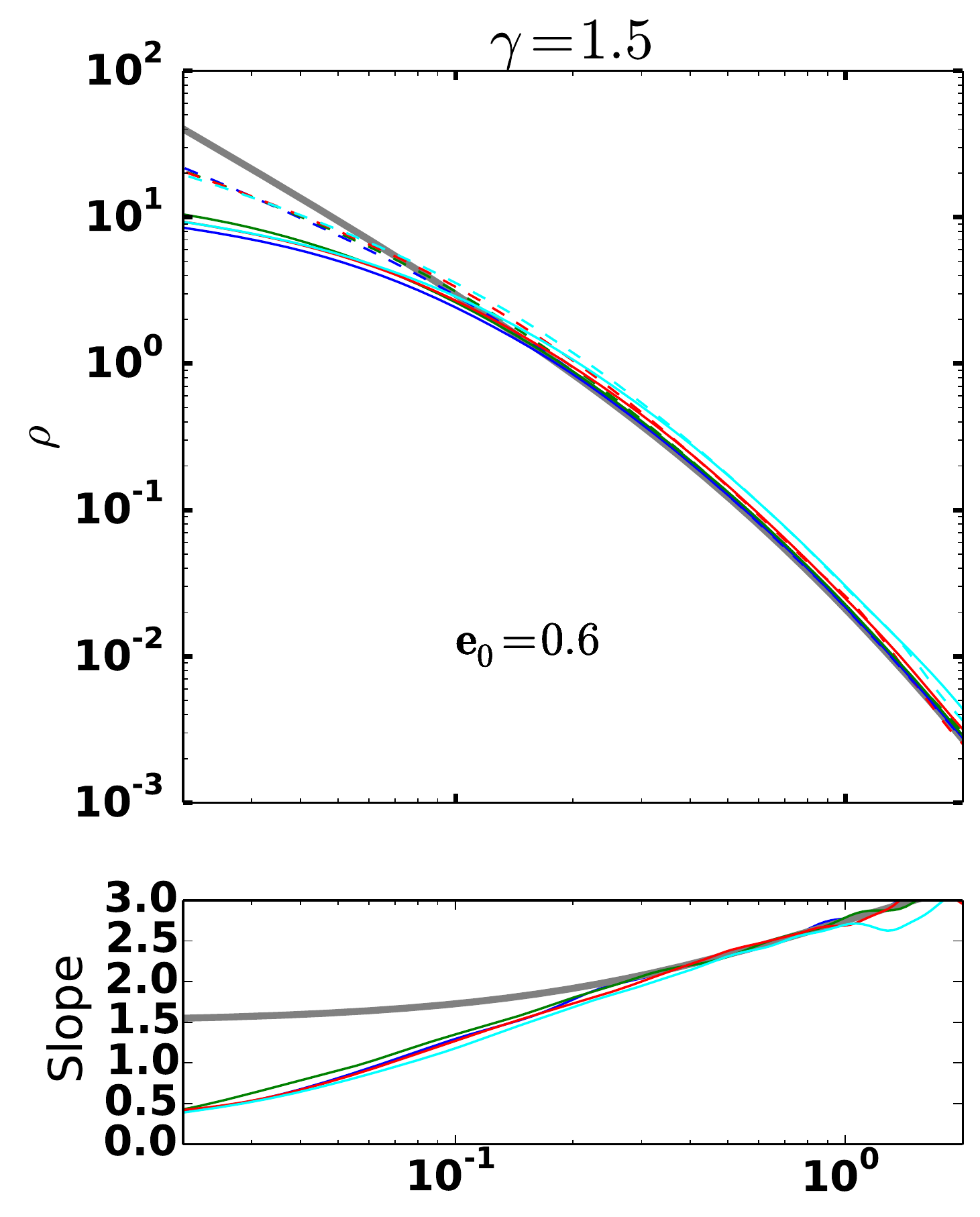}
   \includegraphics[angle=0,width=2.24in]{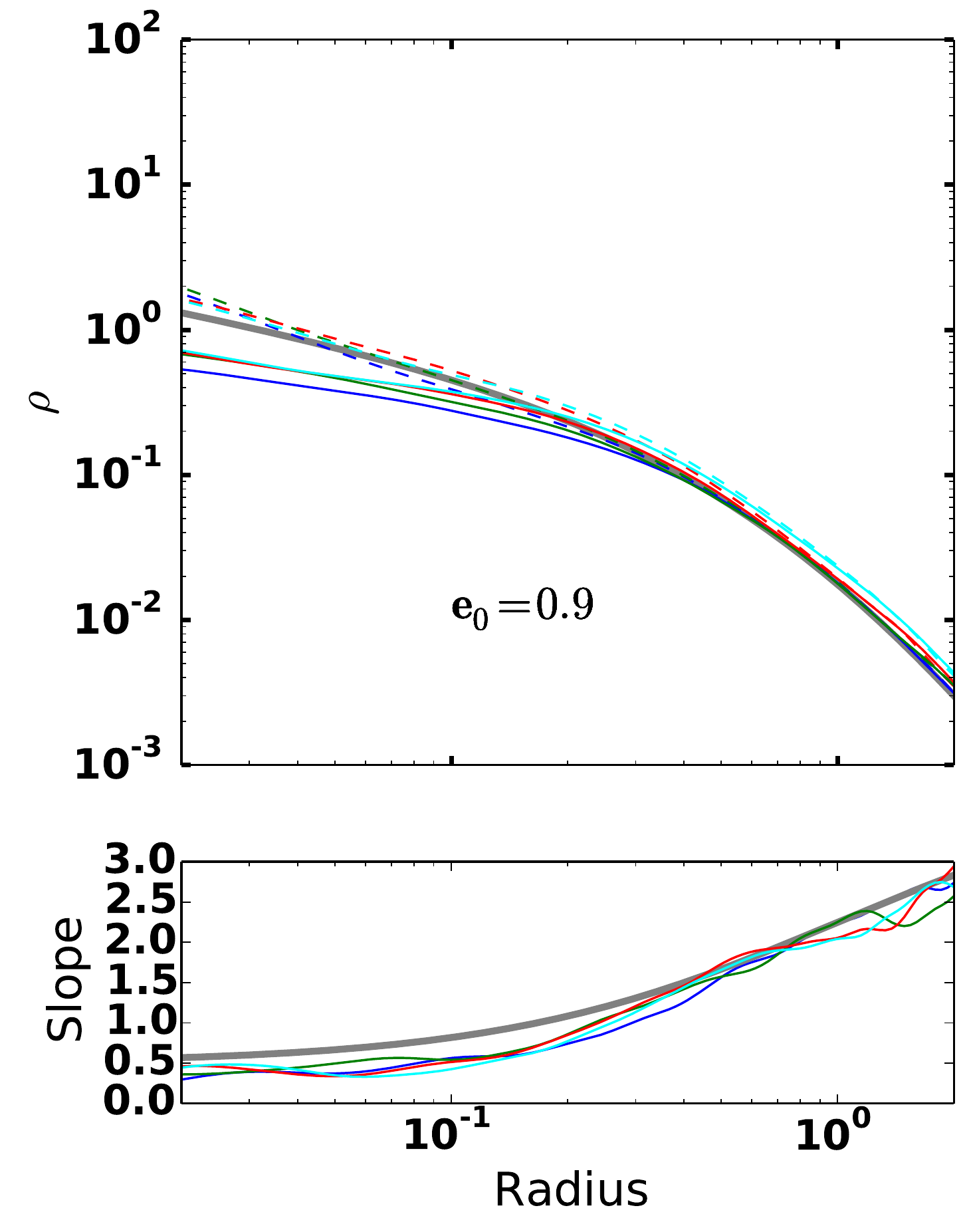}
   \includegraphics[angle=0,width=2.24in]{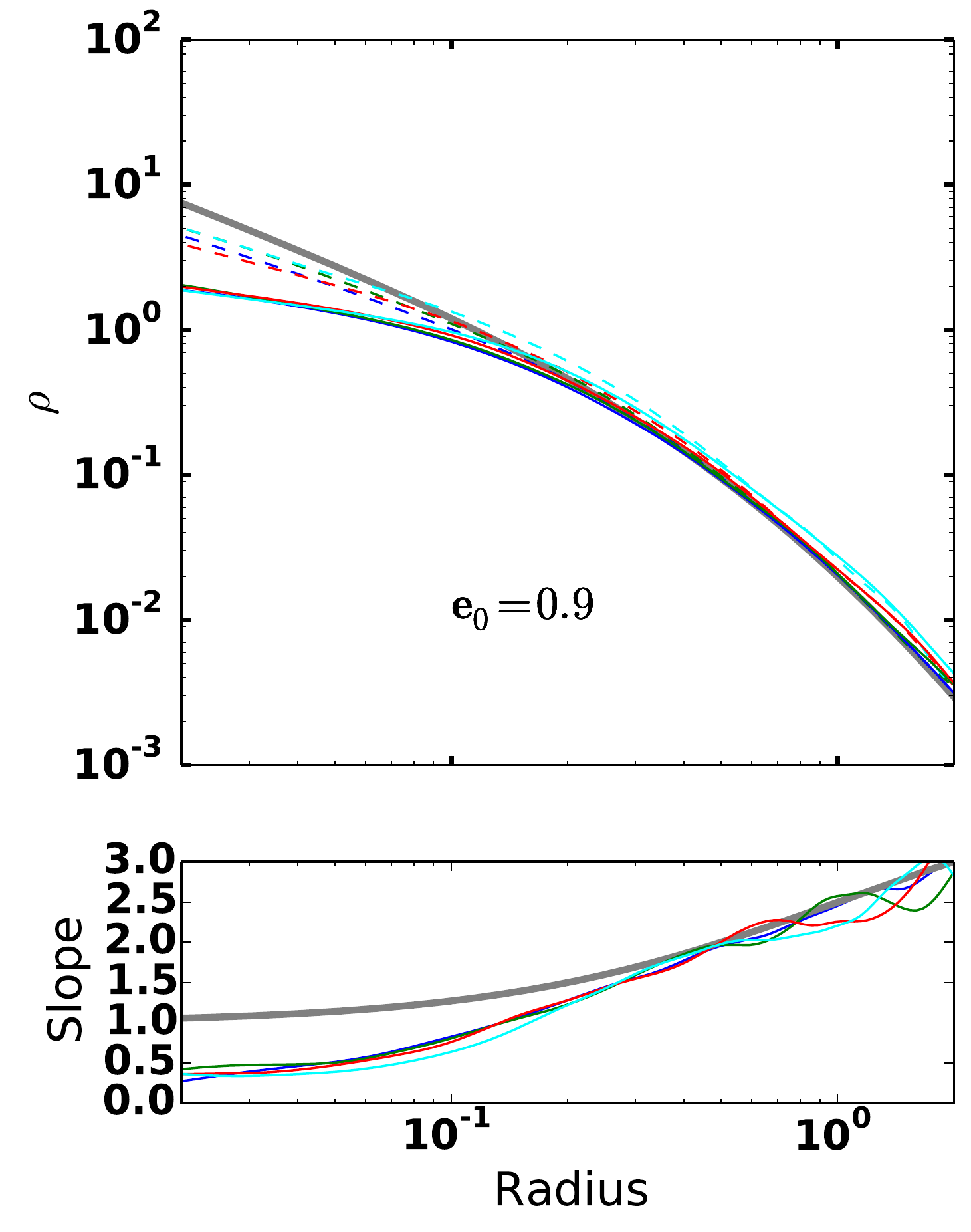}
   \includegraphics[angle=0,width=2.24in]{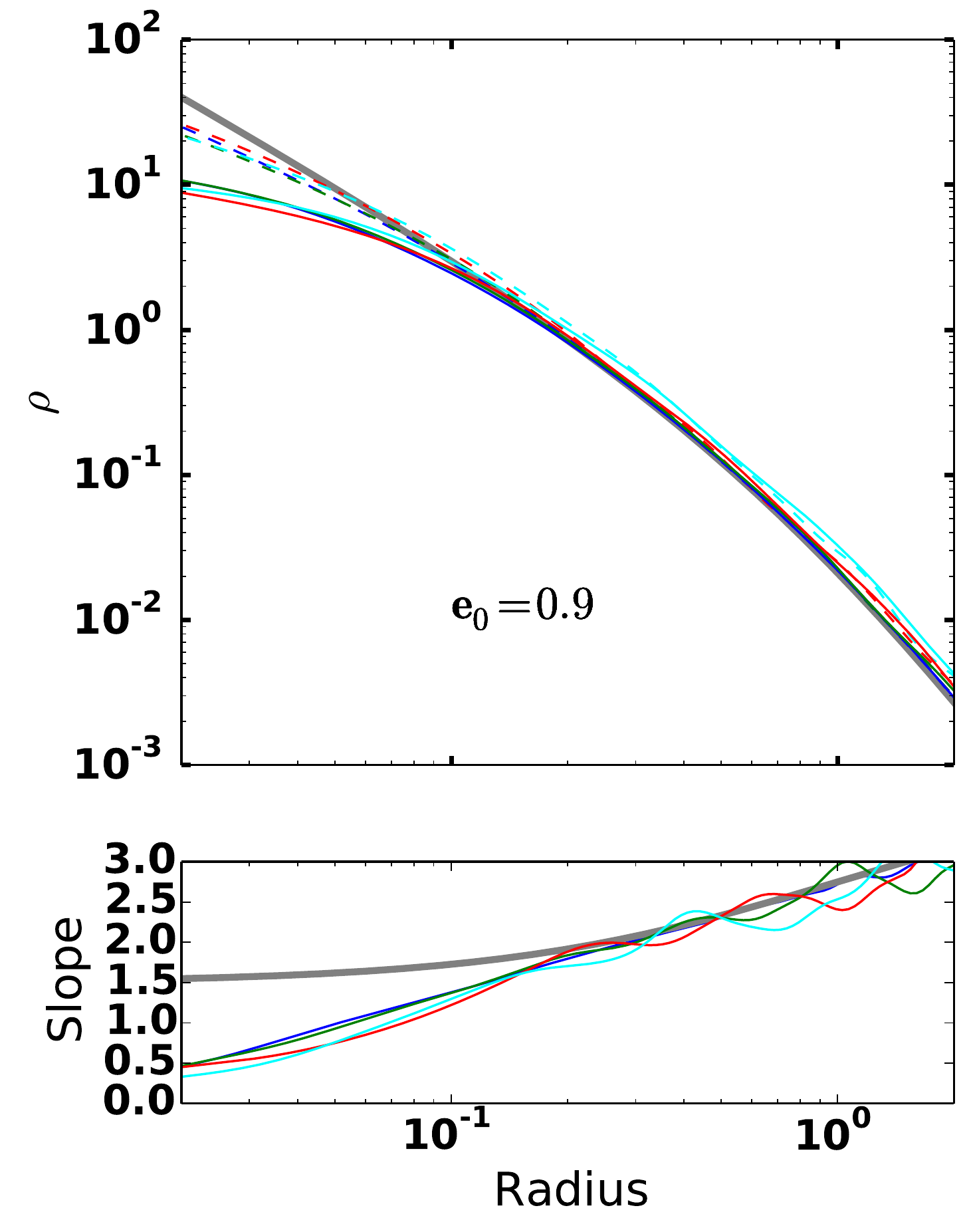}
  \caption{Density profiles ($\rho$) when $a \approx a_{\rm h}/10$ (end of {\tt RAGA} runs) for initial eccentricity $e_{0}=0.6$ (upper panel)  and  $e_{0}=0.9$ (lower panel). Three different initial density slopes, $\gamma=0.5$ (left), 1.0 (middle), and 1.5 (right) are used. For each $\gamma$, four values of $q$ are studied $q=(0.1, 0.25, 0.5, 1.0)$ (color scheme). Dashed lines in the upper plots of each panel represent the density profiles at the end of the {\tt gyrfalcON} runs when a close binary has already been formed and the galaxies have merged. Solid grey lines in the upper plots depict the initial density profile of the host galaxy ($\rho_{i}$).
  Lower panels show the radial dependence of the density slope, $-{\rm d \log \rho/{\rm d}\log r}$ with the grey solid line depicting here the density slope of the primary galaxy profile $-{\rm d \log \rho_{i}/{\rm d}\log r}$.}
\label{com_2}
\end{center}
\end{figure*}

\begin{figure*}
\begin{center}
 \includegraphics[angle=0,width=2.24in,height=2.09in]{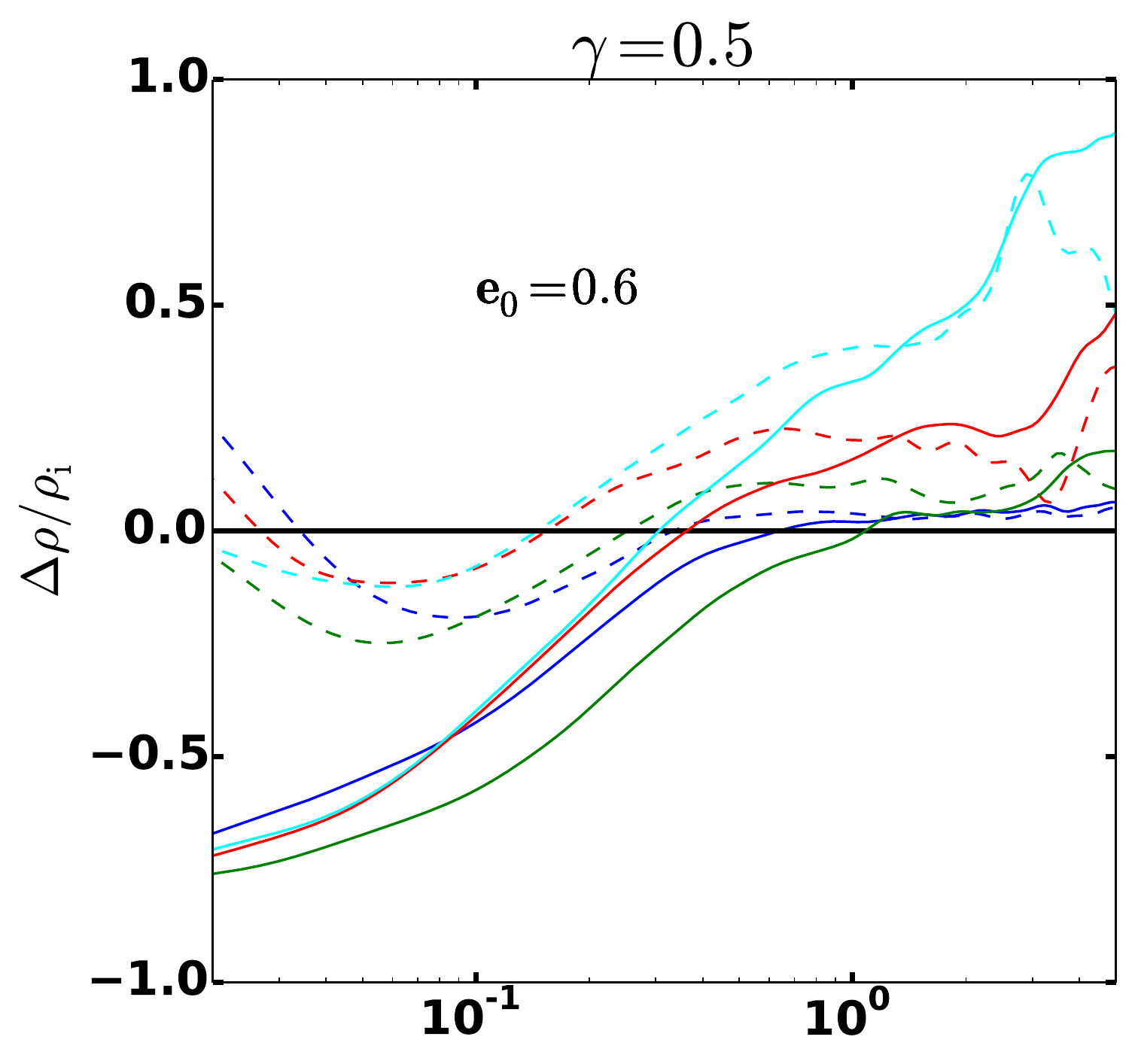}
   \includegraphics[angle=0,width=2.24in]{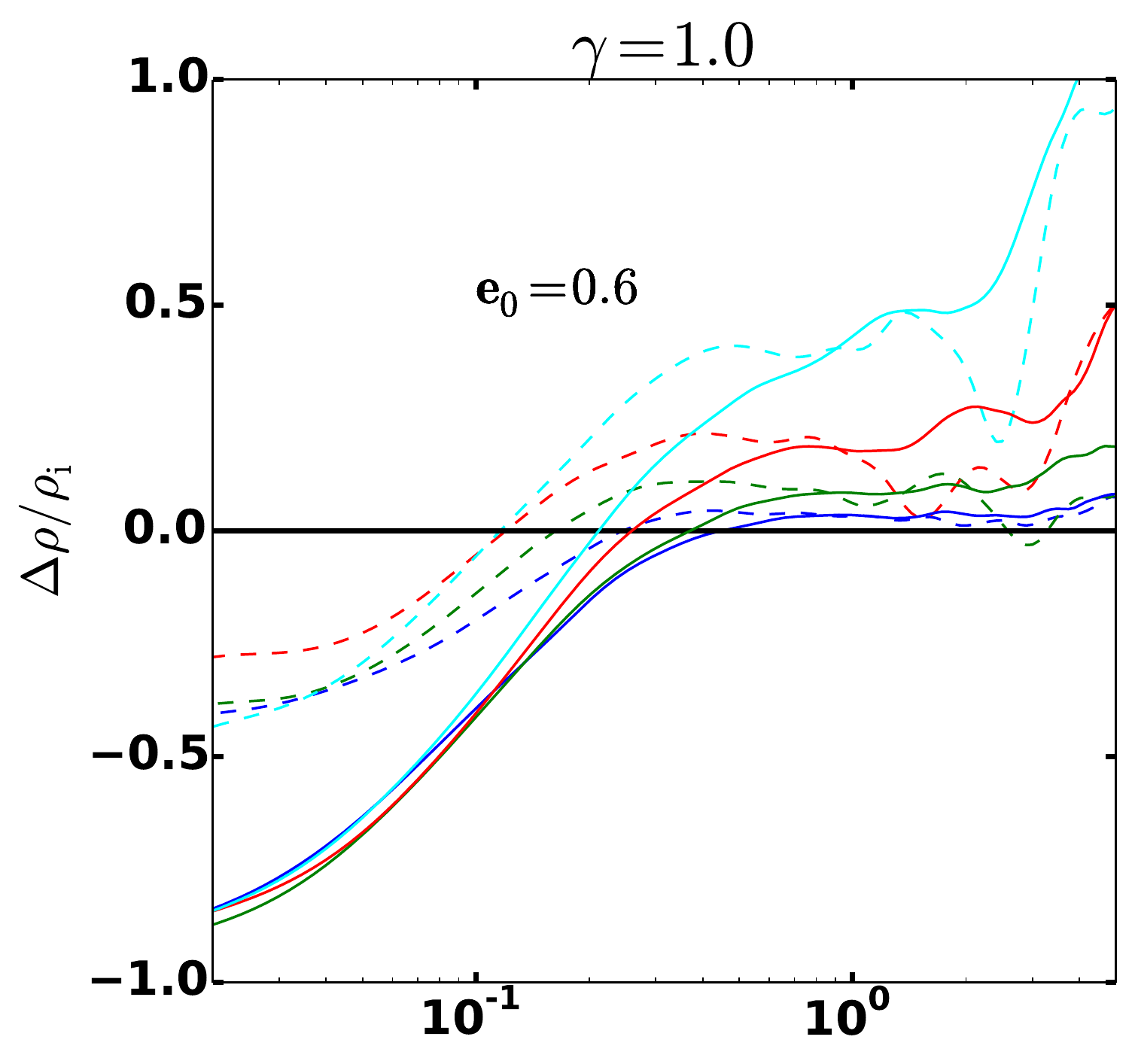}
   \includegraphics[angle=0,width=2.24in]{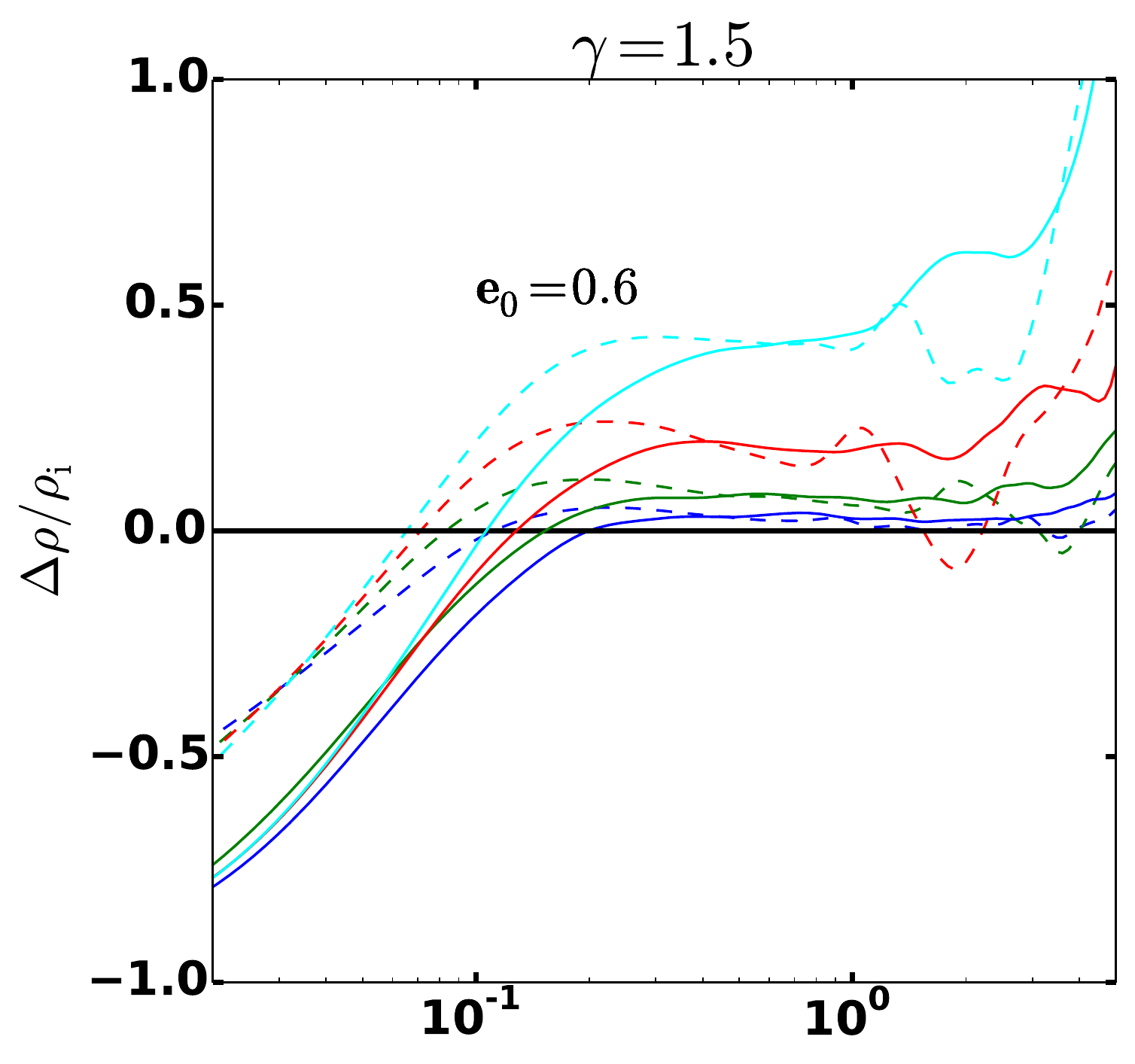}
   \includegraphics[angle=0,width=2.24in,height=2.14in]{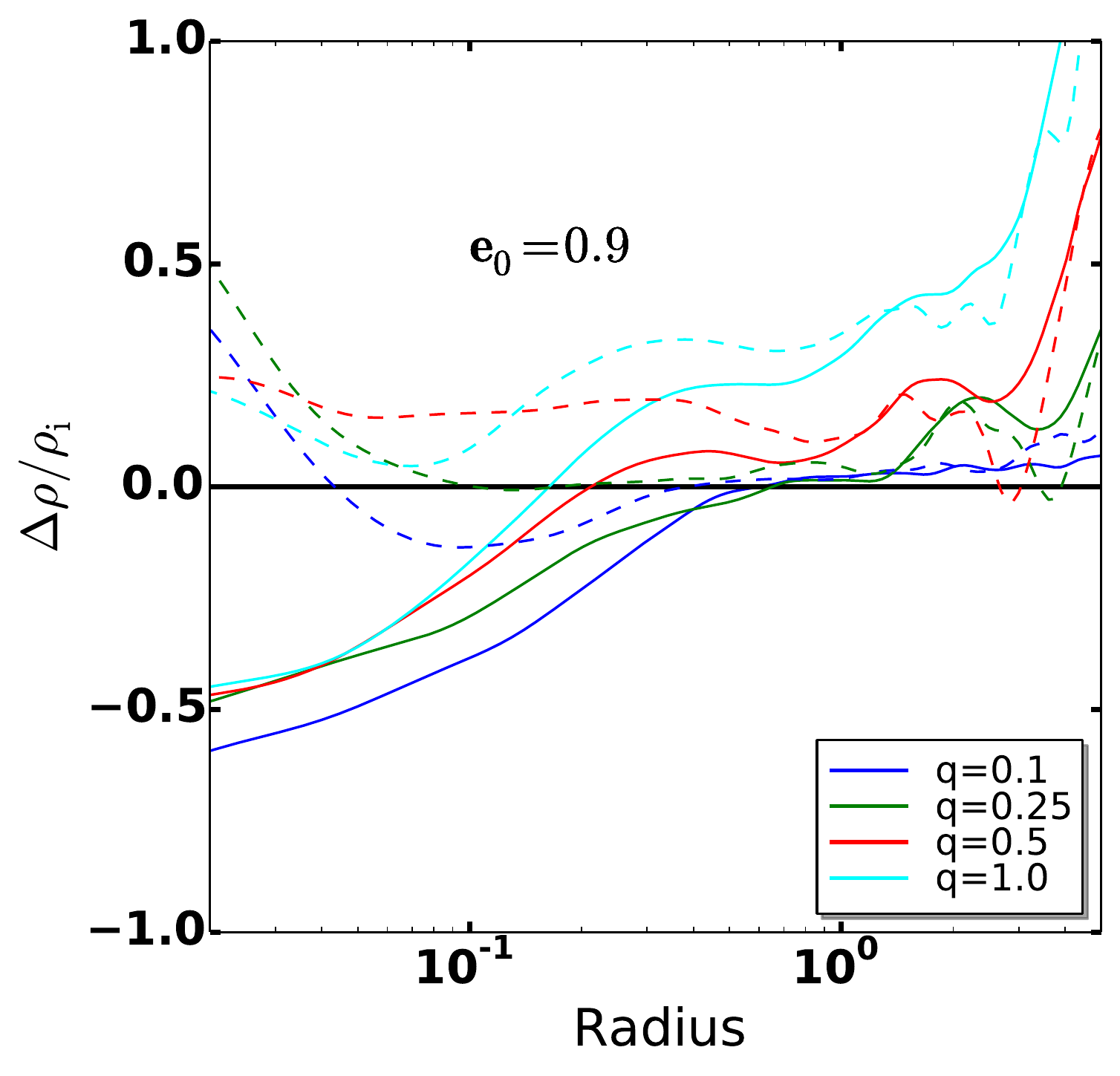}
   \includegraphics[angle=0,width=2.24in]{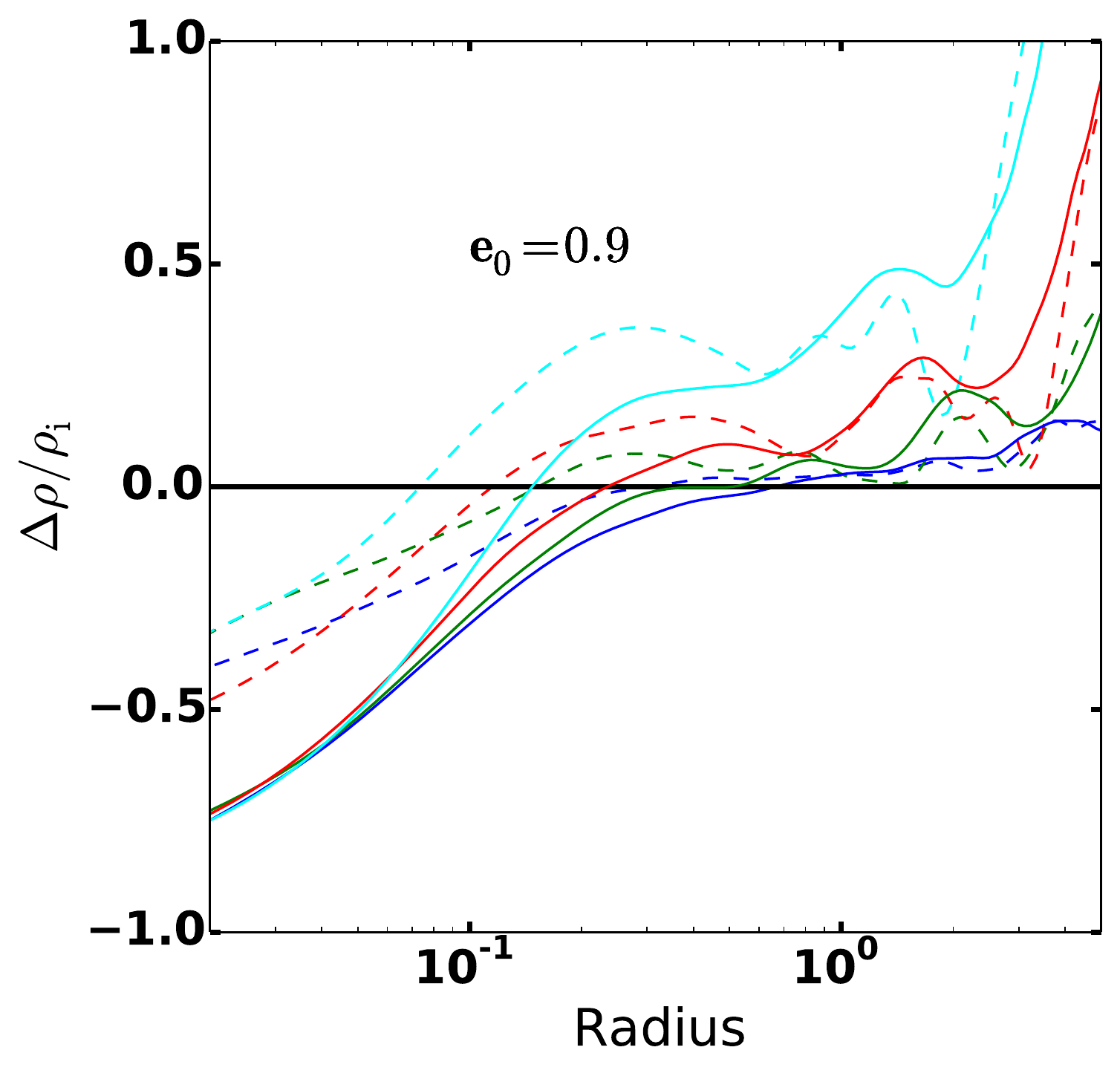}
   \includegraphics[angle=0,width=2.24in]{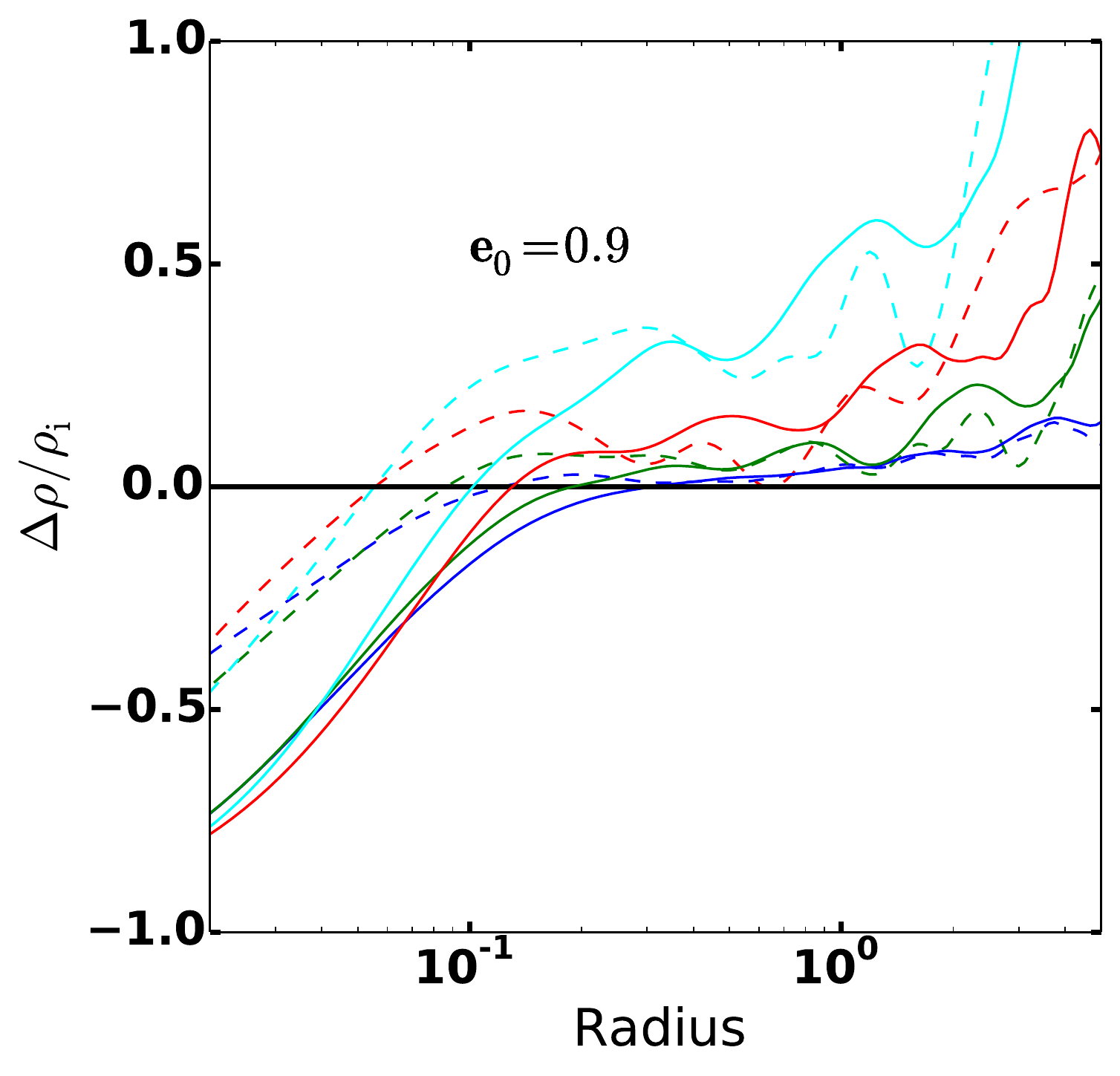}
  \caption{Fractional change in the host galaxy density profile relative to the primary galaxy initial density profile ($\rho_{\rm i}$) for all simulations shown in Figure \ref{com_2}. Dashed lines represent the density profiles at the end of the {\tt gyrfalcON} runs and solid lines refer to the density profiles at the end of the {\tt RAGA} runs. As shown the evolution of the inner density profile within $r_{\rm h}$ is not sensitive to $q$ while the outer profile is significantly affected by the merger mass ratio. }
\label{diff}
\end{center}
\end{figure*}

\subsection{Until binary formation}
\label{sec:binform}
$N$-body simulations of the early merger stages are carried out using the fast tree-code $N$-body integrator  {\tt gyrfalcON}, which employs the $N$-body tool box NEMO \citep{2000ApJ...536L..39D,2002JCoPh.179...27D}. This code features individual adaptive time steps employing a block-step scheme.  The code contains two parameters that affect the speed and accuracy of the calculation, the particle softening length $\epsilon$ and the time-step parameter $k$ such that the timestep is given by $\tau=2^{-k}$. In order to resolve the simulations down to the influence radius  of the primary SMBH we choose $\epsilon=0.008$ while $k=10$ conserves the energy to an accumulated fractional change of order $10^{-3}$.  

Once galaxies are merged and a close massive binary forms, the {\tt gyrfalcON} simulation is terminated.
Practically, this is done when the separation between the two SMBH particles reaches  $r\simeq r'_{\rm h}/10$, with $r'_{\rm h}$ the influence radius of the binary defined as the radius containing a mass in stars twice the mass of the central black hole binary:
\begin{equation}
M(r'_{\rm h})=2M_{12}
\label{hardening}
\end{equation}
where $M_{12}=M_1+M_2$.
The termination time $t_{\rm b}$ for each simulation is given in Table \ref{tab:table1}. 
The hardening radius of the binary at $t=t_{\rm b}$, also shown in Table \ref{tab:table1}, is  given by \citep{2013degn.book.....M}
\begin{equation}\label{hard}
a_{\rm h}=\frac{q}{(1+q)^{2}}\frac{r'_{\rm h}}{4}.
\end{equation}

We note that the softening length chosen is much smaller than the primary black hole influence radius, $\epsilon \lesssim r_{\rm h}/10$, where, roughly speaking, the two SMBHs first become bound to each other  (Table \ref{tab:table1}). This guarantees that the binary evolution  within the influence radius of the SMBH is correctly modelled in our simulations.

\subsection{After binary formation}
\label{sec:afterbinary}

As we mentioned in Section \ref{sec:initial} we continue the merger simulations at $t>t_{\rm b}$ using the Monte Carlo stellar-dynamical code {\tt RAGA} \citep{2015MNRAS.446.3150V}.
This code is designed to model the effect of a central SMBH binary on the stellar distribution, the scattering of stars and the subsequent evolution of the binary orbit. Furthermore, it is a code designed to simulate the dynamical evolution of stellar systems in arbitrary geometry. This unique feature of {\tt RAGA} allows it to start with  galaxy models that are non-spherical, like our merger remnant models at the end of {\tt gyrfalcON}. In
the cosmological setting, binary SMBHs are expected to form via
galaxy mergers, and the merger remnants could well have a
complex structure that does not exhibit any
genuine and particular symmetry \citep{2011ApJ...732...89K,2015ApJ...810...49V}. This means {\tt RAGA} enables us to study core formation using realistic initial conditions for the galaxy model surrounding the central binary, i.e., non-spherical galactic potentials that are the direct product of the merger process.

 {\tt RAGA} represents the system as a collection of particles that move in the smooth time-varying potential due to the stars and central binary. The stellar potential is represented as a multipole expansion with coefficients being regularly recomputed from the particles themselves, and accounts for several dynamical processes, including the interaction between a massive black hole binary and the stars. 
Particle trajectories are computed independently and in parallel, using a high-accuracy adaptive-timestep integrator; the potential expansion and diffusion coefficients are updated at time intervals $\Delta t=5$. By setting $\Delta t=5$ we can properly take into account the change in the potential due to stellar ejections and core formation as this occurs over a much longer  timescale ($t\gtrsim 10^2$). 

To ensure convergence of our results we performed additional runs with {\tt RAGA} for one of our fastest evolving models with $\gamma=1.5,q=0.5$. We used three different time steps for the potential update, $\Delta t=1,3,10$. The
density profile of these models as well as the corresponding break radius and mass deficit values (see Section\ \ref{MDBR}) at the final integration time
were all consistent with each other,
indicating that 
our results are not sensitive to the particular choice of $\Delta t$.

While we redirect the reader to \citet{2015MNRAS.446.3150V} for a much more detailed description of the code, we note here a few key properties.
The effect of two-body gravitational encounters, or two-body  relaxation, is modelled in {\tt RAGA} by local (position-dependent) velocity diffusion coefficients; the magnitude of relaxation can then be adjusted to the actual number of stars in the target system and it is not related to the number of particles in the simulation. Here we turn off the collisional relaxation of the system completely. This  allows us to explore the evolution of the binary and galaxy in the relevant limit  $N\rightarrow \infty$, i.e., the collisionless limit of real galaxies. 
Contrary to an $N$-body code, {\tt RAGA} does not require the calculation of pair-wise force interactions between all the particles in the model. But it integrates the orbits of trace stars in a smooth potential. These features make the code much more efficient and at the same time more accurate
than an $N$-body simulation. On the contrary,
$N$-body integrations are affected by spurious
relaxation and other discreteness effects, which would cause 
the radial distribution of stars to evolve
 even in circumstances
where two-body effects should be negligible.
{\tt RAGA} also relies on a number of approximations and limitations.
First, the treatment of relaxation relies on the local diffusion coefficients  which are computed assuming that the system is isotropic and spherically symmetric. This approximation could break down in a strongly non-spherical or anisotropic system. 
Moreover, {\tt RAGA} assumes that all stars have the same mass, it does not take into account processes that are not described by standard two-body relaxation theory \citep[e.g.,][]{1996NewA....1..149R}, and does not include primordial and dynamically formed binaries and their contribution to the energy budget of the system. Also, updating the multiple expansion of the potential after every episode can induce some spurious relaxation effect \citep{2015ApJ...810...49V}. Finally, {\tt RAGA} does not include relativistic corrections to the motion which might become important in the last stages of binary hardening.

We note here that when we switch to {\tt RAGA} we use as initial conditions the stellar positions and velocities from the final snapshot of the {\tt gyrfalcON} simulation.  Also, the SMBHs are placed and kept at the center of mass of the entire system throughout the {\tt RAGA} simulations. 
This means that the Brownian motion of the binary due to encounters with surrounding stars is neglected. 
This represents a further improvement with respect to $N$-body simulations where the random motion of the binary would be unrealistically large due to the higher SMBH-star mass ratio.  This effect becomes significant in $N$-body simulations  if the mass ratio between the SMBHs and the stars is $\lesssim 10^3$ \citep{2016MNRAS.461.1023B}.
The input parameters for the massive binary are the semi-major axis, and eccentricity measured from the last timestep of the {\tt gyrfalcON} simulation (see Table \ref{tab:tableraga}). {\tt RAGA} simulations are stopped when the binary reaches a separation equal to $a \approx a_{\rm h}/10$. We stop the simulations at $a\approx a_{\rm h}/10$ to ensure that our binaries evolve long enough to study their scouring effect on core formation and simultaneously stop the simulations before the GW wave emission would start affecting the binary evolution. The latter typically happens at $a\sim a_{\rm h}/10$ \citep{2007ApJ...671...53M} for the massive galaxies we are considering, and this is why we stop our simulations at this point. The stopping time is given for all our {\tt RAGA} simulations in Table \ref{tab:tableraga}. Furthermore, we have tried different starting points for the {\tt RAGA} runs and we find that our results (i.e., the evolution of the binary and density profile of the galaxy) are not sensitive to the actual stopping point of the {\tt gyrfalcON} runs, as long as the switch happens at a time where the two galaxies have already merged and the binary has decayed to a separation $a\lesssim r'_{\rm h}$.

\section{Quantifying Changes in Galaxy Cores}
\label{sec:cores}

\subsection{Changes in density profiles}

We show in Figure \ref{com_2} the final radial density profile of the galaxy for different initial density profile slope $\gamma$, mass-ratio $q$ and different initial eccentricity $e_{0}$. We also show the density profile slope, $-d \log \rho/d \log r$, as a function of radius for all the models. Both are given at the end of the {\tt RAGA} simulation when the binary semi-major axis is $a\simeq a_{\rm h}/10$, and also at the end of the {\tt gyrfalcON} run.  Here and in what follows, density profiles and slopes are computed using a (Gaussian) kernel-based algorithm, similar to the algorithm described in \citet{1994AJ....108..514M}.
In using this method, we make the simplifying assumption that the  underlying density is spherically symmetric about the center of mass of the binary.


In Figure~\ref{diff} we plot the fractional change in the galaxy stellar density  relative to the primary galaxy initial density profile ($\rho_{\rm i}$). We plot the relative change both at the end of the {\tt gyrfalcON} runs and at the end of the RAGA runs. As shown in Figure~\ref{diff}, the final core density within $r_{\rm h}$ is not too sensitive to $q$ with all models producing a similar mass depletion at small radii. On the other hand, $q$ affects the density in the outer regions of the galaxy, with larger mass ratios leading to  larger  density enhancements. This suggests that core depletion should be virtually independent of the merger mass ratio, a result that we will verify in detail below.

\begin{table*}
  \begin{center}
    \caption{BREAK RADII AND MASS DEFICITS}
    \label{tab:table3}
    \begin{tabular*}{\textwidth}{@{\extracolsep{\fill}}|lll|lll|lll|}
      \hline
      \textbf{Run} & \textbf{$\gamma$} & \textbf{$q$} & \textbf{$r_{\rm b}$}  & \textbf{$M_{\rm def}$}  & \textbf{$M_{\rm ej}$} & \textbf{$r_{\rm b}$} &  \textbf{$M_{\rm def}$} & \textbf{$M_{\rm ej}$}  \\ 
      \hline 
      & & & & ${e}_{0}=0.9$ &  & & ${e}_{0}=0.6$  & \\
      \hline
      1 & 0.5 & 0.1 & 0.33 & 4.05 & 0.65 (0.68) & 0.33   & 4.05  & 1.02 (0.75) \\
      2 & 0.5 & 0.25 & 0.33 & 3.32   & 1.04 & 0.35 & 3.54 & 1.60\\ 
      3 & 0.5 & 0.5 & 0.29   & 2.23 & 1.07 &  0.32    & 2.82 & 1.50\\
      4 & 0.5 & 1.0 & 0.31 & 1.93 & 0.97 & 0.35 & 2.53  & 1.40 \\
      5 & 1.0 & 0.1 & 0.18   & 2.85 & 1.00 (1.00) & 0.19  & 2.85 & 1.08 (1.05)\\
      6 & 1.0 & 0.25 & 0.18  & 2.37 & 1.14 & 0.19 & 2.37 & 1.53\\
      7 & 1.0 & 0.5 & 0.17  & 2.02 & 0.95 & 0.20  &  2.02 & 1.50\\
      8 & 1.0 & 1.0 &  0.18  & 1.75  & 1.02 & 0.21 & 1.75 & 1.36\\
      9 & 1.5 & 0.1 & 0.08 &  0.91 & 1.08 (1.17) & 0.09 & 1.10 & 1.12 (1.20)\\
      10 & 1.5 & 0.25 & 0.08 & 0.68 & 1.03 & 0.08 & 0.75  & 1.03\\
      11 & 1.5 & 0.5 &  0.09  & 0.70 & 1.19 & 0.08  & 0.72 & 1.19\\
      12 & 1.5 & 1.0 & 0.08  & 0.51 & 0.95 & 0.09 & 0.68 & 1.13\\
      \hline
    \end{tabular*}

 \end{center}
 \small Break radius and mass deficit values for all  simulations given in Table \ref{tab:table1}. We use the core-S\'{e}rsic fit method described in Section \ref{sec:cores} to estimate the break radius $r_{\rm b}$ and mass deficit $M_{\rm def}$. The ejected mass relative to the merger remnant profile at the end of the {\tt gyrfalcON} run is given as $M_{\rm ej}$. The values in parentheses refer to the ejected mass relative to the initial density profile of the primary galaxy  and are only given for $q=0.1$. Mass deficits as well as the ejected mass are always given in units of the total binary mass $M_{12}$.
\end{table*}

\subsection{Ejected mass} \label{discr}

One physical quantity that can be estimated from our simulations is the actual mass ejected by the binary during the merger process. 
We use two methods here to estimate this ejected mass.

In the first method, we calculate the ejected mass relative to the remnant density profile at the formation of the close binary.  We do this by computing the difference between the density of the models at the end of the {\tt gyrfalcON} run and at the end of  the {\tt RAGA} run.  We 
compute the difference within the radius at which the density in the two models is the same.
 As shown in Table \ref{tab:table3}, {the ejected mass in all simulations is quite close to the binary mass} and lies in the range $ M_{\rm ej} = 0.7$ to $1.6M_{\rm 12}$. 
 When interpreting these values one should note, however, that 
 some  scouring happens even  before  the  start  of  the  slingshot  ejection  phase. 
Fig. \ref{com_2} and \ref{diff} show that the  density  is  lower  than  the  one  of  the  original  profile even  just  at  the  start  of  the  {\tt RAGA}  run,  implying  that  the  effect  of  the merger  itself  is  slightly  heating  up  the  central  regions. This is consistent with
\citet{2018MNRAS.477.2310B} who  performed  very  similar  simulations  and  explored  the  merger  between  equal mass  galaxies  including  0,  1  or  2  SMBHs  in  the  simulations  (see  their  Fig. 13).   They find  that  even  without  any  SMBH,  or  even  with  just  one,  the central density slightly decreases.

{In the second method, we}
calculate the difference in integrated mass between the final density profile of the galaxy remnant and the initial density profile of the primary galaxy. 
Because this calculation is more accurate for smaller mass-ratios, expected to affect less the outer density profile of the primary galaxy, we perform it only for  $q=0.1$. 
The values we find in these cases are given in Table \ref{tab:table3} and are all of the order $\sim M_{12}$.

{We can estimate the order of magnitude of the expected ejected mass as follows. As the binary evolves from  $r'_{\rm h}$ where the two SMBHs become a close pair to $a\approx a_{\rm h}$, it will have given up an energy 
\begin{eqnarray}\label{Denergy}
\Delta \mathcal{E}&\approx& \frac{GM_1M_2}{2a_{\rm h}}-\frac{GM_1M_2}{2r'_{\rm h}}\approx \nonumber\\
&&-\frac{1}{2}M_2\sigma^2+2M_{12}\sigma^2 
\approx 2M_{12}\sigma^2
\label{energy}
\end{eqnarray}
to the surrounding stellar nucleus \citep[e.g.,][]{2006ApJ...648..976M,2013degn.book.....M}, with $\sigma$ the stellar velocity dispersion. 
Thus, the energy transferred from the binary to the stars should be roughly proportional to
the combined mass of the binary. This result suggests that
the binary will displace a mass in stars of order its own mass,
roughly independent of the mass of the secondary SMBH. Indeed, this expectation is born out by the ejected mass that we find in the simulations, showing that $M_{\rm ej}$ is approximately independent of $q$.}

\subsection{Approximating ejected mass with the deficit mass and break radius}\label{MDBR}

The light profiles of luminous $M_{\rm B} \leq -20.5 \: \, \rm mag$ elliptical galaxies are often observed to depart systematically from a simple S\'{e}rsic model near their center \citep{1997AJ....114.1771F,1994AJ....108.1598F,1995AJ....110.2622L,2003AJ....125.2951G,2004AJ....127.1917T}. This departure, a downward deviation
with respect to the inward extrapolation of the outer S\'{e}rsic profile, emanates from a central starlight deficit and happens at a radius that can be characterized by a core or break radius, $r_b$.

{
We quantify $r_b$ of a merger remnant by fitting its projected surface density to a five-parameter core-S\'{e}rsic function

\begin{equation}
\mu_{\rm Ser}(r)=\mu_{0} \left[ 1+ \left(\frac{r_{\rm b}}{r}\right)^{\alpha}\right]^\frac{\gamma}{\alpha}
e^{\left[ -b\left( \frac{r^{\alpha}+r_{\rm b}^{\alpha}}{R_{\rm e}^{\alpha}} \right) ^\frac{1}{n\alpha} \right]}.
\end{equation}

We follow \citet{2018ApJ...864..113R}
and fix the S\'{e}rsic index $n=4$ to minimize degeneracy between $R_e$ and $n$.
We compute the break radius from the surface density profile of the model projected on the $x-y$, $x-z$, and $y-z$ planes and average the three values. 
However, because our models are characterized only by a small degree of 
flattening the values of the fitted parameters are not too sensitive to the direction of projection adopted; the projection on the $x-y$ and the other planes give  results that  differ  by  $\lesssim 10\%$ from each other.
In Figure \ref{fits} we show two characteristic examples of the S\'{e}rsic fits to the surface density profiles at the end of the RAGA runs projected on the $x-y$ plane. 
Table~\ref{tab:table3} lists the resulting averaged core radii for the merger remnants in our simulations. 

\begin{figure*}
\centering
 \includegraphics[angle=0,width=2.67in]{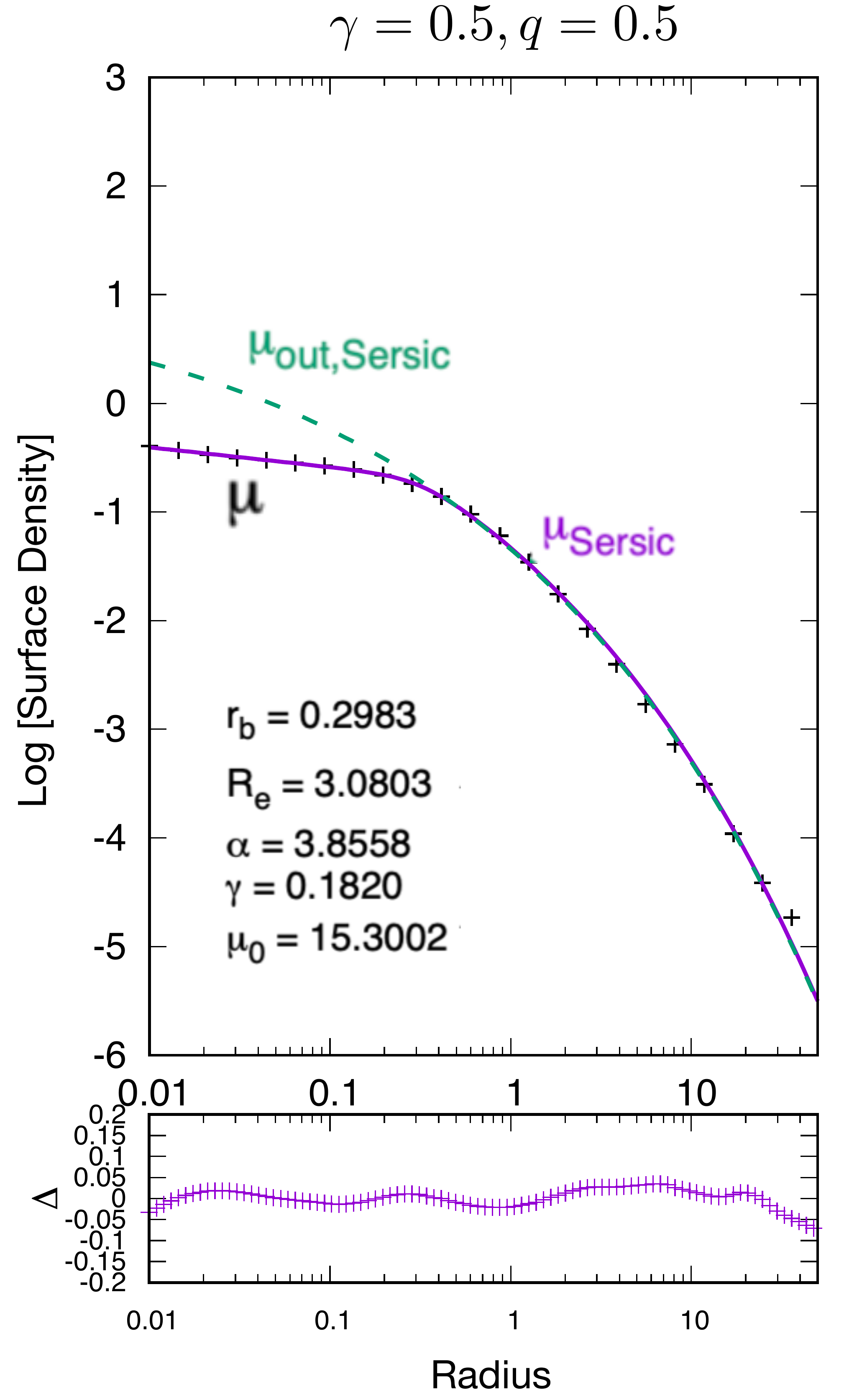}
 \includegraphics[angle=0,width=2.8in,height=4.43in]{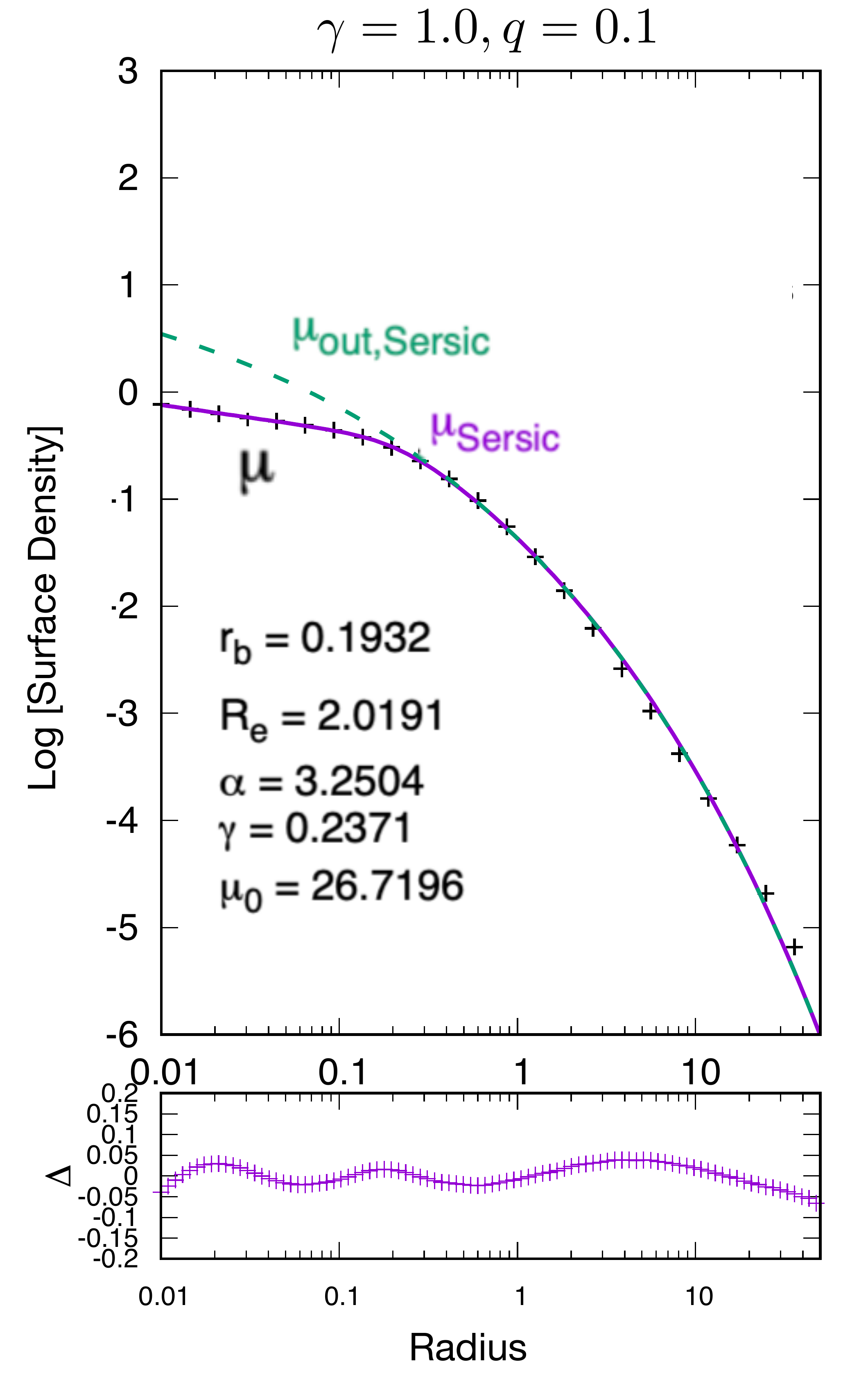}
\caption{S\'{e}rsic fits (core $\mu_{\rm Ser}$ and outer $\mu_{\rm out,Ser}$) to the surface density profiles $\mu$ at the end of the {\tt RAGA} runs  projected on the $x-y$ plane for the cases $\gamma=0.5,q=0.5$ and $\gamma=1.0,q=0.1$ ($e_{0}=0.6$). The fractional change $\Delta$ is defined as $\Delta=(\log{\mu}-\log{\mu_{\rm Ser}})/\log{\mu_{\rm Ser}}$. }
\label{fits}
\end{figure*}

We estimate the mass deficit by comparing the core-S\'{e}rsic function with the inward extrapolation of a (non-cored) S\'{e}rsic function fitted to the profile at $r > r_b$. 
The central deficit is then calculated as the difference in the cumulative mass within $r_{\rm b}$ given by the two functions:
\begin{equation}
   M_{\rm def}=4\pi\int_{0}^{r_{\rm b}} \left[\mu_{\rm out, Ser}(r)-\mu_{\rm Ser}(r)\right]r^{2}dr.
\label{deficitfirst}
\end{equation}
We compute $M_{\rm def}$ for the three projections on the $x-y$, $x-z$, and $y-z$ planes.
Table~\ref{tab:table3} presents the corresponding averaged values of the mass deficit (in units of the total binary mass). 
 We note here that that systematics due to radial range of fitting, fixing and freeing of $\alpha$ and $n$, can impact the fits and change $r_{\rm b}$ by  $\sim 50\%$ level. The corresponding deviations of $M_{\rm def}$ are at the factor of 2-3 level, so larger than that. Moreover, it is also worth noting that observational estimates of core radii and mass-deficits are affected by  systematics such as PSF, noise, radial coverage and assumptions about the radial dependence of the mass-to-light ratio which are not present in the simulations.

Table~\ref{tab:table3} shows that 
unlike $M_{\rm ej}$, $M_{\rm def}$ increases with decreasing $q$ and $\gamma$.
This behaviour is because the S\'{e}rsic fits are sensitive to the outer galaxy density profile, which is affected by 
 both $q$ and $\gamma$.}

\begin{figure}
\begin{center}
 \includegraphics[angle=0,width=2.2in]{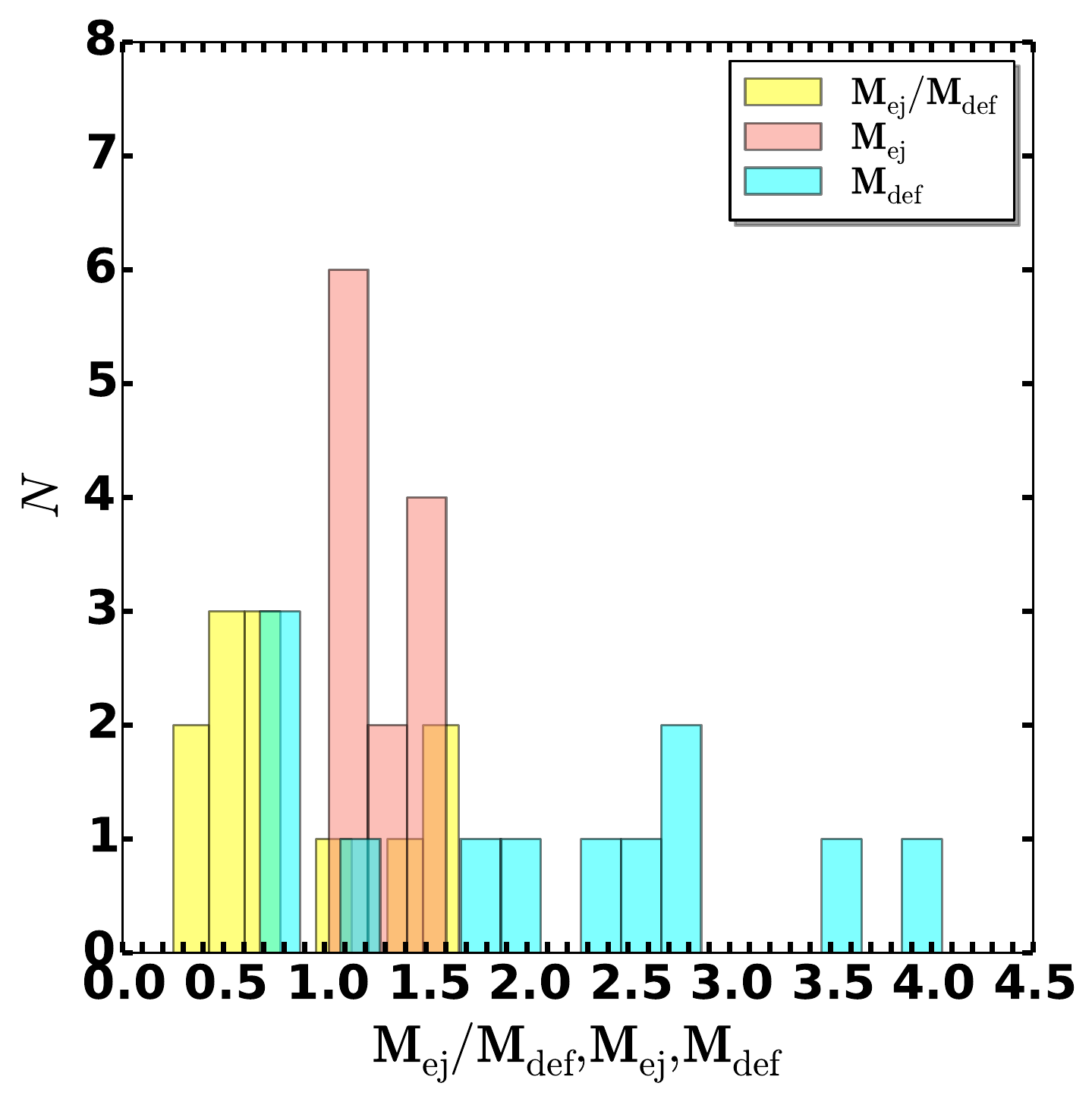}
 \includegraphics[angle=0,width=2.2in]{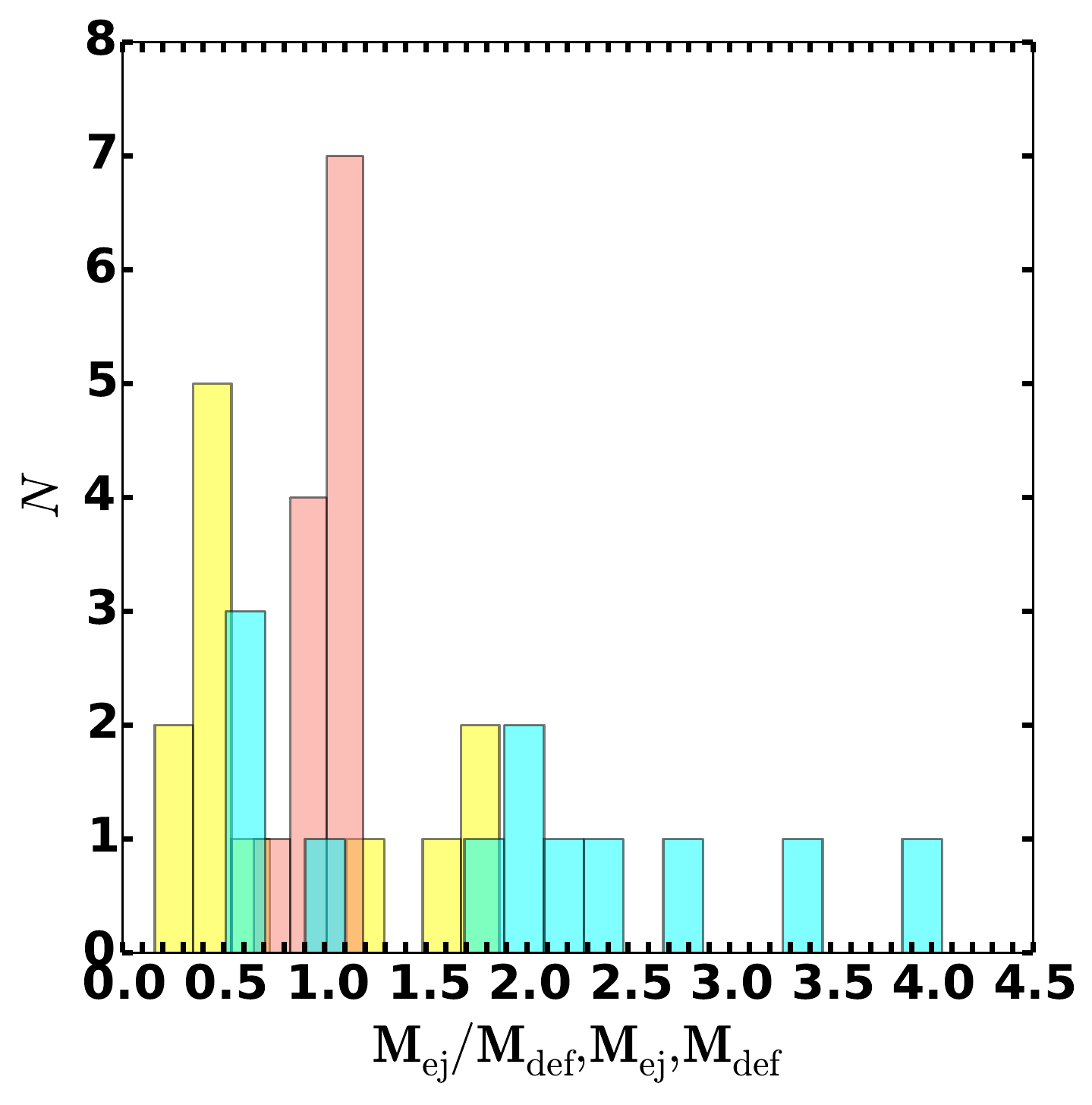}
\caption{Histogram of $M_{\rm ej}$, $M_{\rm def}$ and $M_{\rm ej}/M_{\rm def}$ for the two different eccentricities $e_{0}=0.6$ (left panel) and $e_{0}=0.9$ (right panel) shown in Table \ref{tab:table3}. The ratio $M_{\rm ej}/M_{\rm def}$ spans the range $\sim 0.2$ to $2$.}
\label{eje_hist}
\end{center}
\end{figure}

{ Finally, in Figure~\ref{eje_hist} 
we show the distributions of $M_{\rm def}$, $M_{\rm ej}$ and the ratio 
$M_{\rm ej}/M_{\rm def}$. We see that the ratio of the deficit mass and the total binary mass ranges from 0.5 to 4 for the 12 merger remnants. 
As also displayed in Figure \ref{eje_hist}, 
the ratio of the  mass deficit and the ejected mass spans a similarly large range. Deficit masses therefore 
provide only an order-of-magnitude estimate of the ejected mass. 
We find this result unsurprising because the deficit mass is computed from an assumed progenitor profile (a non-cored S\'{e}rsic).
It is overly simplistic to expect all progenitor galaxies to have a universal 
inner density profile.}

\section{The Roles of Triaxiality and Multiple Mergers on Core Formation}
\label{sec:results}

{In this section we look at (i) the evolution of the mass deficit and core radius with time during the hardening of the binary and (ii) how the mass deficit and core radius evolve under multiple mergers. Prior investigations of these  issues have typically relied on spherical models, and as we will show, introducing more realistic triaxiality will change the behavior of the apparent mass deficit and break radius relative to the prior literature. Then we can use our new results to draw conclusions about whether the observed mass deficits and break radii can be used to infer any information about the merger history of galaxies. We can also check the consistency of our results with the suggestion
that cores are made by SMBH binary scouring mechanisms acting over the course of one
or more successive dry mergers.}

\begin{figure}
\begin{center}
  \includegraphics[angle=0,width=2.4in]{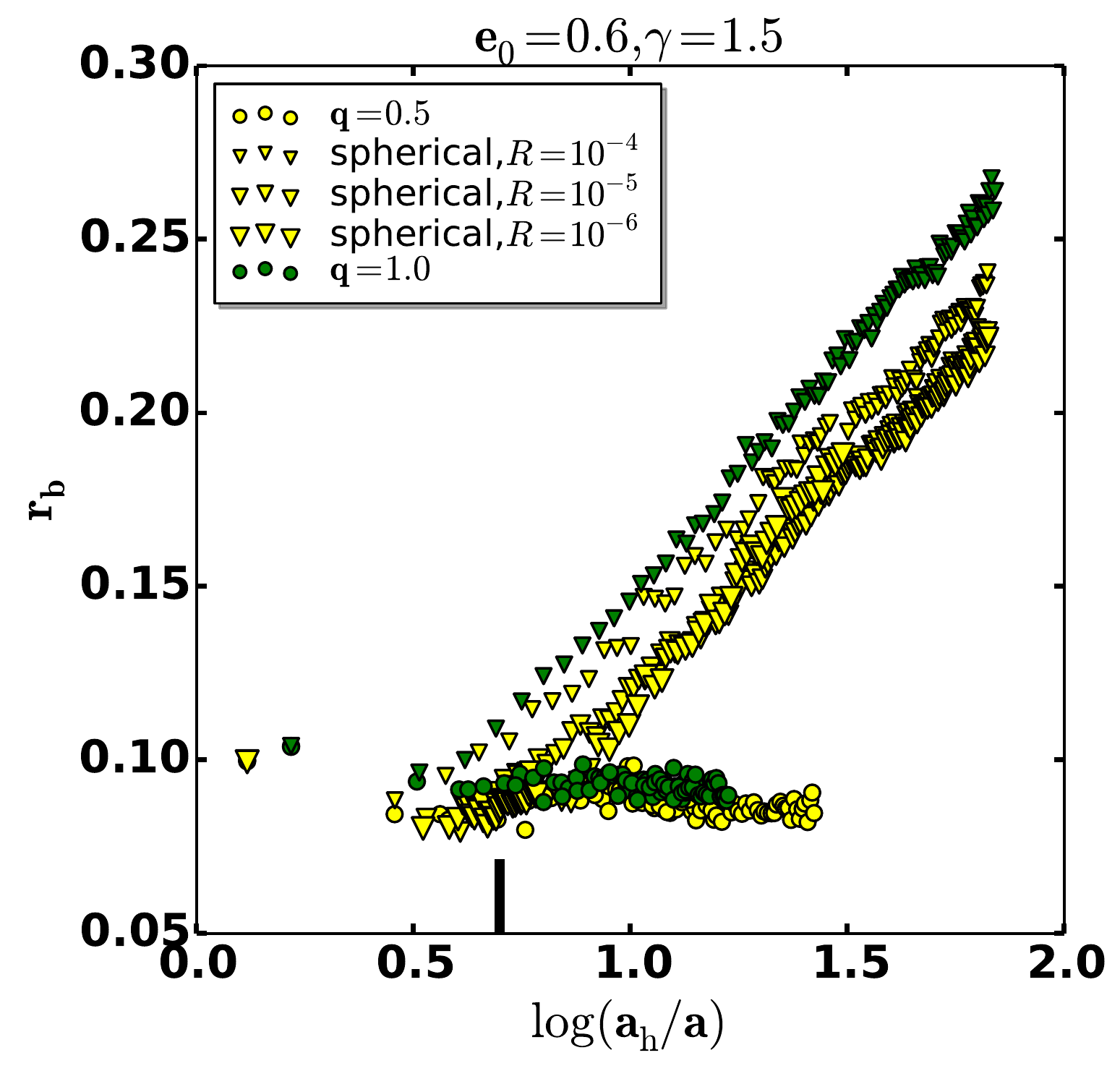}
  \includegraphics[angle=0,width=2.4in]{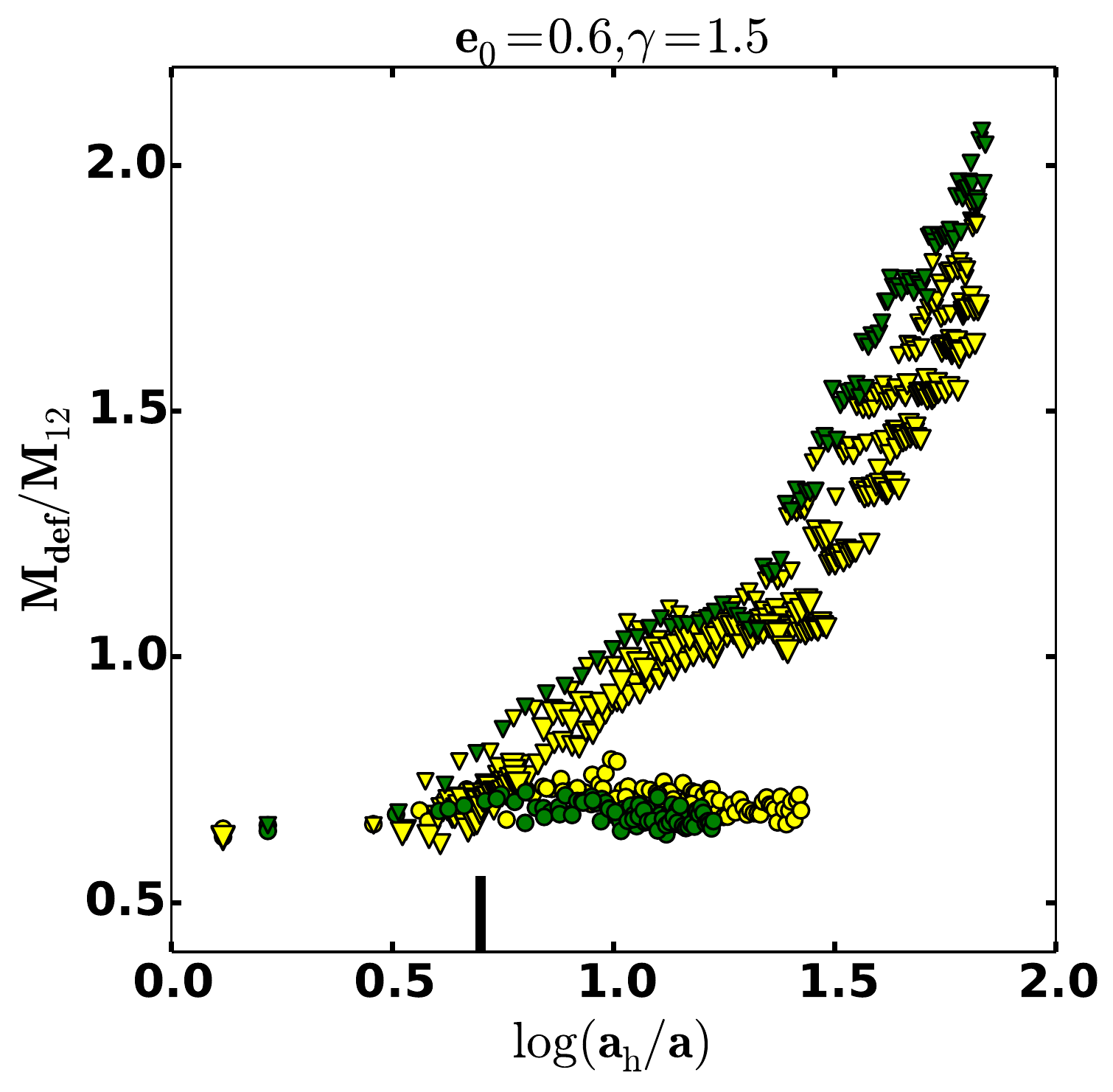}
  \caption{Mass deficit $M_{\rm def}/M_{12}$ and break radius $r_{\rm b}$ as a function of the hardening radius $a_{\rm h}/a $ for $\gamma=1.5,e_{0}=0.6$ and $q=0.5,1.0$. The filled circles are for the actual non-spherical model while the triangles for its spherical equivalent. For the spherical models, the size of the triangles increase with the decrease in the relaxation rate $R=10^{-4},10^{-5},10^{-6}$. While in the actual models the mass deficit and break radius increases until approximately $a_{\rm h}/a\approx 5$ and after that remains nearly constant, in the equivalent-spherical model they keep increasing as the binary orbit shrinks. As shown, this result is independent of the two-body relaxation rate used in the spherical models.
  Vertical thick black tick mark depicts the point $a_{\rm h}/a=5$, where the behaviour in the two models start to differ significantly from each other.}
\label{deficit_ev2}
\end{center}
\end{figure}

\begin{figure*}
\begin{center}
   
   \includegraphics[angle=0,width=1.72in]{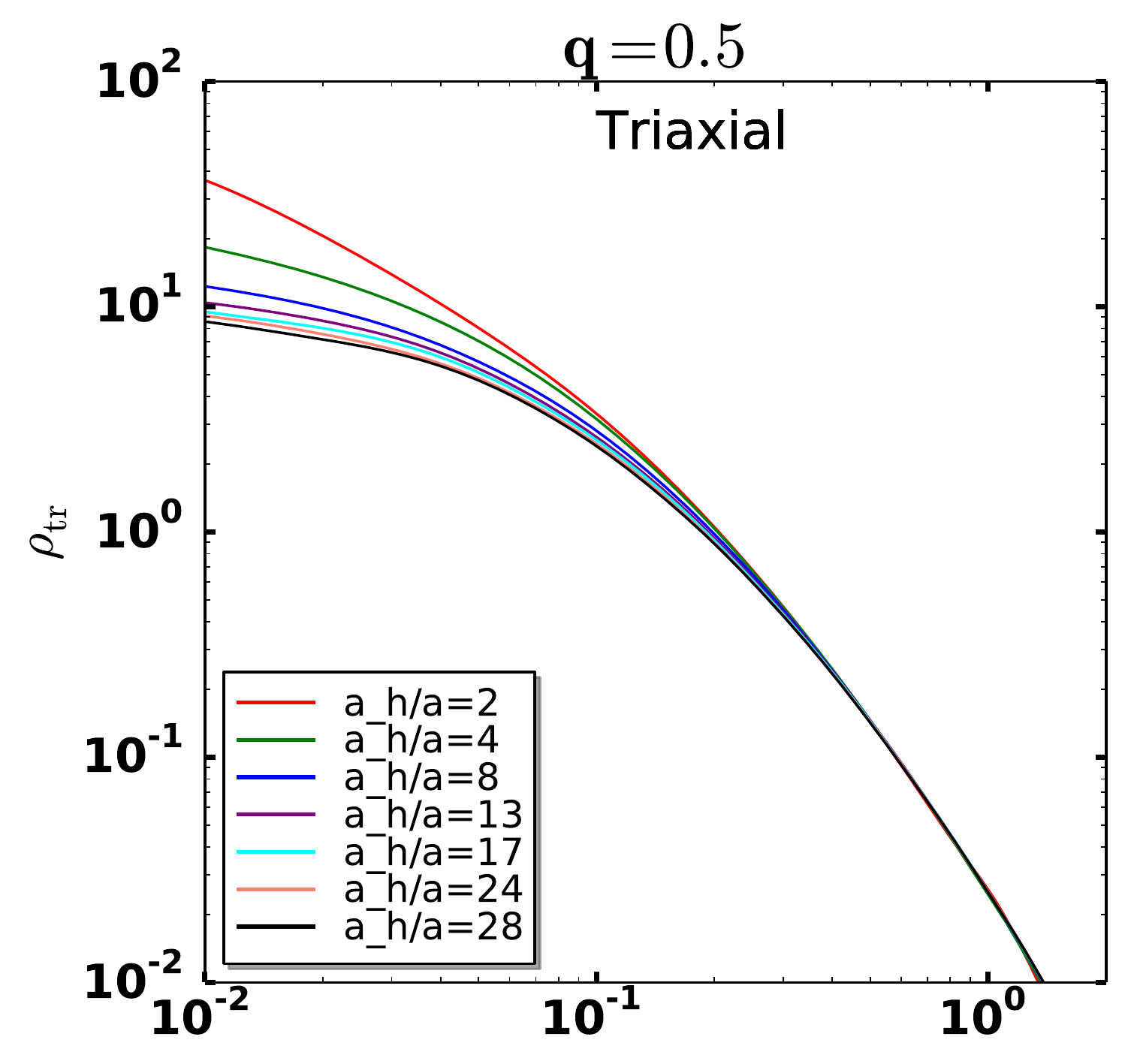}
   \includegraphics[angle=0,width=1.72in]{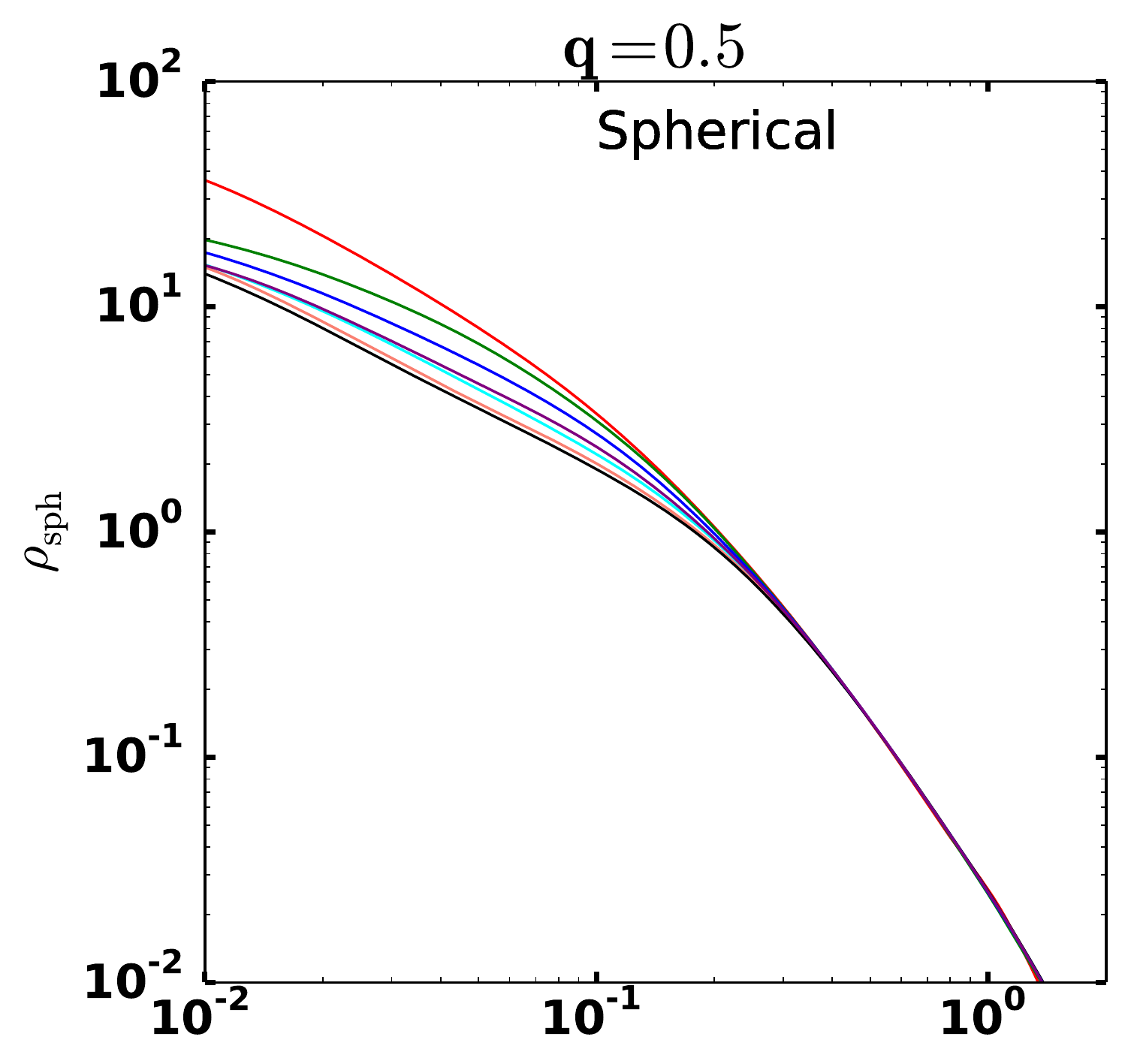}
   \includegraphics[angle=0,width=1.72in]{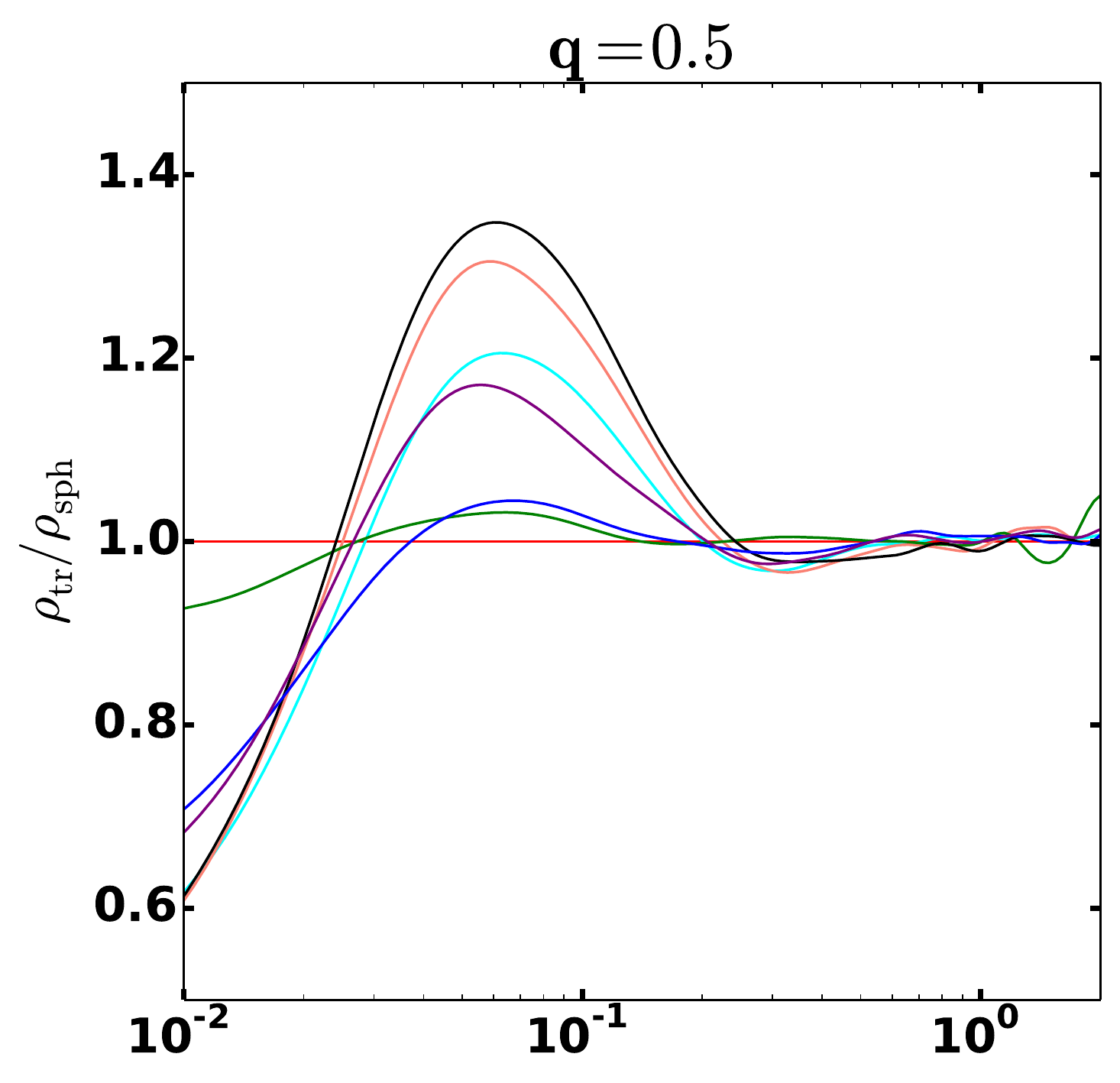}

   \includegraphics[angle=0,width=1.72in]{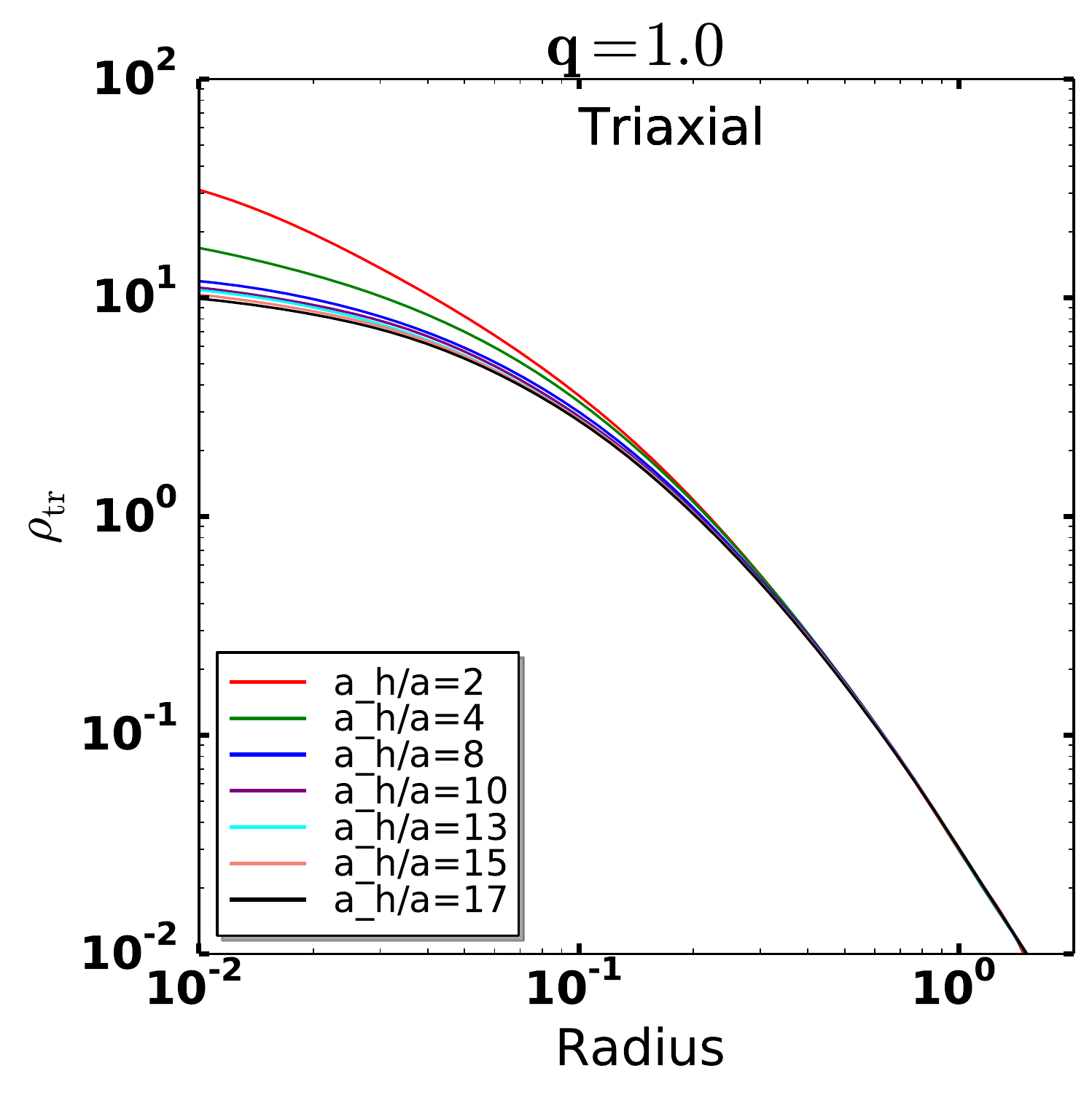}
    \includegraphics[angle=0,width=1.72in]{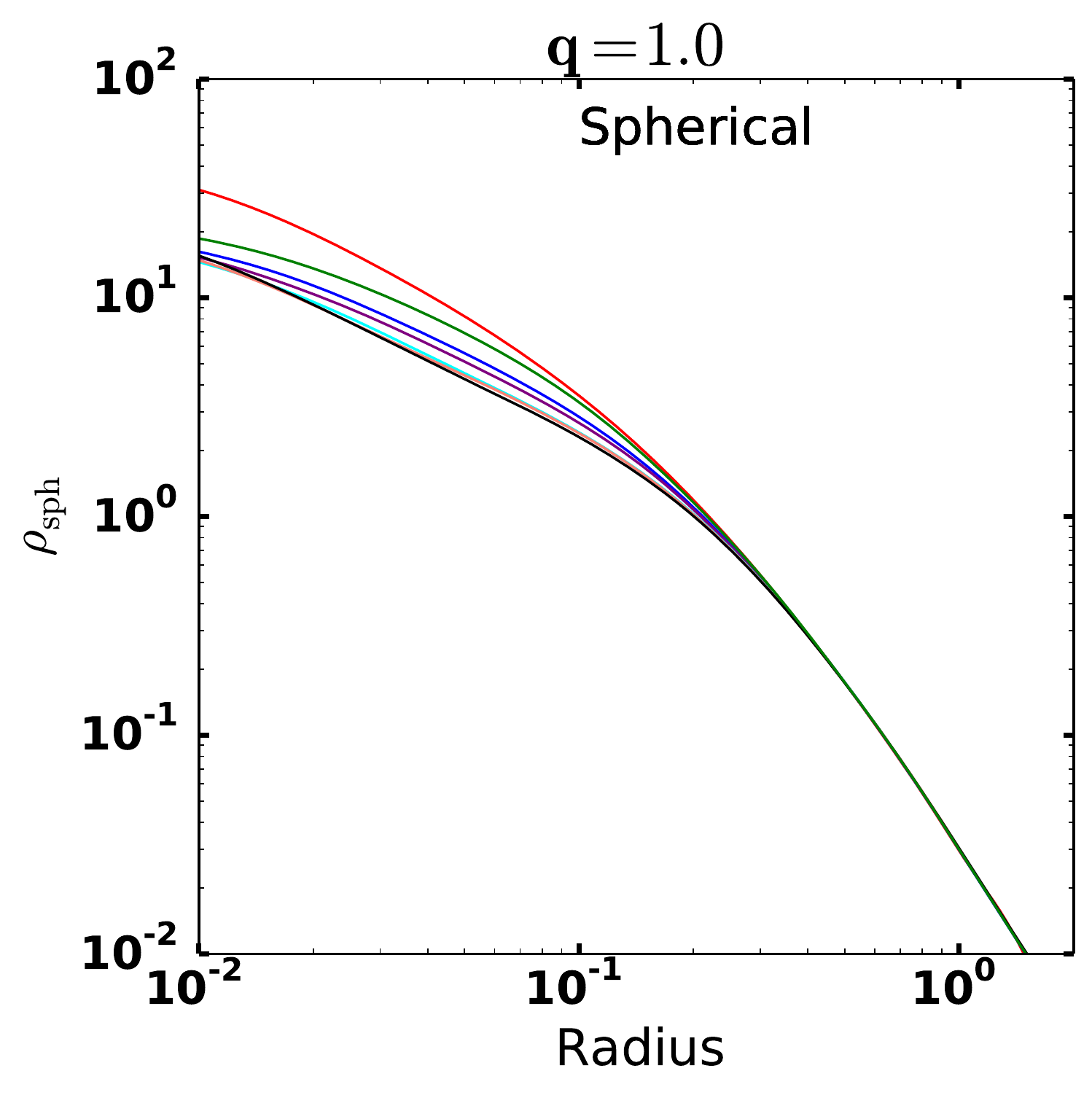}
   \includegraphics[angle=0,width=1.72in]{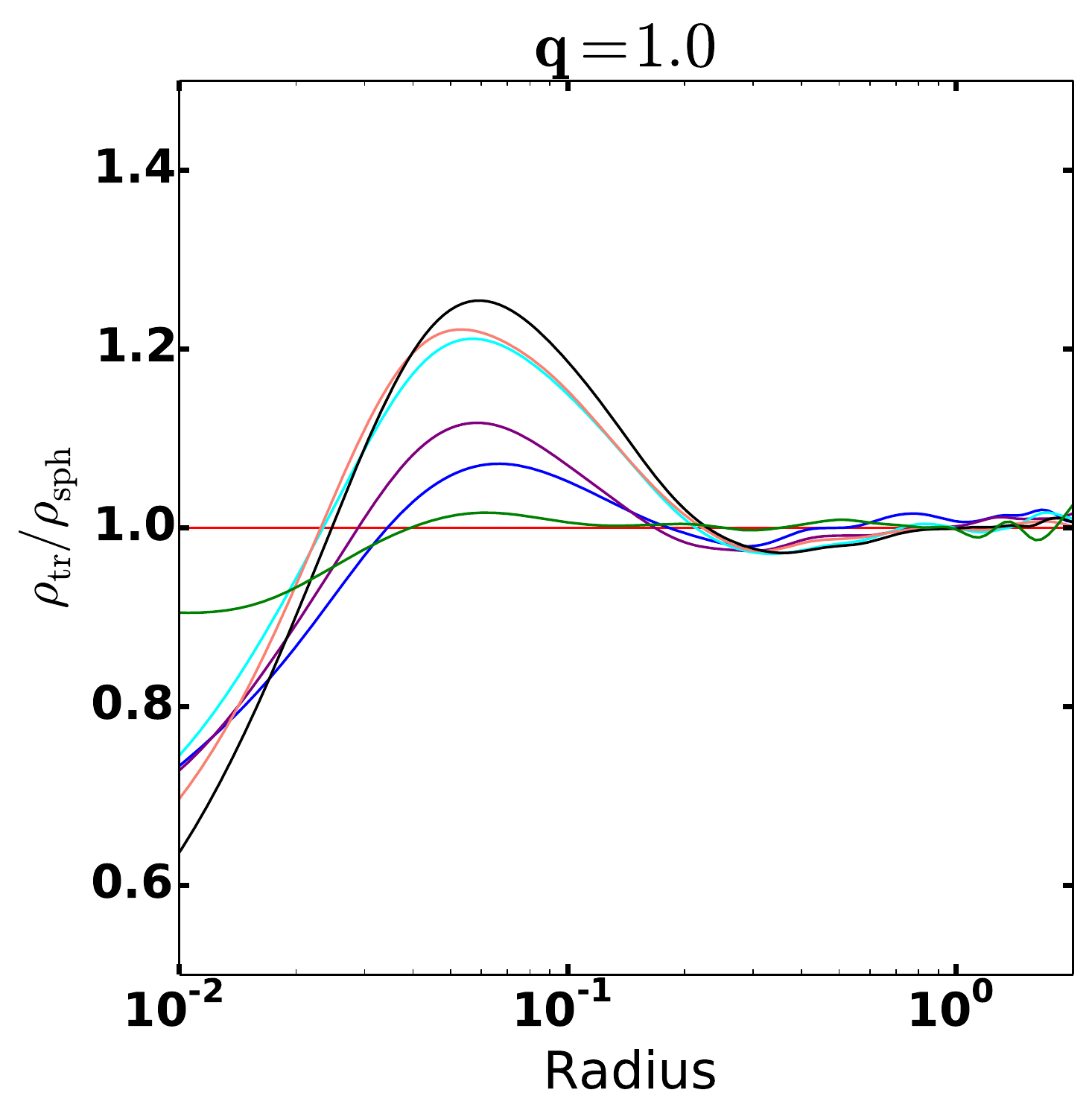}

\caption{Evolution of the density profile in the merger  (i.e., triaxial) models and their equivalent spherical models shown in Figure \ref{deficit_ev2} during the hardening of the binary. As indicated in the legends, different colors refer to different times in the binary evolution, until the binary reaches a separation of $a_{h}/a=2,4,8,13,17,24,28$ (upper panel) and $a_{h}/a=2,4,8,13,17,24,28$ (lower panel). The right panels show the ratio between the density in the triaxial and spherical models as a function of radius, clearly showing the different evolution of the central density in the two models.}
\label{dens-ts}
\end{center}
\end{figure*}

\begin{figure}
\begin{center}
  \includegraphics[angle=0,width=2.4in]{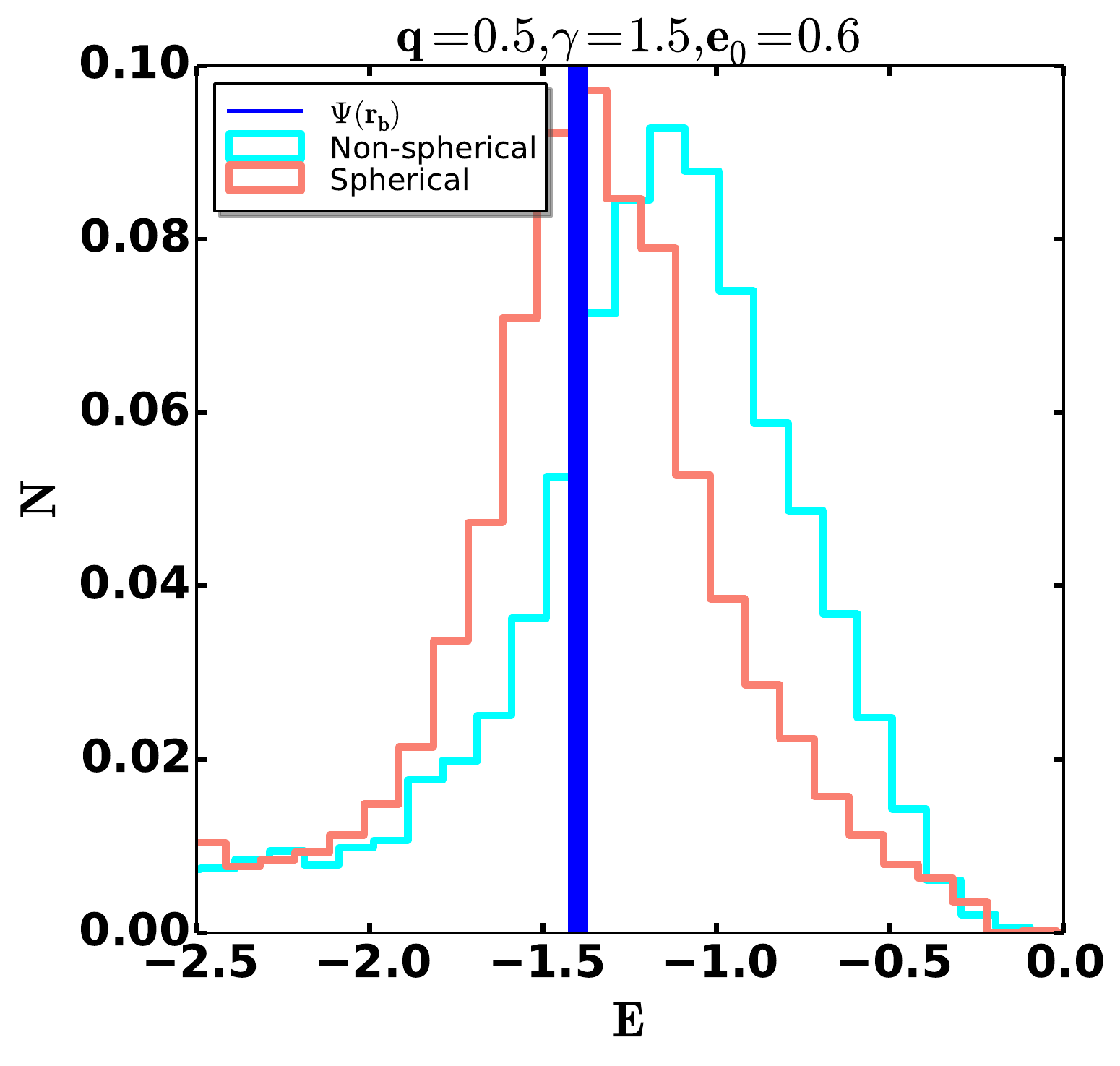}
  \includegraphics[angle=0,width=2.4in]{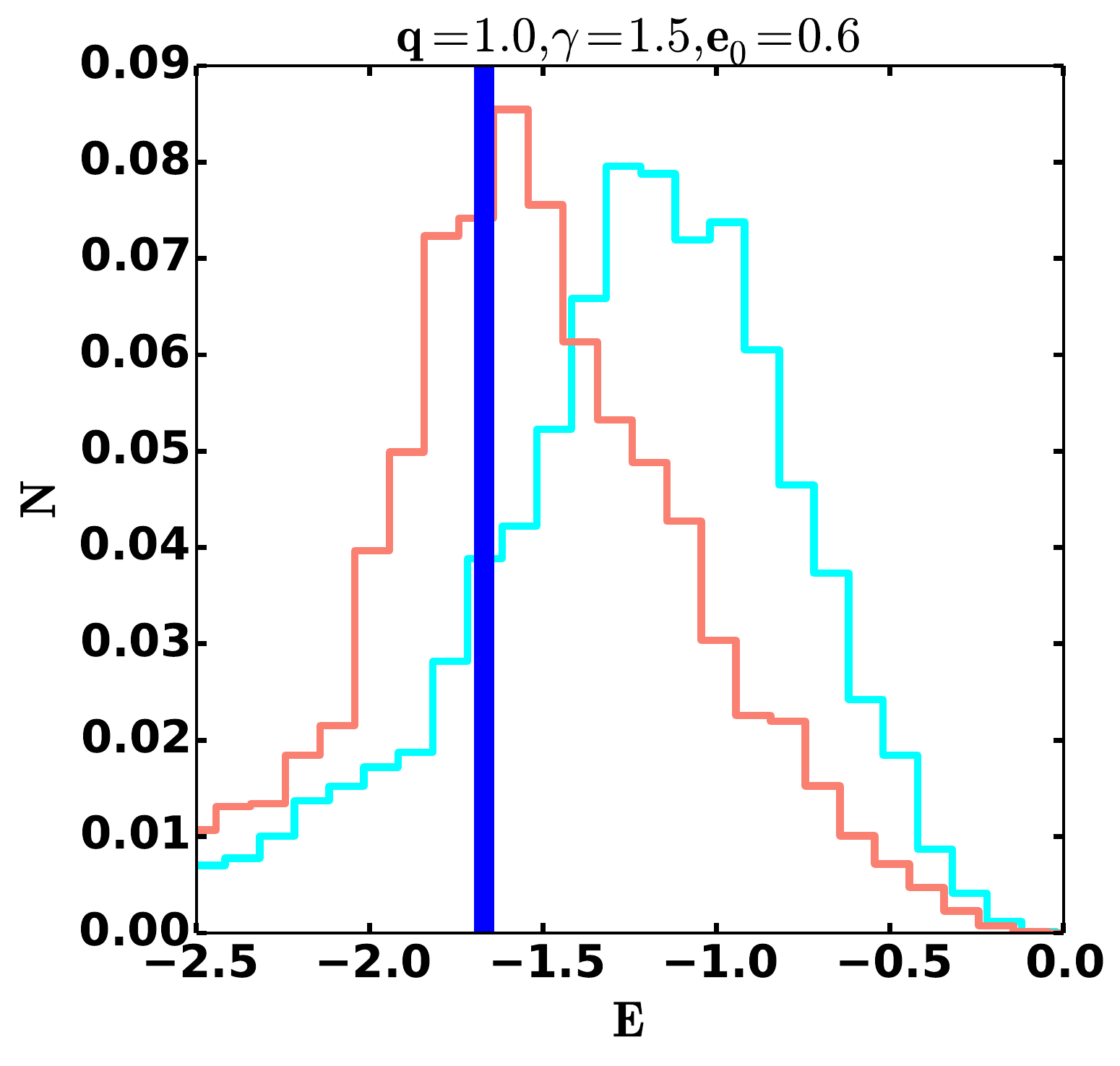}
  \caption{Normalized initial energy distribution of ejected stars in the actual (non-spherical) and spherical model for $\gamma=1.5,e_{0}=0.6$ and $q=0.5$ (top), $q=1.0$ (bottom), during the {\tt RAGA} simulations. 
  The ``core energy``, $\Psi(r_{\rm b})=-GM_{12}/r_{\rm b}+ \Psi_\star(r_{\rm b})$ with $r_{\rm b}$ the final core radius in the triaxial models,
  is given by the thick blue line. While in the equivalent spherical models the majority of ejected particles come from energies $E\sim \Psi(r_{\rm b})$, in the actual merger models they have energies much higher than the core value, i.e., they come from $r\gg r_{\rm b}$. 
  }
\label{energy_triaxial}
\end{center}
\end{figure}

\begin{figure*}
\begin{center}
 \includegraphics[angle=0,width=1.9in,height=1.9in]{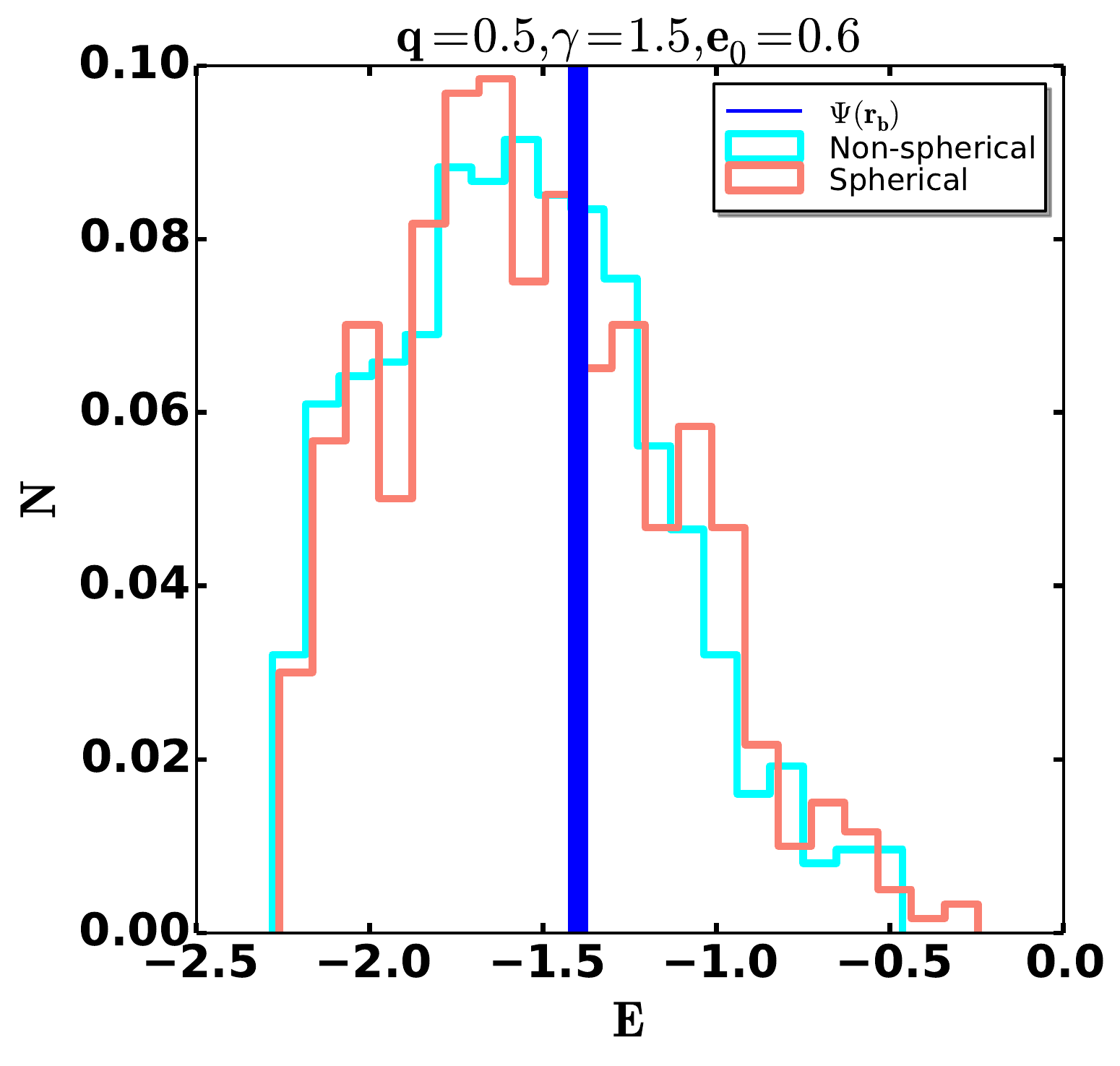}
   \includegraphics[angle=0,width=1.9in]{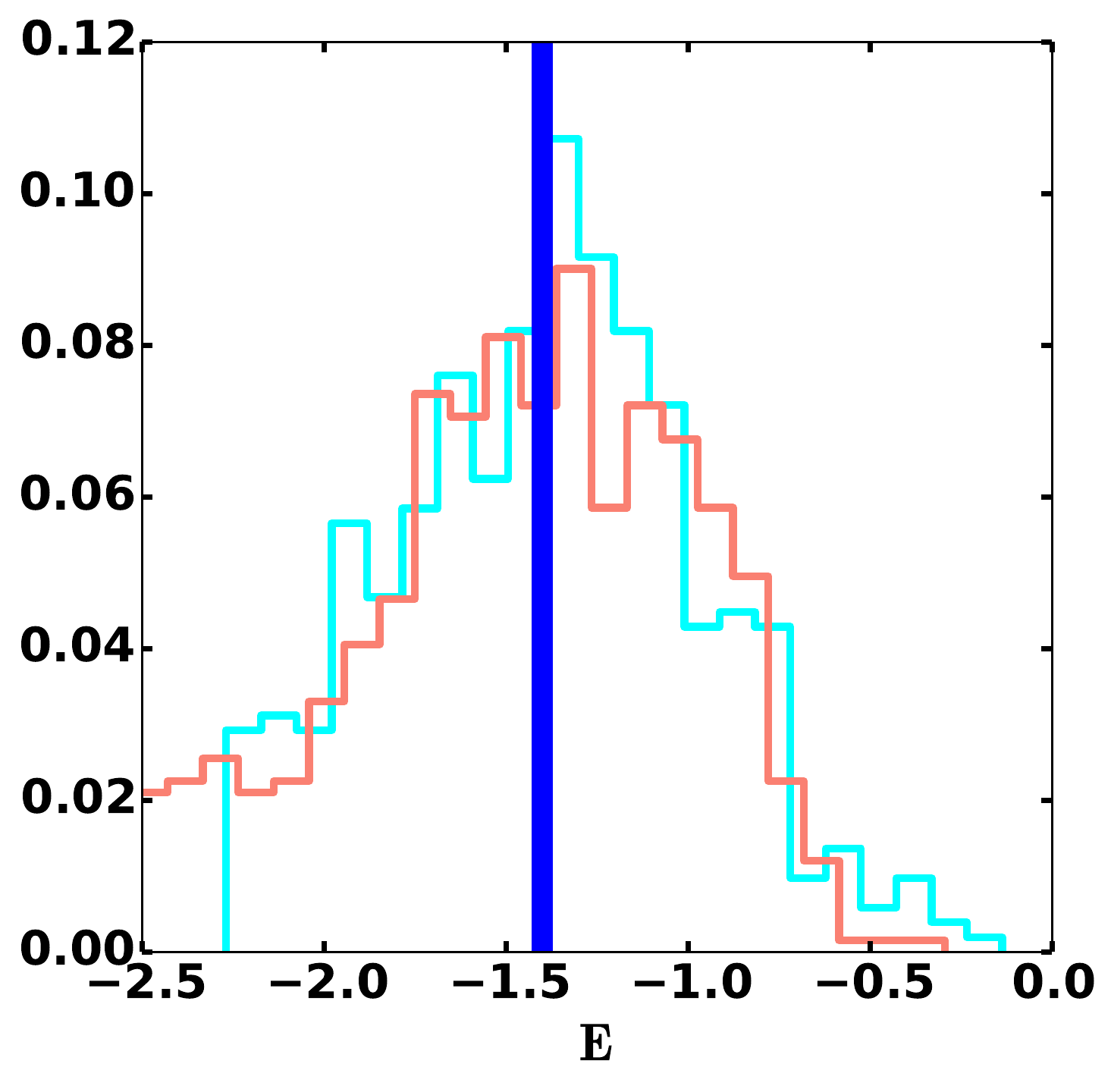}
   \includegraphics[angle=0,width=1.9in]{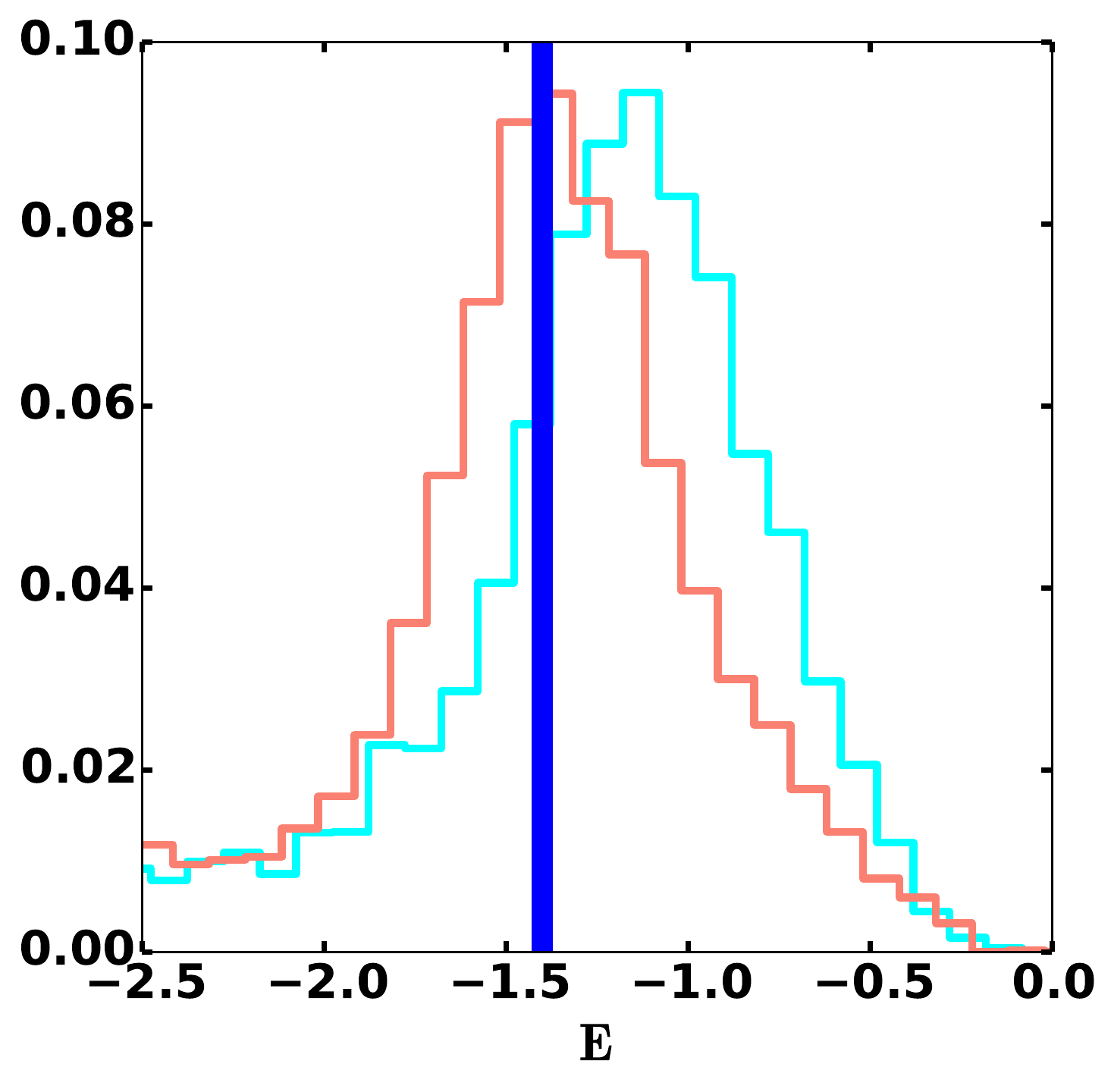}
   
   \includegraphics[angle=0,width=1.9in,height=1.9in]{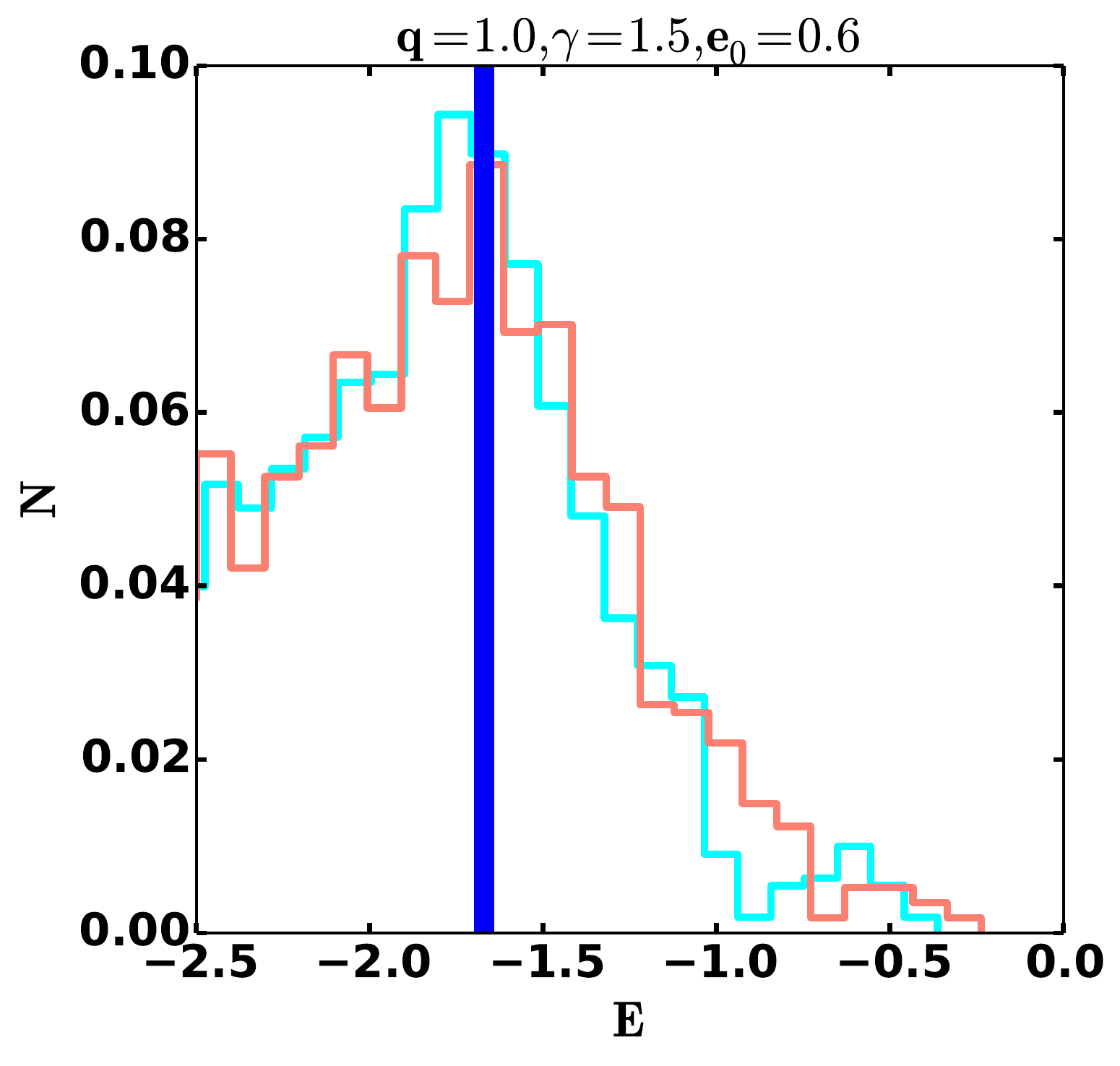}
   \includegraphics[angle=0,width=1.9in]{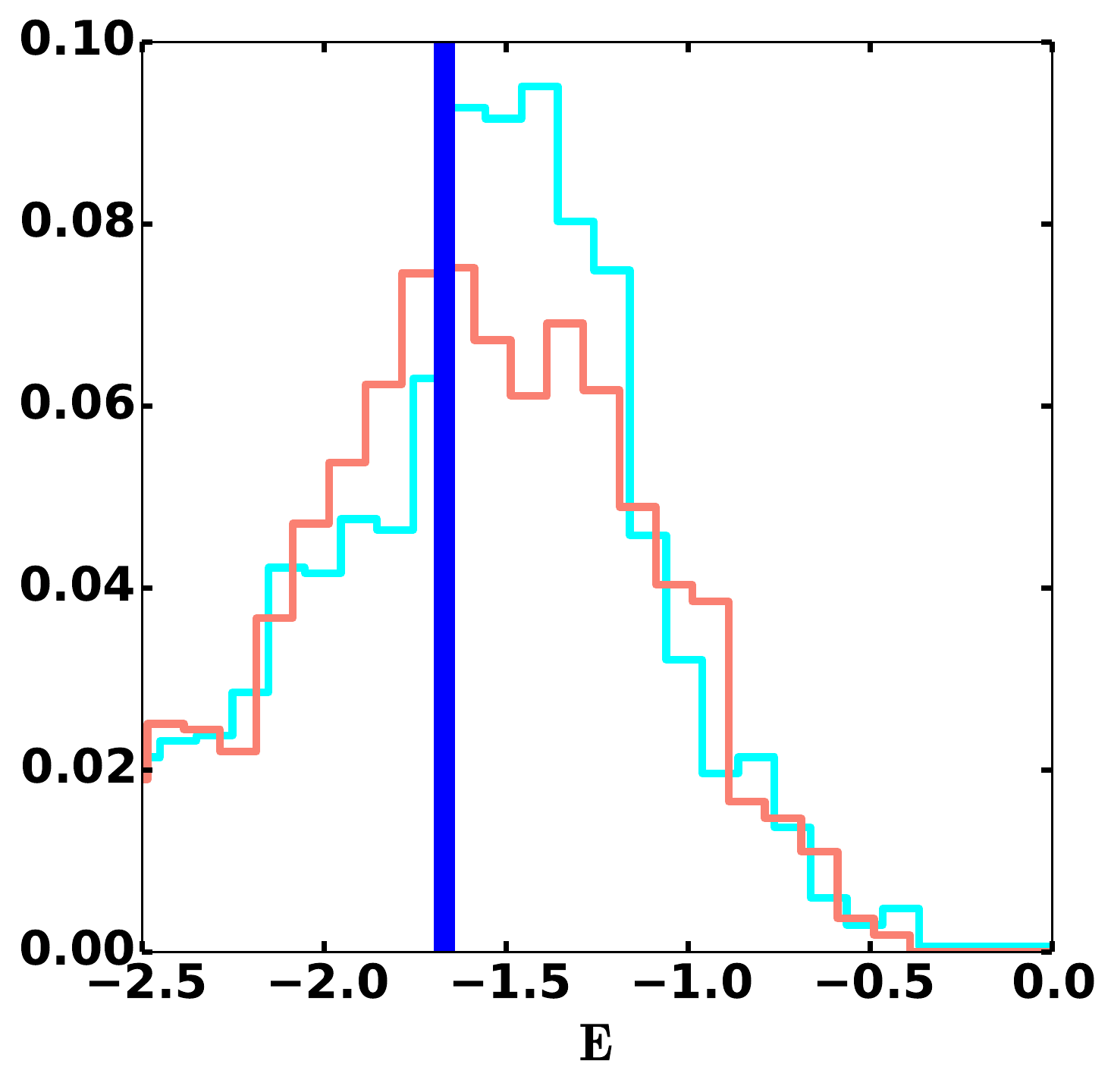}
   \includegraphics[angle=0,width=1.9in]{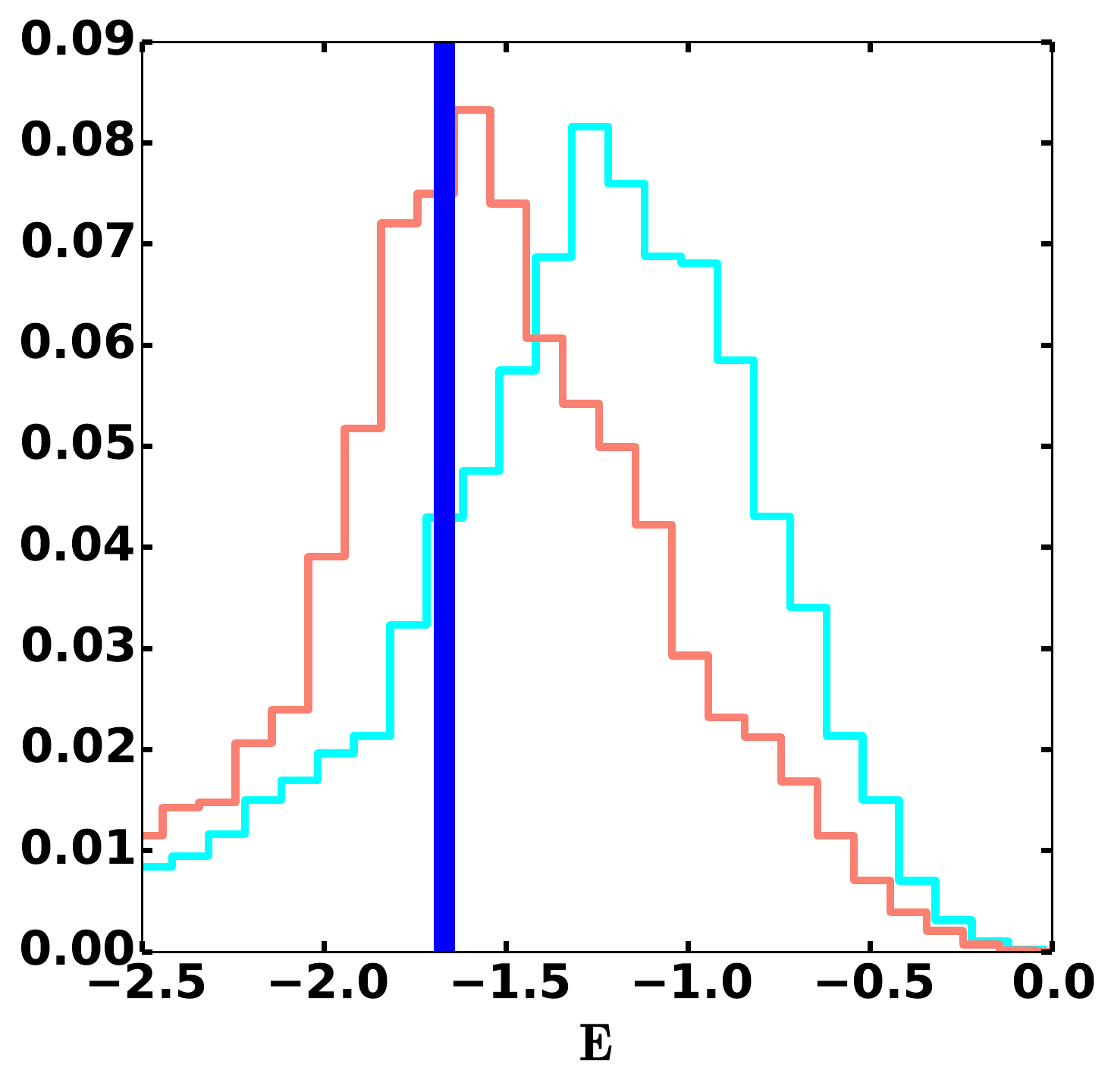}

  \caption{ Evolution of the normalized initial energy distribution of ejected stars in the merger (non-spherical) and equivalent spherical models. We show this distribution for $\gamma=1.5,e_{0}=0.6$ and $q=0.5$ (top), $q=1.0$ (bottom) in different phases of evolution during the hardening of the binary.
  We included all stars that were ejected by the binary during the time interval in which its semi-major axis evolves from $a_{\rm h}$ to $a_{\rm h}/3$ (left panels), from $a_{\rm h}/3$ to $a_{\rm h}/6$ (middle panels) and from  $a_{\rm h}/6$ to $a_{\rm h}/20$ (right panels). The ``core energy``, $\Psi(r_{\rm b})=-GM_{12}/r_{\rm b}+ \Psi_\star(r_{\rm b})$ with $r_{\rm b}$ the final core radius in the triaxial models, is given by the thick blue line. At early stages of evolution and in all models, the majority of ejected particles come from energies smaller or equal than the core energy $E\sim \Psi(r_{\rm b})$. At later times, however, the evolution is different.
While in the equivalent spherical models the peak of the distribution remains near $E\sim \Psi(r_{\rm b})$, in the non-spherical models the peak continues to  move towards higher energies. 
}
\label{check}
\end{center}
\end{figure*}

\subsection{Time evolution of core properties and the effect of triaxiality}

First, we consider the  effect of the binary on the structure
of the galaxy's core as a function of time. {In previous work \citep[e.g.,][]{2007ApJ...671...53M}, the mass deficit and core radius have increased continuously as the binary hardens, nearly as the square of the binary separation. In turn, this can lead to measured mass deficits as large as four times the black hole mass. As we will show, this behavior does not hold in our more realistic merger models.}

Our mass deficits and core radii are plotted in Figure \ref{deficit_ev2} as a function of binary hardness $a_{\rm h}/a$ for the cases of $q=0.5$ and $q=1.0$ with $\gamma=1.5,e_{0}=0.6$.
Simulations with a smaller binary mass ratio showed a very similar behaviour.
A particularly striking feature is that,  after the binary reaches a separation $a \approx a_{\rm h}/5$ there is little to no increase in the value of both $M_{\rm def}$ and $r_{\rm b}$. 
In the left panels of Figure \ref{dens-ts} we show the evolution of the density profile, where we see that at later times the density profile in the merger models remains nearly unchanged.
{The saturation of $M_{\rm def}$ and $r_{\rm b}$ at small $a$ is in contrast with prior findings \citep[e.g.,][]{2007ApJ...671...53M}. We suspect this is due to our use of non-spherical models.
Fortunately, we can  check with {\tt RAGA} what dominates the differences between the spherical and triaxial models.}

At the beginning of the simulation, we use {\tt RAGA} to construct the equivalent spherical model by averaging the density profile $\rho(r,\theta,\phi)$
of the actual merger model over the angles $\theta$, and  $\phi$, retaining only the radial dependence of the galaxy mass profile and potential $M(r)$ and $\Psi(r)$.
Thus, we enforce the model to remain spherical by setting to zero all the angle-dependent terms in the multipole expansion of the potential.
Then we simply integrate the stellar orbits in this equivalent spherical model, and compare the evolution of the mass deficit and core radius in these models to that found in the  non-spherical case. 
In a perfectly spherical galaxy with two-body relaxation turned off, the hardening of the binary towards smaller separations will cease at $a\sim a_{\rm h}$. The stalling happens because no other process exists that can repopulate 
the binary loss-cone after it has been fully depleted by dynamical ejections \citep{2002MNRAS.331..935Y,2015ApJ...810...49V}.
To avoid the stalling of the binary in our equivalent spherical models, we therefore use a non-zero relaxation rate of $R= \ln \Lambda / N=10^{-4}$, with $\ln \Lambda$ the Coulomb logarithm. Practically, $R$ determines the rate at which stars will interact with the central binary and consequently the binary hardening rate. On the other hand,   the values of mass deficit and break radius at a given $a$ during the hardening sequence are nearly independent of $R$ -- the binary will have  to eject the same number of stars in order to experience the same fractional change in $a$ \citep{2007ApJ...671...53M}.
Thus, the evolution of $M_{\rm def}$ during this phase is not sensitive to the exact value of $R$ adopted, as we  confirmed by running models with different values of $R$ (in the range $10^{-6}$ to $10^{-4}$ for $q=0.5$ and $e_0=0.6$).

The comparison between the merger model and its spherical equivalent model is shown in Figure \ref{deficit_ev2} for $\gamma=1.5,e_{0}=0.6$, and for $q=0.5,\ 1.0$. The evolution in the two cases is different, with both the mass deficit and break radius evolving much more rapidly and towards higher values in the spherical equivalent models. 
In Figure \ref{dens-ts} we show the evolution of the density profiles in the two geometries. While after some point the inner density profile in the merger models remains nearly unchanged,  in the equivalent-spherical models the density keeps evolving and $r_{b}$ increases with time.
In order to explain the different evolution seen between collisional spherical systems  
and collisionless triaxial ones,  we note some key differences between the two models.

First, the continued hardening of the binary requires a repopulation of the depleted orbits, and the mechanism of loss-cone repopulation depends on galaxy morpholgy.
In the triaxial models, where relaxation is assumed to be negligible, the time evolution of the central density is solely determined  by the ejection of stars 
on low angular momentum orbits. In the spherical models instead,
the evolution of the central density reflects the imbalance between the  loss of stars that are ejected by the binary and the repopulation of the loss region due to stars that diffuse in energy from  large radii. Because the relaxation time is so long, the loss term is generally much larger than the latter term and the mass deficit increases with time.
For a careful study of core depletion in collisional spherical nuclei we redirect the reader to \citet{2007ApJ...671...53M}.
The other key difference between the two models is that morphology affects the ``loss region'':
the behaviour seen in Figure \ref{deficit_ev2} 
could be at least in part attributed to the
 differences in the orbital distributions of stars used by the binary to harden between the spherical and triaxial cases. We will demonstrate this point in what follows.

The results observed in Figure \ref{deficit_ev2}   can be understood by studying  the initial orbits of particles that are eventually ejected by the central binary in the two models.
We analyze the properties of stars that interact with the binary in
the following way. For each particle entering or leaving the sphere
of radius $5a$ from the binary centre-of-mass, we record the initial
and final values of energy. Particles that leave the interaction
zone with a positive total energy are ejected from the galaxy. The normalized distribution of {\it initial} energies of these ejected particles is displayed in Figure \ref{energy_triaxial} for the time interval in which $a$ evolves from $a_{\rm h}$ to its final value in the non-spherical models: $a=2.9\times 10^{-4}$  for $q=0.5$ and $6.1\times 10^{-4}$ for $q=1.0$. We see that in the spherical case most ejected particles  come from energies that are near the core energy, approximated as $\Psi(r_{\rm b})=-GM_{12}/r_{\rm b}+ \Psi_\star(r_{\rm b})$ with  $\Psi_\star(r_{\rm b})$ the stellar potential. In the non-spherical case instead, the ejected particles come from energies that are much higher than the core energy, i.e. from much larger radii. Thus, while the number of stars ejected in the two models is approximately the same,  in the non-spherical case the mass deficit and break radius remain relatively constant  because the ejected stars mostly come from $r\gg r_{\rm b}$. It is clear from this analysis that mass deficits in triaxial galaxies are smaller than in the spherical geometry, since most stars ejected by the binary are from orbits with very large characteristic radii.
The fact that for triaxial systems the hardening of a massive binary is almost entirely driven by draining of centrophilic orbits that arrive from outside the SMBH binary sphere  of influence was already noted before \citep[e.g.,][]{2015MNRAS.446.3150V,2019MNRAS.484.2851L}.
Here, we have shown that the morphology of a galaxy has a direct impact on the evolution of its nuclear density profile.

In Figure \ref{check} we look in more detail at the normalized initial energy distribution of ejected stars in the non-spherical and equivalent spherical models in different phases of evolution during the hardening of the binary. The first phase includes all stars that were ejected by the binary during the time interval in which its semi-major axis evolves from $a_{\rm h}$ to $a_{\rm h}/3$, the second phase from $a_{\rm h}/3$ to $a_{\rm h}/6$ and the third phase from  $a_{\rm h}/6$ to $a_{\rm h}/20$. As shown in Figure \ref{check} at early stages of evolution and in all models (left and middle panels), the distributions are all similar to each other, with the majority of ejected particles coming from energies near the core energy $E\sim \Psi(r_{\rm b})$. Thus, in this stage of evolution we expect $M_{\rm def}$ and $r_{\rm b}$ to evolve similarly in all models. This explains why in Figure \ref{deficit_ev2} at $a\gtrsim a_{\rm h}/5$ the non-spherical and spherical models give similar results regarding the time evolution of $M_{\rm def}$ and $r_{\rm b}$.
At later times, however, the behaviour is different.
In the equivalent spherical models the peak of the distribution remains near $E\sim \Psi(r_{\rm b})$, while in the non-spherical models the peak continues to  move towards higher energies. This 
explains why  at later times, $a\lesssim a_{\rm h}/5$, the evolution of $M_{\rm def}$ and $r_{\rm b}$ depends on the model morphology and
appear to proceed faster in the spherical case  (see Figure \ref{deficit_ev2}).

\subsection{Multi-stage mergers}

\begin{table*}
  \begin{center}
    \caption{MULTI-STAGE MERGERS}
    \label{tab:table4}
    \begin{tabular*}{\textwidth}{@{\extracolsep{\fill}}|lll|lll|lll|}
      \hline 
      \textbf{Run} & \textbf{$\gamma$} & \textbf{$M_2$} & \textbf{$r_{\rm b, 2}$}  & \textbf{$r_{\rm b,3}$}  & \textbf{$r_{\rm b,4}$}  & \textbf{$M_{\rm def,2}$}  & \textbf{$M_{\rm def,3}$} & \textbf{$M_{\rm def,4}$}  \\ 
      \hline 
      1 & 1.5 & 0.001 & 0.11  & 0.13  &  0.14 & 0.66  & 0.69 & 0.67 \\
      2 & 1.5 & 0.0025 & 0.10   & 0.11  & 0.12  & 0.48  & 0.51 & 0.46 \\
      3 & 1.5 & 0.005 & 0.11   &  0.13 & 0.14 &  0.49  & 0.47 & 0.41  \\
      4 & 1.5 & 0.01 & 0.12  &  0.14 &  0.16 & 0.51  & 0.42  & 0.38 \\
      \hline
    \end{tabular*}

 \end{center}
 \small Break radii and mass deficits (scaled to $M_{12}$) for the multiple merger simulations. Break radii and mass-deficits of the form $r_{\rm b, N},M_{\rm def, N}$ with $N=2,3,4$ refer to the core-S\'{e}rsic fit method value for the break radius and the mass-defict after the second, third and fourth merger respectively. 
  
\end{table*}

\begin{figure*}
\begin{center}
 \includegraphics[angle=0,width=1.82in]{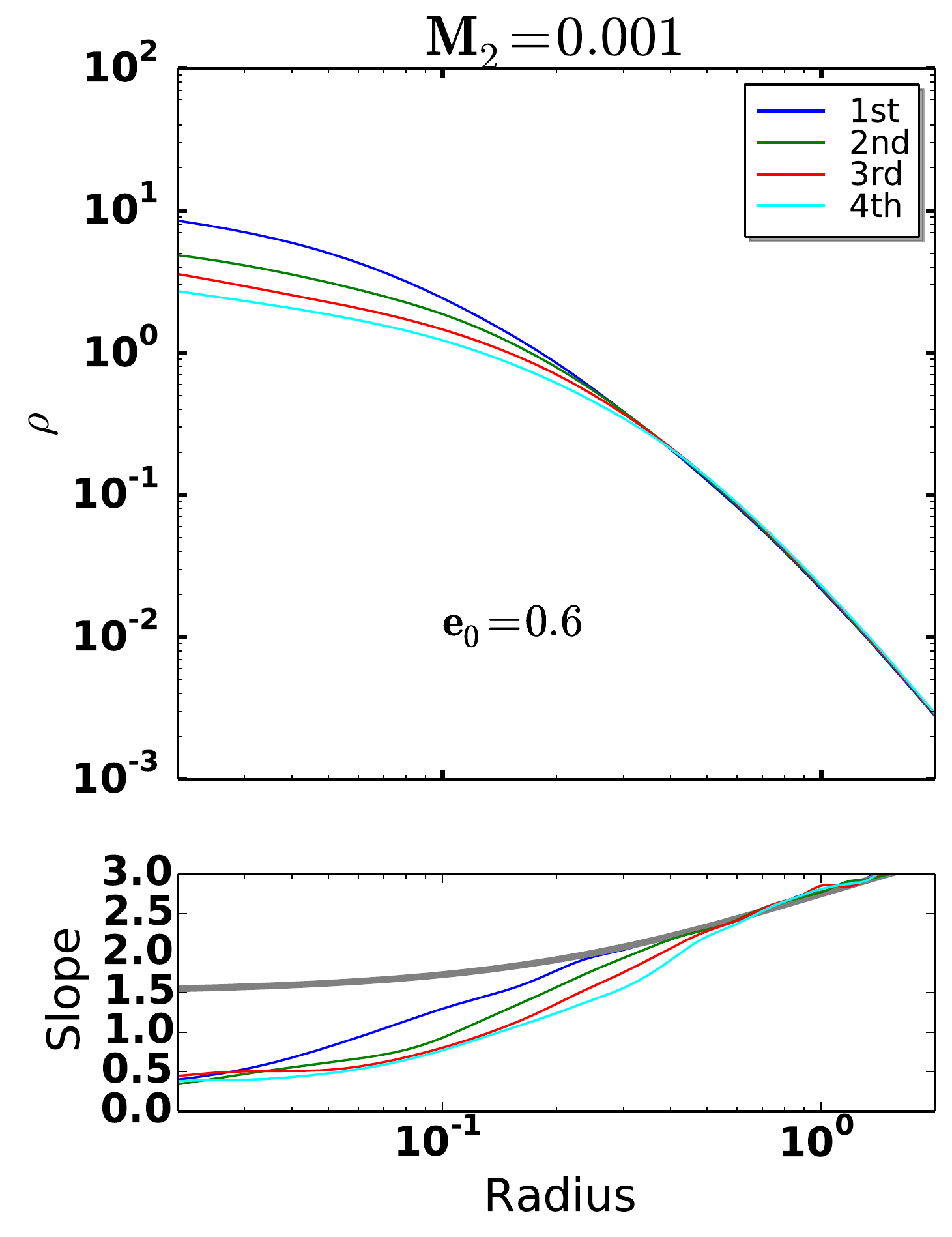}
   \includegraphics[angle=0,width=1.72in]{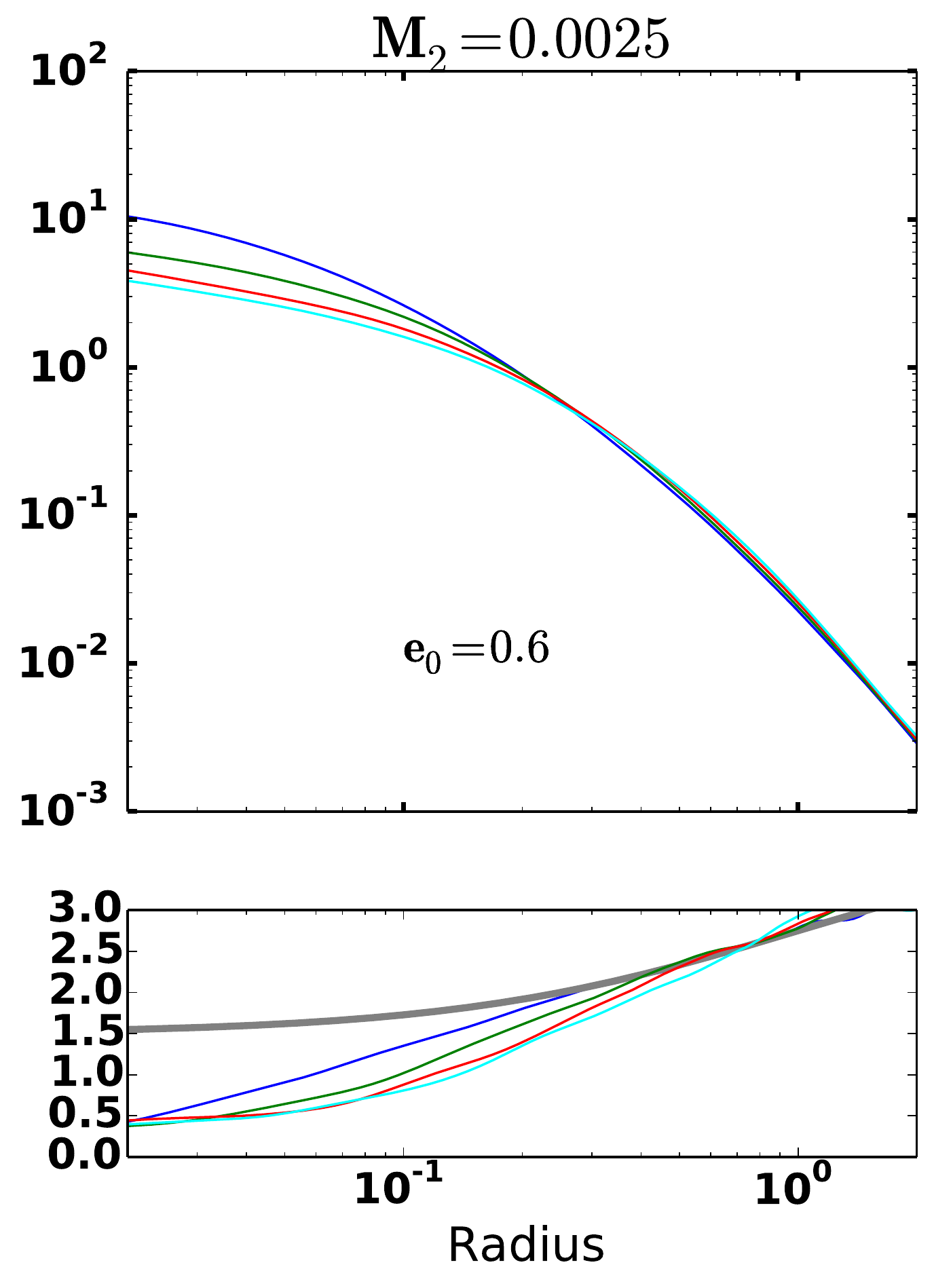}
   \includegraphics[angle=0,width=1.72in]{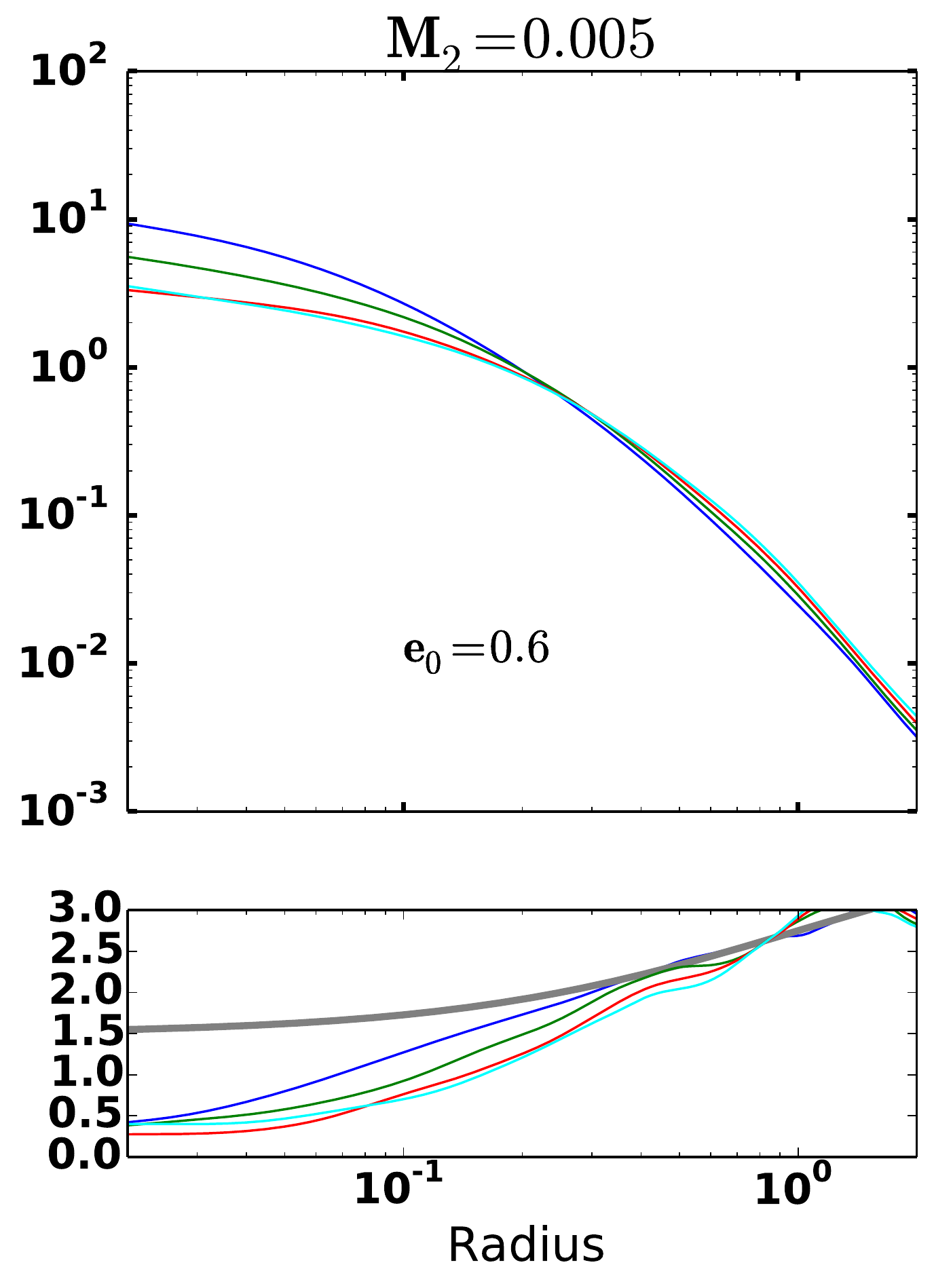}
   \includegraphics[angle=0,width=1.72in]{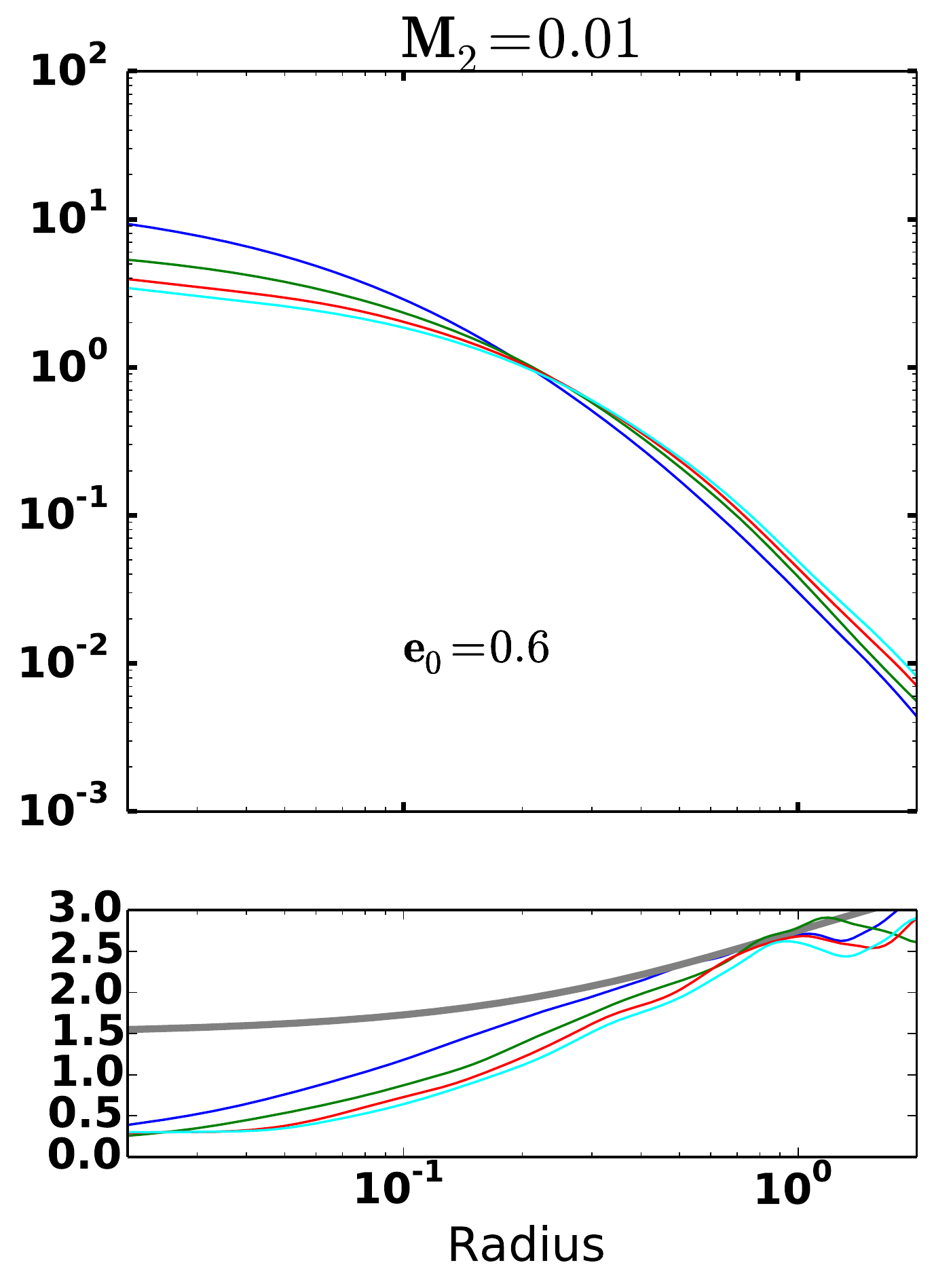}

   \includegraphics[angle=0,width=1.7in,height=1.78in]{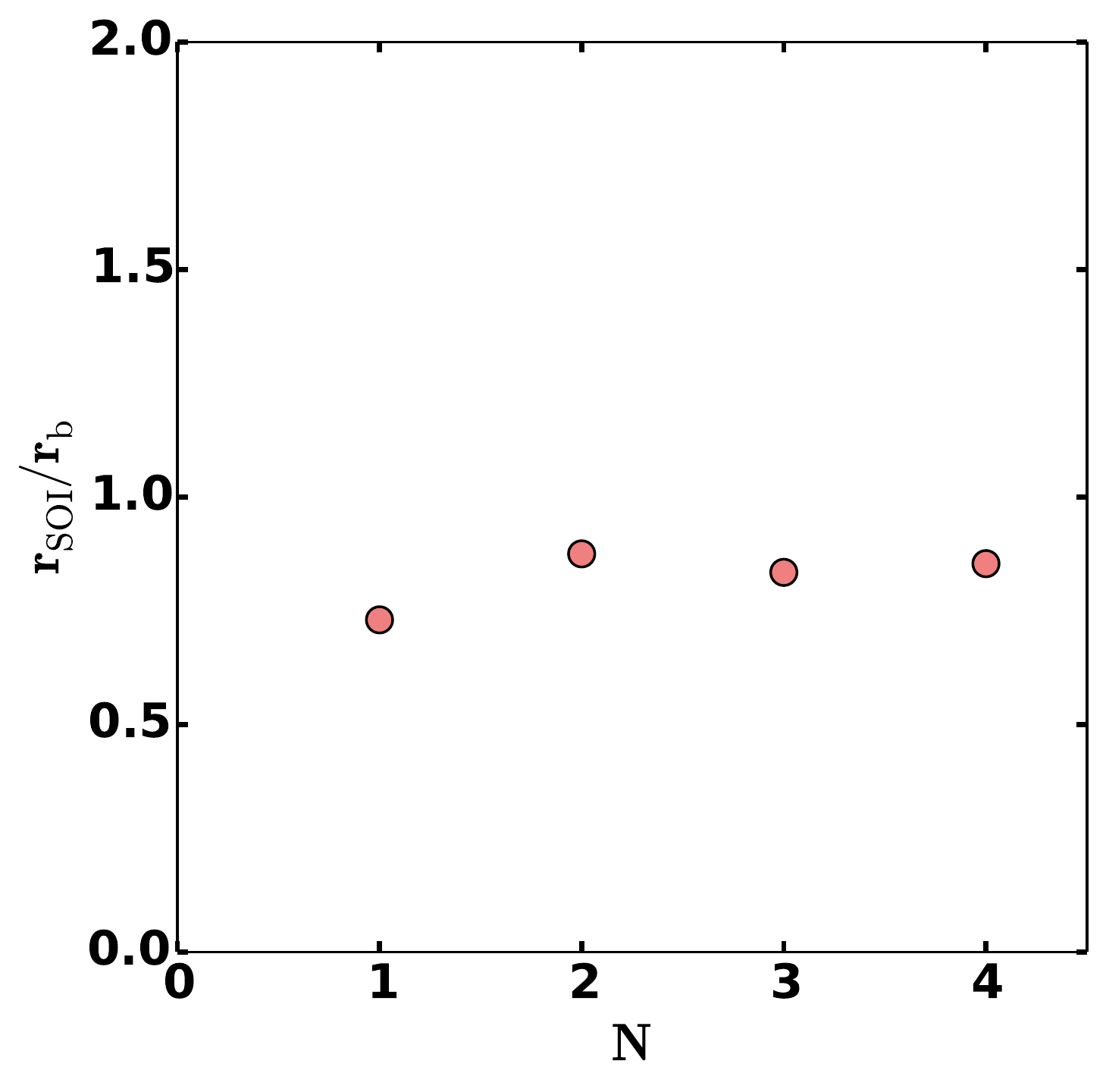}
   \includegraphics[angle=0,width=1.7in]{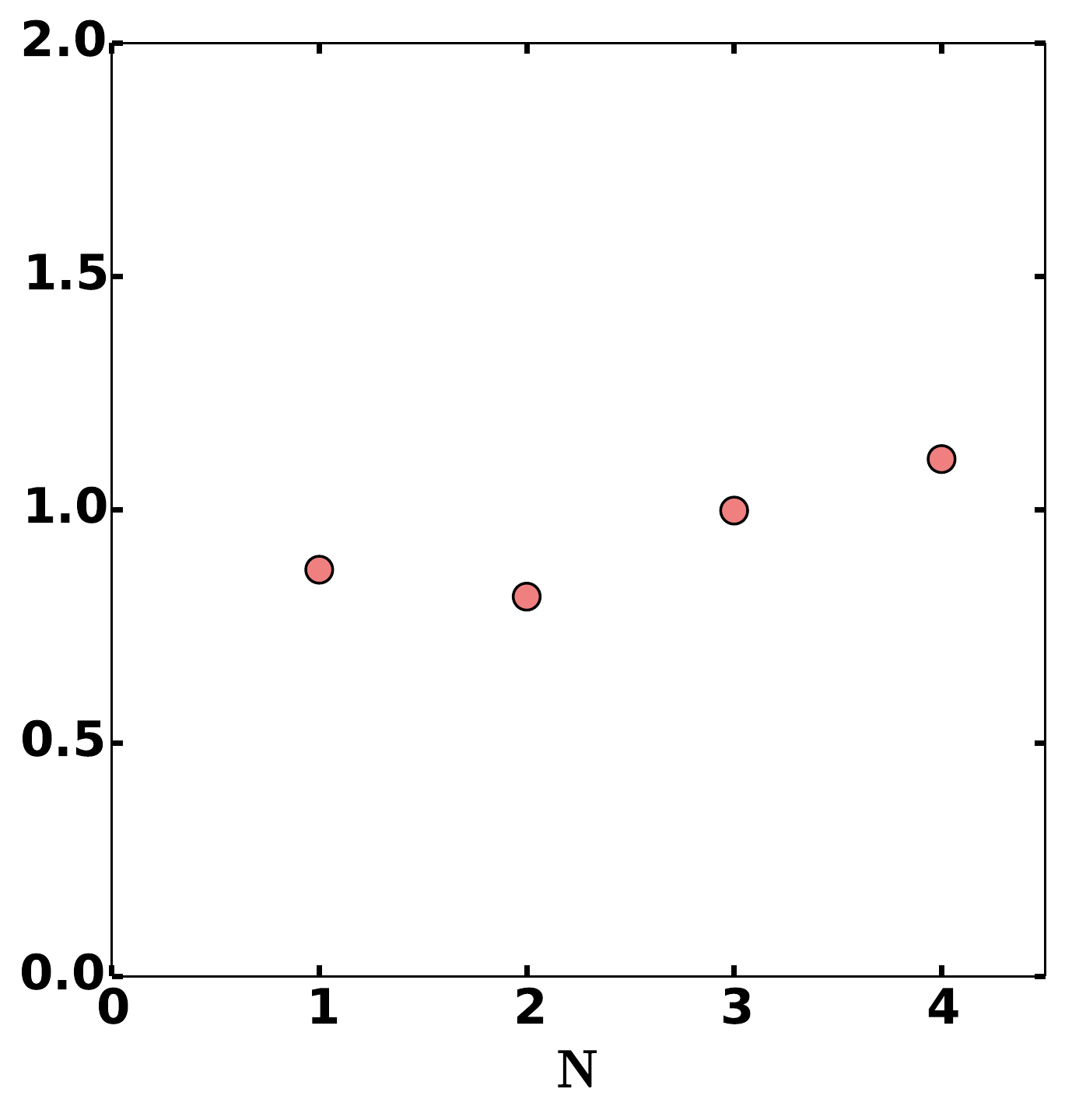}
   \includegraphics[angle=0,width=1.7in]{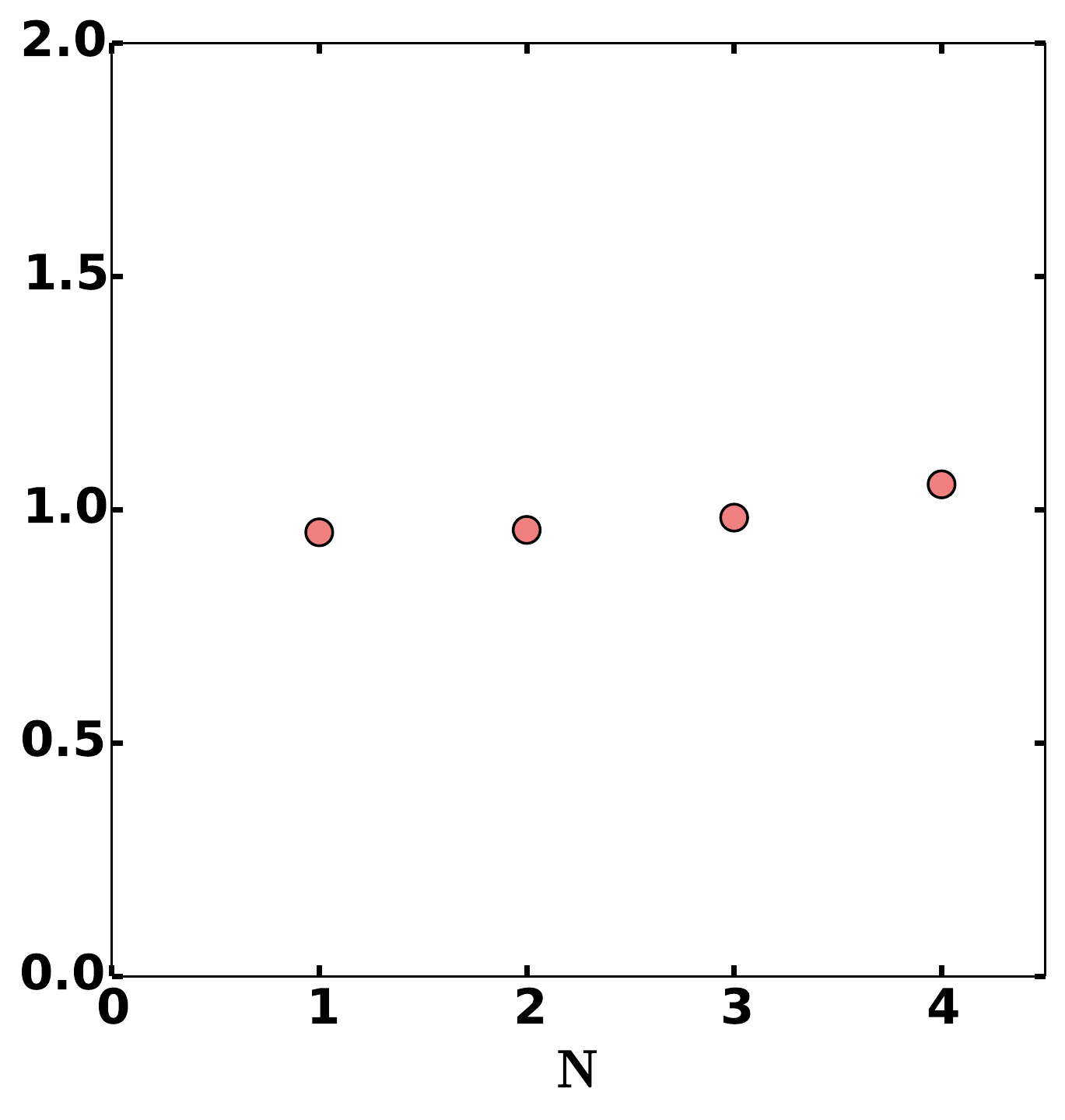}
   \includegraphics[angle=0,width=1.7in]{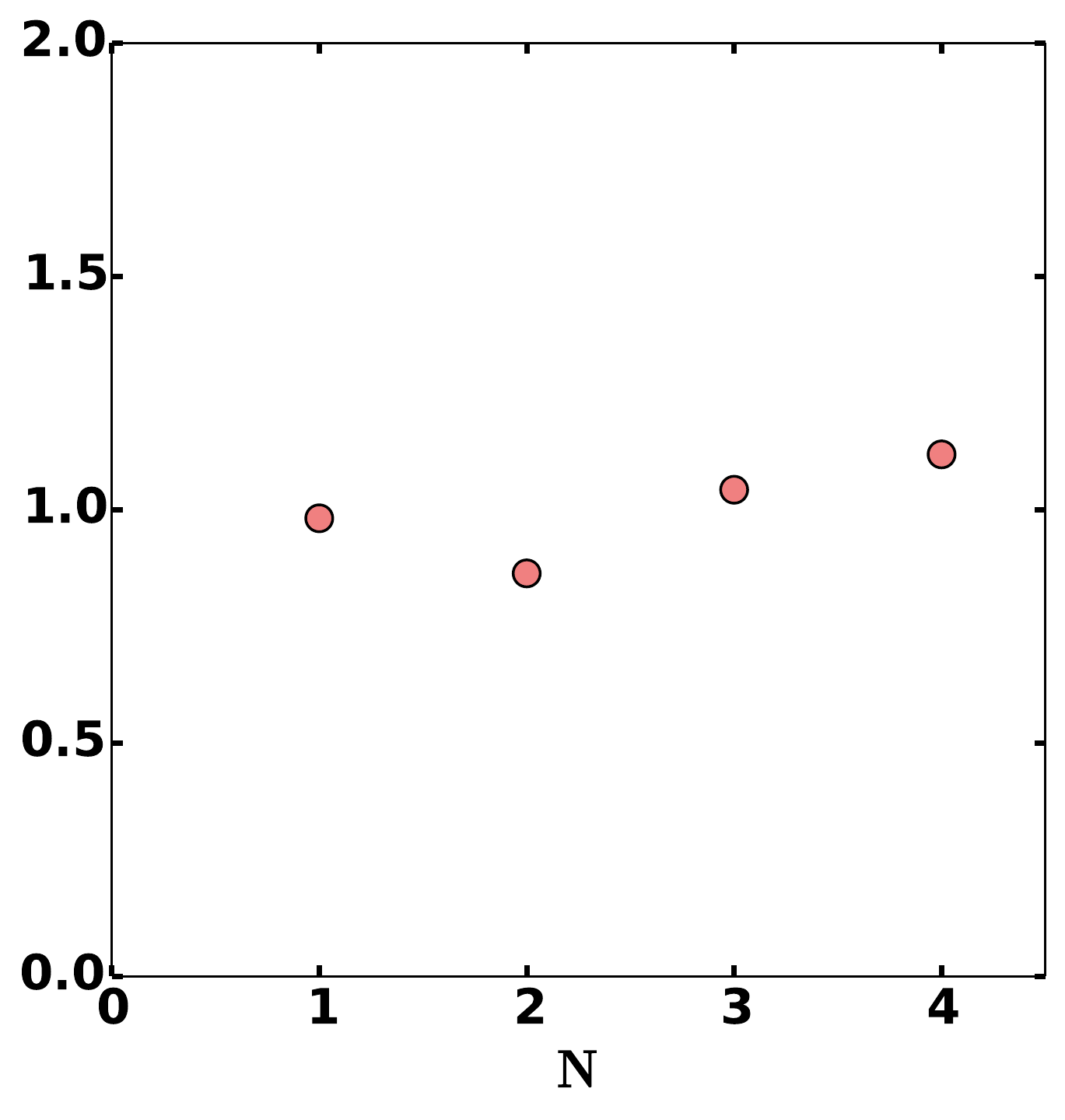}
   
   \includegraphics[angle=0,width=1.7in,height=1.78in]{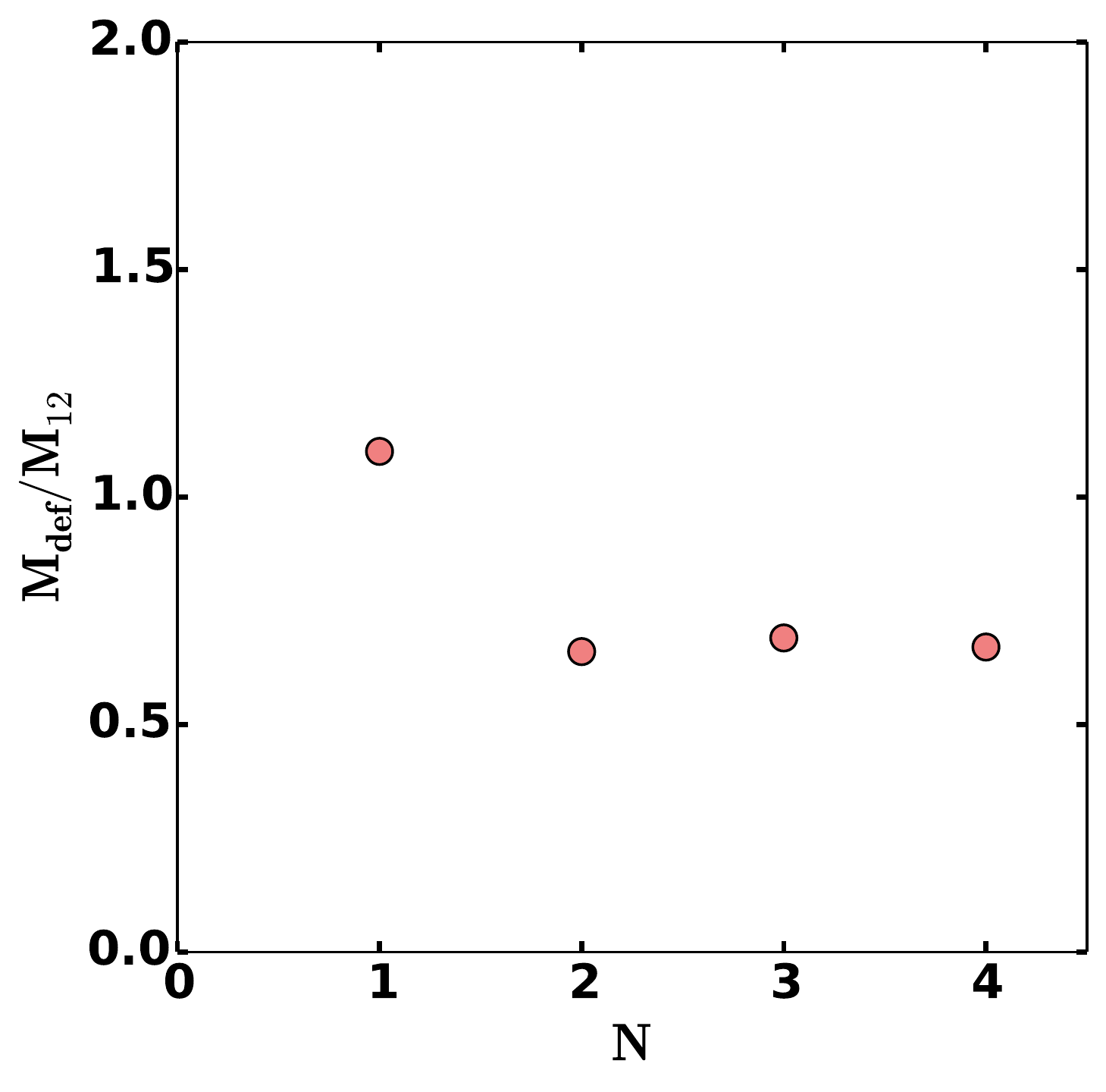}
   \includegraphics[angle=0,width=1.7in]{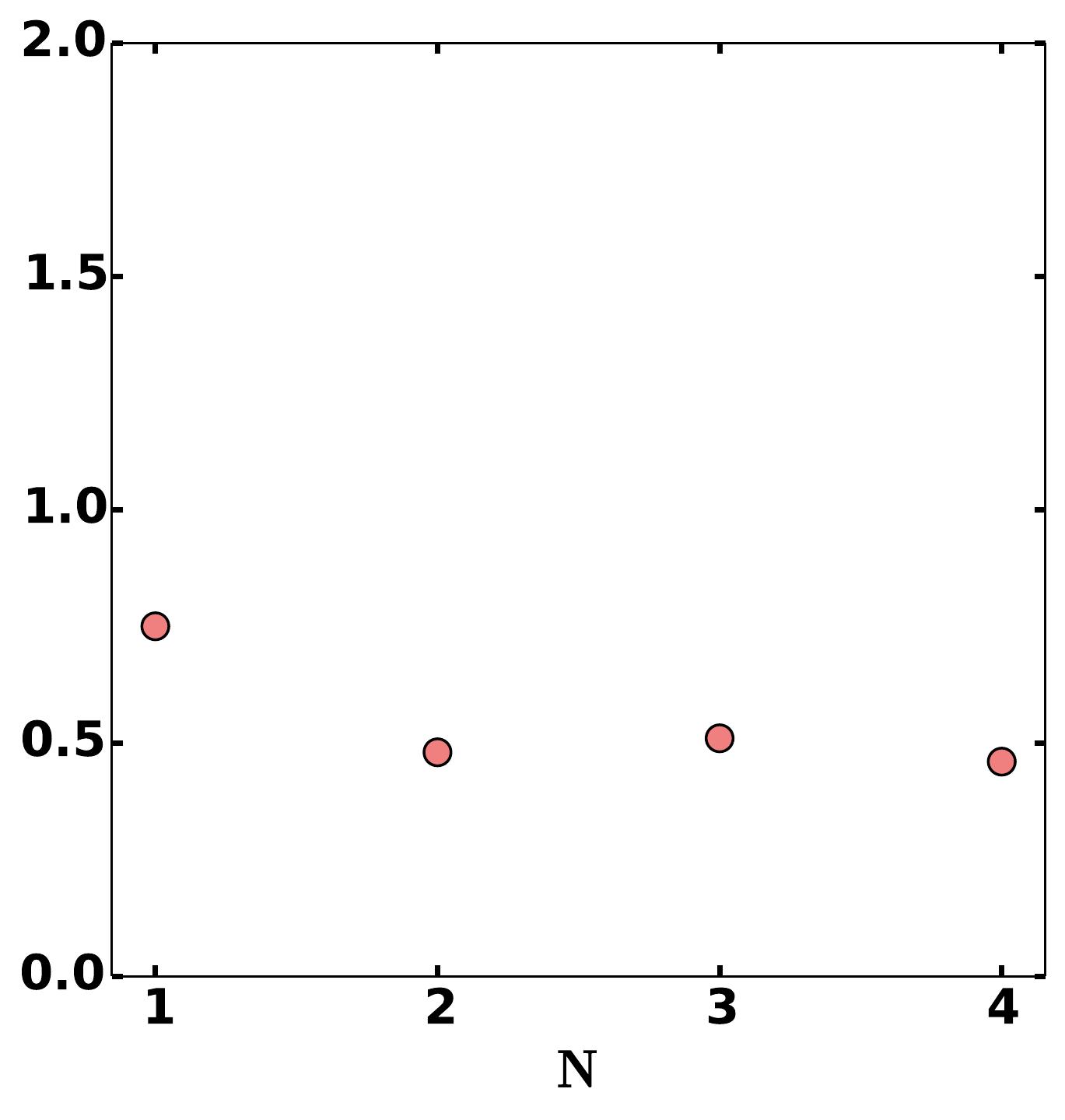}
   \includegraphics[angle=0,width=1.7in]{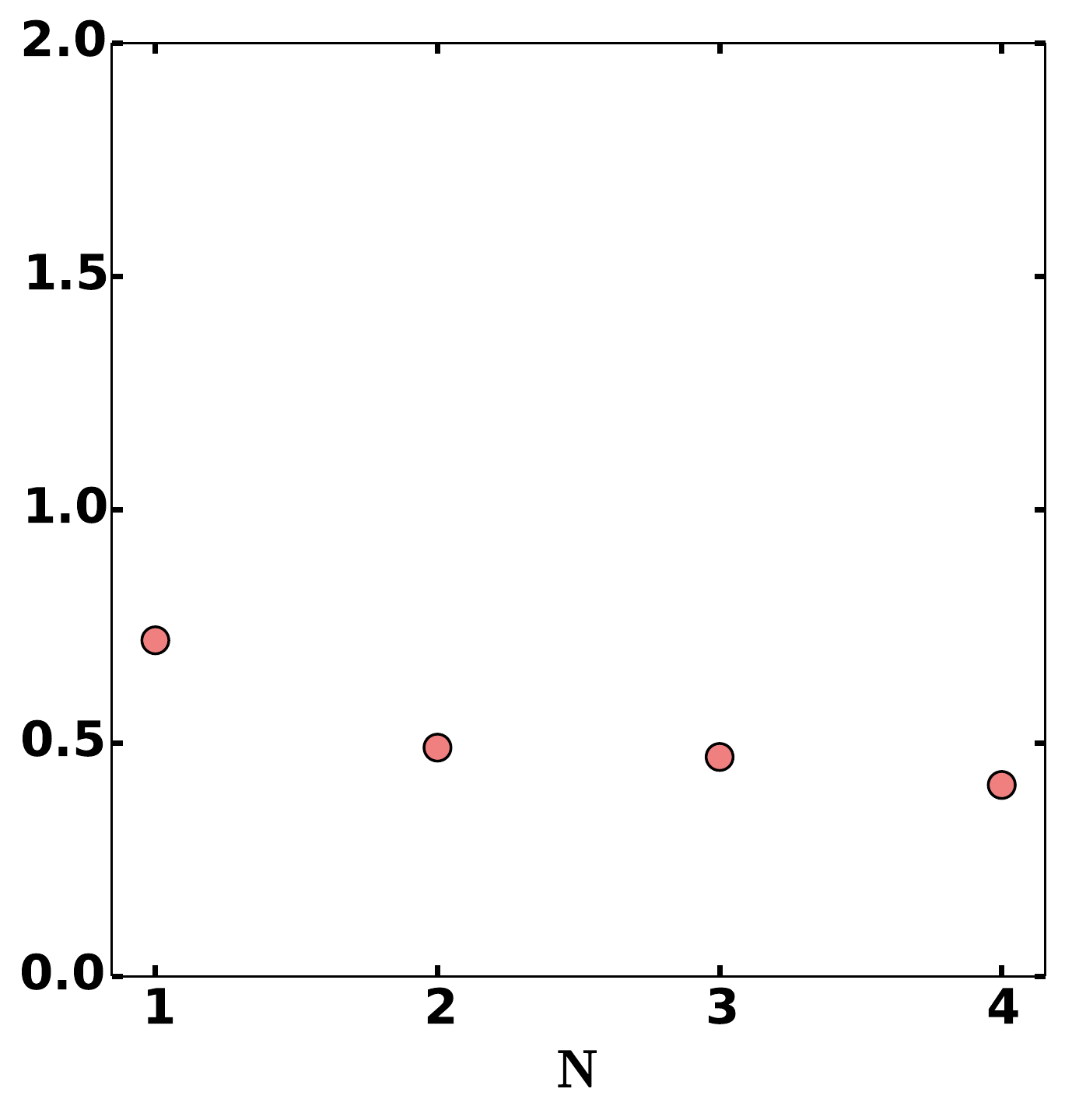}
   \includegraphics[angle=0,width=1.7in]{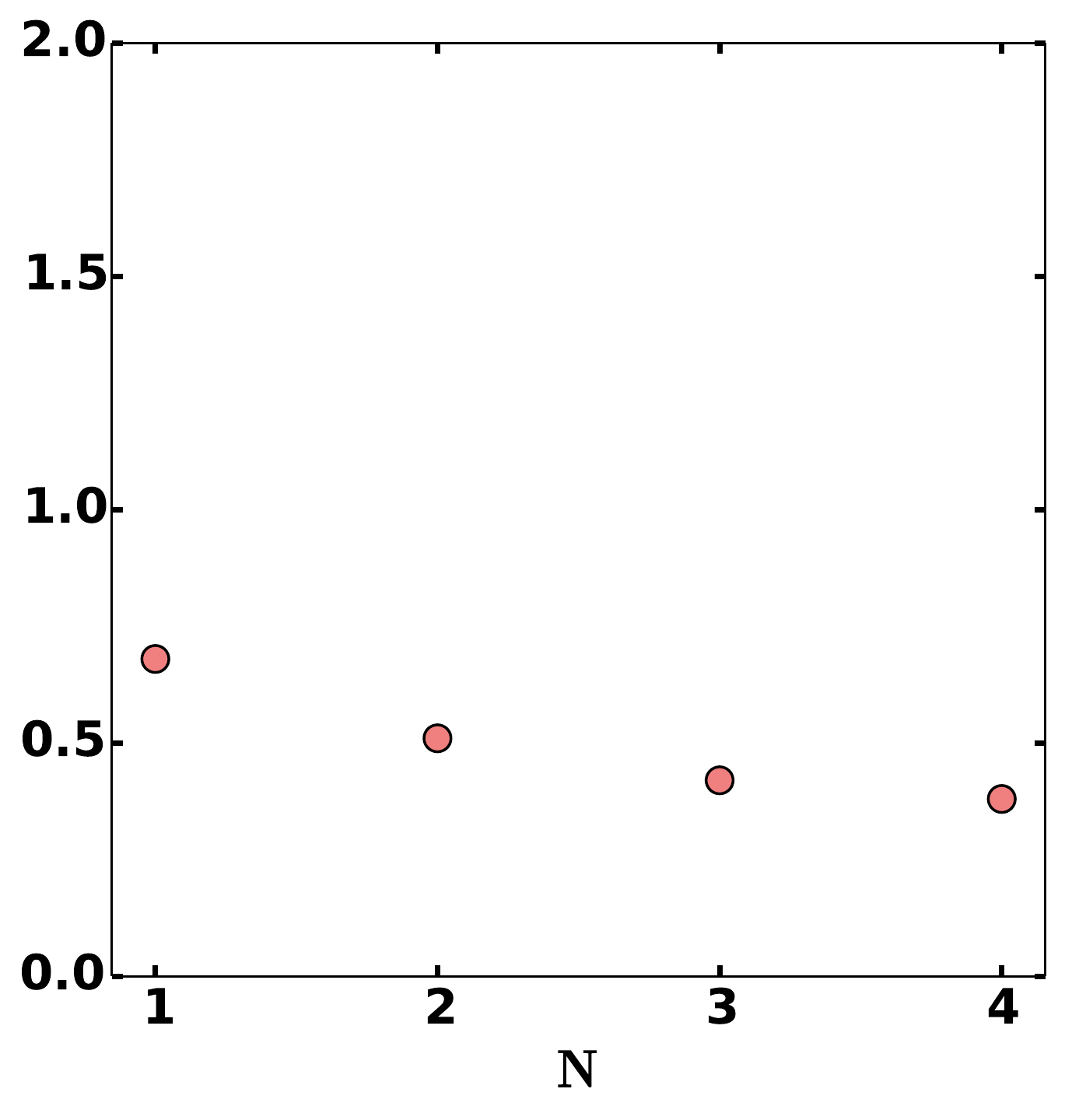}
   
  \caption{ Upper panels: density profiles and slopes of the remnant galaxy at the end of each repeated merger (color scheme). As in Figure \ref{com_2}, the solid grey line depicts the density slope of the primary galaxy initial profile. Middle panel: ratio of sphere-of-influence radius to break radius after each merger.  For all mass-ratios the ratio $r_{\rm SOI}/r_{\rm b}$ remains $\approx 1.0$  after $N$ mergers.
  Bottom panels: mass-deficit in units of the combined binary mass after each subsequent merger. The mass-deficit of the merger remnant remains $M_{\rm def} \approx  M_{12}$ even after $N=4$ repeated mergers and for all mass-ratios.
  All secondary galaxies  have initial eccentricity $e_{0}=0.6$ and $\gamma=1.5$.
  }
\label{multiple}
\end{center}
\end{figure*}

In this section we study the evolution of the break radius and mass deficit through consecutive galaxy mergers. More specifically,
we start each multi-stage merger simulation from each of the $\gamma=1.5,e_{0}=0.6$ $N$-body merger models described above. First,
 the two massive particles are  replaced by a single particle with
mass equal to their combined mass, and with position and
velocity equal to the center-of-mass values for the binary.
We then merge the remnant galaxy with  a new secondary galaxy with the same value of $M_2$ as in the first merger and $e_{0}=0.6$, until the stopping radius $a \approx a_{\rm h}/10$ is reached and the two SMBHs are again combined together in one particle.   We continue the same process for three mergers. We note here that all multiple mergers are performed in the x-y plane.
We consider four values for $M_2=(0.1, 0.25, 0.5, 1.0)\times 10^{-2}$. 
 Our results at the end of each merger are given in Table \ref{tab:table4} and Figure \ref{multiple}. We note here that the mass deficit is always expressed as a multiple of the combined mass of the two SMBHs, $M_{12}$, at the end of each  merger.

Figure \ref{multiple} shows the final density profiles, break radii and  mass-deficits of the galaxy remnants at the end of each  merger. Break radii and mass-deficits are computed using the core-S\'{e}rsic fit method described in Section \ref{sec:cores}. In the middle panel of Figure \ref{multiple} we plot the ratio $r_{\rm SOI}/r_{\rm b}$ of the remnant galaxies at the end of each merger computed in the same way as described in Section \ref{sec:scaling}. We see that the ratio  $r_{\rm SOI}/r_{\rm b}$   remains almost constant $r_{\rm SOI}/r_{\rm b} \approx 1$  after $N=4$ mergers. Similarly,  in the bottom panel of Figure \ref{multiple} we do not find the linear dependence of the mass-deficit on the number $N$ of repeated mergers predicted by \citet{2006ApJ...648..976M}. Instead, we find that the mass deficit of the merger remnant grows together with $M_{12}$ such that $M_{\rm def} \approx  M_{12}$ even after $N=4$ mergers. The same behaviour seems to hold for all mass ratios.
Our results imply that in real galaxies $M_{\rm def}\approx M_{12}$ and $r_{\rm SOI}\approx r_{\rm b}$, independently of the merger history of the galaxy and in good agreement with the observations \citep{2003AJ....125.2951G,2006ApJ...644L..21F,2007ApJ...662..808L,2016Natur.532..340T}.

A similar analysis  was recently considered by \cite{2019ApJ...872L..17R} and \cite{2021MNRAS.502.4794N}. These authors  perform repeated 
 galaxy-galaxy merger simulations similar to ours and study the evolution of the mass deficit and core radius, but find contrasting results.  \cite{2019ApJ...872L..17R} find that both the break radius and mass deficit increase cumulatively during multiple mergers, thus similarly but even more steeply than the predicted relation $M_{\rm def} \approx 0.5 N M_{12}$ by \citet{2006ApJ...648..976M}. {This strong dependence on the number of mergers led some authors to argue that the larger mass deficit values found in some galaxies were produced by a large number of repeated mergers \citep[e.g.,][]{2013ARA&A..51..511K,2014MNRAS.444.2700D}.}

 The difference with \cite{2019ApJ...872L..17R} can be explained by the method they use to compute the mass deficits. They consider repeated mergers with $q=1/9$ and estimate the central mass deficits from the surface
brightness profiles in the core region ($r < r_{\rm b}$) of the merger remnants with respect to a non-scoured merger model in which the progenitor galaxies initially had no SMBHs. This is similar to our definition of $M_{\rm ej}$ in Section \ref{discr} and  explains why their reported mass deficit values increase more rapidly than our $M_{\rm def}$.
To confirm this, we compute $M_{\rm ej}$ in our minor merger models ($M_2=0.001$) from the difference between the central surface brightness profiles  at  the  end  of the {\tt gyrfalcON} run of the first merger and at the end of the {\tt RAGA} run for each subsequent merger. We find that $M_{\rm ej}$ increases as $M_{\rm ej}/M_{12}$  = 1.12, 1.91, 2.62, and 3.22.

 On the other hand, 
 \cite{2021MNRAS.502.4794N} compute $M_{\rm def}$ from Equation (\ref{deficitfirst}) and
 find  that  core depletion and core size are mainly set by the initial major merger and are preserved
through subsequent mergers, similar to our conclusions. However, we do see that the mass deficit and the core radius increase  through multiple mergers as $M_{\rm def}\propto M_{12}$ and $r_{\rm b}\propto r_{\rm SOI}$.

In conclusion, the differences reported above might be simply explained by the different definitions of $M_{\rm def}$ adopted in the various studies. In this work we have 
given a preference to the ``observational'' approach, and  computed the mass deficit from the core-S\'{e}rsic fit model. This allows us to compare more directly with obervations as we do in the next section.

\section{Comparison to observed correlations}
\label{sec:scaling}

In this section we use our numerical models  to test the scouring  hypothesis that the central regions of bright galaxies were carved out by SMBHs. 
We do this by 
testing that our models reproduce the following observational results: 
(i) the
depleted stellar density in the cores of massive elliptical galaxies extends over the same radius as the gravitational sphere of influence of the central black hole \citep{2016Natur.532..340T};
and (ii) the observed mass deficit is comparable to the mass of the central SMBH \citep{2006ApJS..164..334F,2007ApJ...671.1456C,2013ApJ...768...36D}.

\citet{2016Natur.532..340T} (hereafter T16) studied a sample of 21 
core galaxies with reliable black hole measurements. They find that the region of depleted stellar density in the
cores of massive elliptical galaxies extends over the same radius as
the gravitational sphere of influence of the central black holes. They
define the SMBH radius of influence as the radius at which the enclosed stellar mass equals the central black hole mass, $r_{\rm SOI}$, and find  
a best-fit relation consistent with $r_{\rm b}=r_{\rm SOI}$, while the intrinsic scatter in the $r_{\rm b} - r_{\rm SOI}$ relation is
a factor of two smaller than that in the known scaling relations between
black hole mass and galaxy properties. The fact that this small scatter holds over a
wide range of galaxy environments sampled in the 21 core galaxies leads the authors to claim that the core-formation mechanism is homogeneous and a signature of the dynamical imprint of the central SMBHs. 

\begin{figure}
\begin{center}
  \includegraphics[angle=0,width=2.7in]{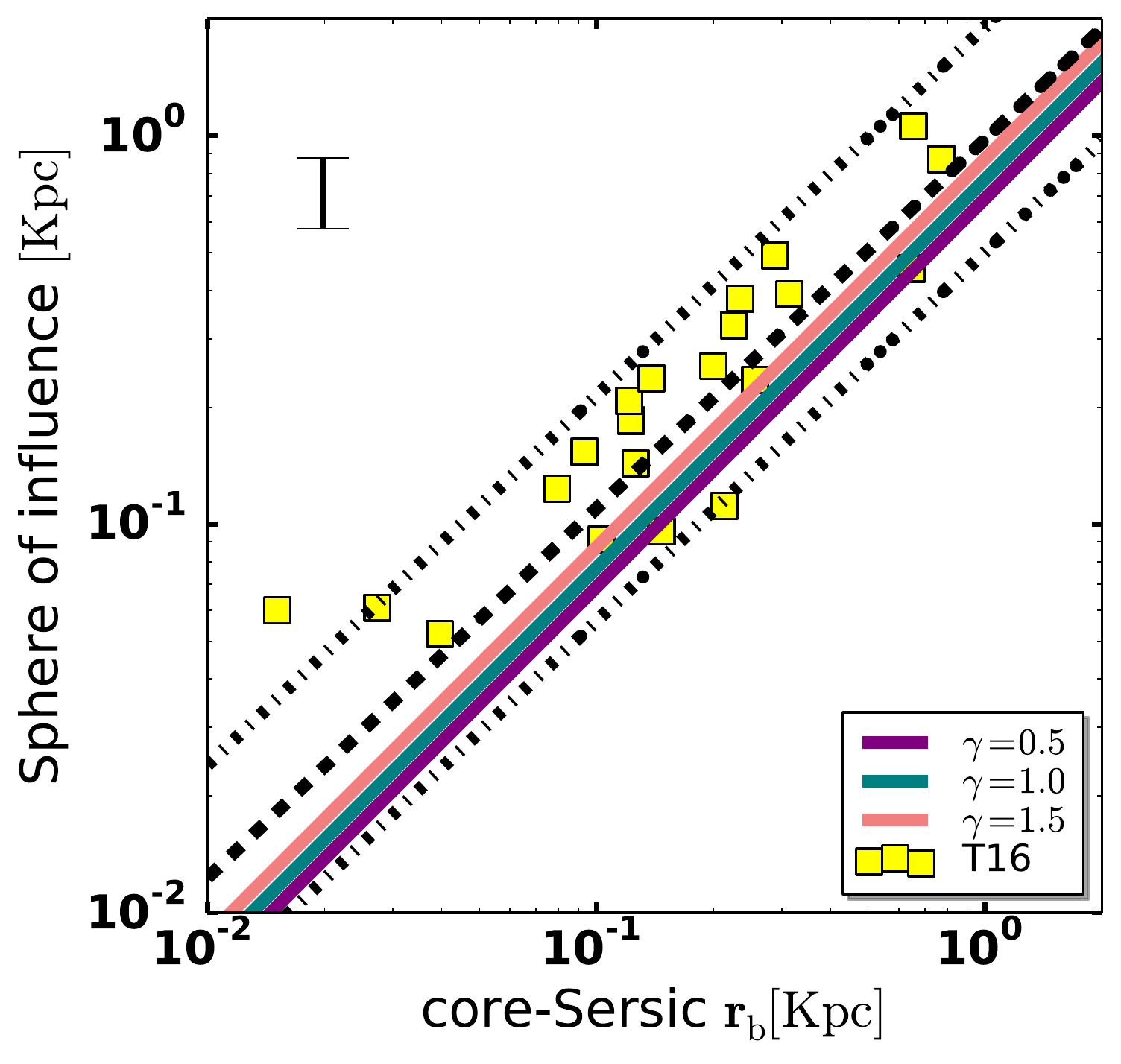}
  \caption{Sphere of influence and core radius scaling relation. Solid lines are from our numerical models and correspond to
   $\langle r_{\rm SOI}/r_{\rm b}\rangle =(0.69,0.77,0.87)$. The 21 core galaxies in Thomas et al. 2016 (T16) are marked as yellow squares. The best-fit linear correlations for the observed systems and confidence intervals are shown as a black dashed lines and are $\log (r_{\rm SOI}/\rm Kpc) = (- 0.01 \pm 0.29)+
(0.95\pm 0.08)\log (r_{\rm b}/\rm Kpc)$. Typical error bar is shown  upper-left.}

\label{scaling}
\end{center}
\end{figure}

We compare the results of our models to T16 by computing for each of our binaries the
ratio
\begin{equation}
 r_{\rm SOI}/r_{\rm b}
\end{equation}
where, in analogy to T16 we define here
\begin{equation}
M(r_{\rm SOI})=M_{12} ~.
\end{equation}
We find that $r_{\rm SOI}/r_{\rm b}$ does not depend significantly on the binary mass ratio and eccentricity. Therefore for each density profile slope, we average $r_{\rm b}/r_{\rm SOI}$ over the models with different values of $q$ and $e$.
 We find that $\langle r_{\rm SOI}/r_{\rm b}\rangle =(0.69,0.77,0.87)$. 
For comparison, the best-fit linear correlation for the observed 21 core galaxies in T16 is $\log (r_{\rm SOI}/\rm kpc)=(-0.01\pm 0.29)+(0.95\pm 0.08)\log (r_{\rm b}/\rm kpc)$.
This comparison is also illustrated in Figure \ref{scaling}.
Our results are well consistent with the observed homology result that $r_{\rm b}\approx r_{\rm SOI}$.

A similar analysis is performed in \cite{2018ApJ...864..113R}.
They also use large-scale $N$-body simulations and show that their merger models
produce a relation that is steeper than 
the observed scaling relation $r_{\rm b} - r_{\rm SOI}$. However, for  steeper initial stellar density
profiles $(\gamma = 1.5)$, their results match reasonably well the observed correlation, while it is significantly steeper than observed
for  $\gamma=1$.  
They use this result to support the scenario in which the progenitors of
the most massive core ellipticals are cuspy power-law
galaxies ($\gamma > 3/2$) with central SMBHs. Our results are only partly consistent with these findings.
While, we are also able to match the observed  $r_{\rm b} - r_{\rm SOI}$ scaling relation,  we do not find a large 
difference between models with different initial density profile slope $\gamma$.
We note here that  in our models the SMBH mass
is a fixed fraction, $M_1/M_{\rm tot}=0.01$, of the galaxy mass.  On the other hand, \cite{2018ApJ...864..113R} used the same galaxy model and
varied the SMBH mass over one order of magnitude range. 
The trend demonstrated by their models (their Figure 11) is clearly that the ratio $r_{\rm SOI}/r_{\rm b}$ increases slightly with the 
SMBH to galaxy mass ratio, $M_1/M_{\rm tot}$.
 This suggests that the exact value of 
$r_{\rm SOI}/r_{\rm b}$ might be related to the value of $M_1/M_{\rm tot}$, which is
in turn determined by the $M-\sigma$ correlation.

{At the order of magnitude level, we expect} $r_{\rm b}\simeq r_{\rm SOI}$. In fact, Equation~(\ref{Denergy}) shows that the binary injects in the nucleus an energy $\Delta \mathcal{E}\approx 2M_{12}\sigma^2$. This energy is comparable to the total energy of stars within the sphere of influence of the binary. Thus, by absorbing such energy the stars within $r_{\rm SOI}$ must undergo a substantial redistribution in phase space. If there was a pre-existing cusp, it will be disrupted out to a  radius $r_{\rm b}\approx r_{\rm SOI} $ \citep[e.g.,][]{2013degn.book.....M}, in agreement with our results.

{Observers have also investigated the ratio $M_{\rm def}/M_\bullet$, finding values of order unity, and in some cases significantly higher \citep[e.g.,][]{2006ApJS..164..334F,2007ApJ...671.1456C,2008MNRAS.391.1559H,2013ARA&A..51..511K,2013ApJ...768...36D}. We have shown above that $M_{\rm def}$ can be factors of a few discrepant from the actual ejected mass, both because of the unknown shape of the progenitors and due to triaxiality allowing the binary to access stars at large radius on centrophilic orbits. Moreover, its value depends significantly on the different fitting adopted. Thus, we do not compare quantitatively with observations here, however we note that our results do reproduce the range of values for $M_{\rm def}/ M_{\bullet}$ found in most observational work \citep{2013ApJ...768...36D,2013ARA&A..51..511K}.}

\section{Summary}

In this paper we use large scale $N$-body simulations of galaxy mergers to study the core formation process in large elliptical galaxies. 
We explore various parameters that affect the dynamical imprint that a massive black hole binary leaves on the structural parameters of the galaxy remnant. We quantify the cores formed in our models by measuring the break radius and mass deficit at the end of our simulations and study the way these quantities depend on the initialization parameters of our simulations. We run multistage mergers for some of our models to test the scenario, presented commonly in previous work, that both the core radius and mass deficit increase cumulatively with the number of mergers. Finally, we use these models as a test to the scouring hypothesis by comparing our results to the observed $M_{\rm def} - M_{\bullet}$ and $r_{\rm b} - r_{\rm SOI}$ correlations.

We simulated the entire galaxy merger process from its early stages. Our method combined a tree code until the binary becomes a close pair, and a Fokker-Planck Monte Carlo code for the later evolution of the binary. This method allowed us to explore the formation of depleted stellar cores in a realistic set up. 
Our main results are:

\begin{itemize}

\item[1)] The mass ejected, $M_{\rm ej}$, depends weekly on the mass-ratio of the SMBH binary and the initial density profile slope of the merging galaxies,
and it is of the order the SMBH binary mass.

\item[2)] In all of our models, after the binary reaches a separation $a \approx a_{\rm h}/5$, there is little to no change in the value of both the mass-deficit $M_{\rm def}$ and core radius, $r_{\rm b}$.
Thus, the estimated value for the  mass deficit is not sensitive to the actual stopping point of the integration.
This result is in contrast with previous calculations where simplified spherical initial conditions were used, and where  the mass deficit and core radius were predicted to 
increase much more rapidly during the hardening phase of the binary and to reach values as high as $M_{\rm def}/M_{12}\approx 5$ by the time of coalescence  \citep[e.g.,][]{2002MNRAS.331L..51M,2007ApJ...671...53M}.

\item[3)] 
In order to explain the different evolution observed by us and in the spherical models used by others, we evolve the SMBH binaries in 
a spherical-equivalent  model of the actual merger model.  The spherical-equivalent model is constructed  by retaining only the radial dependence of the galaxy mass profile and potential of the actual model.
 We find that the mass deficit and break radius increase much more rapidly  in  the  spherical-equivalent  models than in the  merger models, and in similar fashion to that observed in previous work.
We show that in the spherical models the ejected stars come from energies that are near the core energy, while in the non-spherical case they come from energies that are much higher than the core energy, i.e. from much larger radii. Thus, while the number of stars ejected in the two models is approximately the same, the damage done to the inner regions of the galaxy in the non-spherical case is less
 because the ejected stars come from radii $r\gg r_{\rm b}$.

\item[4)] Multi-stage merger simulations  show that the core radius and the mass deficit of a galaxy grow  such that the ratios $M_{\rm def}/M_{12}$ and $r_{\rm b}/r_{\rm SOI}$ remain approximately constant and near unity after the first merger. 
We conclude that the inner core  of a galaxy does not grow linearly with the number of mergers as often claimed in the literature \citep{2006ApJ...648..976M}. On the other hand, our results naturally
explains the small  scatter in the $r_{\rm b}$-$r_{\rm SOI}$ 
relation found by \citet{2016Natur.532..340T}.

\item[5)] Our models reproduce the observational results that the region of depleted stellar density extends over the same radius  as  the  gravitational  sphere  of  influence  of the central black hole, $r_{\rm b}\approx r_{\rm SOI}$ \citep{2016Natur.532..340T}, 
as well as the range of values  $0.5\lesssim M_{\rm def}/M_{\bullet}\lesssim 4$ found in  elliptical galaxies \citep[e.g.,][]{2006ApJS..164..334F,2007ApJ...671.1456C,2013ApJ...768...36D,2014MNRAS.444.2700D}.
We conclude that a SMBH binary scouring hypothesis provides a natural explanation for the observed central properties of bright galaxies and their extended cores.

\end{itemize}

In this paper we have performed $N$-body simulations to  explore the formation of galaxy cores
via mergers and the dependence on the galaxy progenitors masses, density profiles and orbits. 
This approach was useful to understand the main ingredients that affect the final central core properties of the merger remnant. In the future, we plan of extending  this work and consider more realistic sets of initial conditions where the galaxy progenitor orbits, masses and merger trees are extracted directly from cosmological simulations of galaxy formation. Moreover, we will consider the effect of gravitational wave recoil kicks on the kinematics and core scouring.
This will help to have a more complete picture of
the effect of SMBH binary formation on the central structure of galaxies.

\bigskip

Acknowledgments: We thank the anonymous referee for their constructive comments that helped us to improve the paper.
FD is supported by a  PCTS fellowship and
a Lyman Spitzer Jr. fellowship at Princeton University.

\end{document}